\documentclass{article}

\usepackage{amsmath,amsthm,amsfonts}

\usepackage{graphicx}
\usepackage{epstopdf}
\usepackage{rotating}
\usepackage{pdflscape}

\usepackage{multirow}
\usepackage{xcolor}

\usepackage[section]{placeins}
\usepackage[labelfont=bf]{caption, subcaption} 

\usepackage{authblk}

\title{A Multispecies Cross-Diffusion Model for Territorial Development}
\author[1]{Abdulaziz Alsenafi \thanks{abdulaziz.alsenafi@ku.edu.kw }}
\author[2]{Alethea B. T. Barbaro \thanks{a.b.t.barbaro@tudelft.nl}}
\affil[1]{Department of Mathematics, Kuwait University, Kuwait}
\affil[2]{Delft Institute of Applied Mathematics, Faculty of Electrical Engineering, Mathematics and Computer Science, Delft University of Technology, The Netherlands}

\begin{document}

\maketitle

\section{Abstract}

We develop an agent-based model on a lattice to investigate territorial development motivated by markings such as graffiti, generalizing a previously-published model to account for $K$ groups instead of two groups. We then analyze this model and present two novel variations. Our model assumes that agents' movement is a biased random walk and that the agents' movement is away from rival groups' markings. All interactions between agents is indirect, mediated through the markings. We numerically demonstrate that in a system of three groups, the groups segregate in certain parameter regimes. Starting from the discrete model, we also formally derive the continuum system of $2K$ convection-diffusion equations for our model.  These equations exhibit cross-diffusion due to the avoidance of the rival groups' markings. Linear stability analysis is performed to study the phase transition that the system undergoes as parameters are varied.  Two new variations of the model are introduced and numerically studied.  Both exhibit a phase transition, with segregation dynamics that are distinct both from each other and from the original model.\\

\noindent \textbf{Keywords:} phase transition, cross-diffusion, segregation model, pattern formation, agent-based model

\section{Introduction}

Many types of organisms are known to exhibit territorially. Examples include insects, fish, amphibians, reptiles, birds, and mammals \cite{temeles1994role}, and, of course, human beings \cite{sack1986human}.
Even plants could be considered to display this trait \cite{schenk1999spatial}, with some such as \emph{Eucalyptus} excreting a chemical that inhibits the growth of other species \cite{may1990assessment}.  Reasons for territorial behavior include protection of breeding sites and access to resources, and territorial organisms have several ways of claiming their territory.  Two common ways are through some sort of marking, either chemical or physical, and through direct confrontation.

In this paper, we focus on the case of territory formation for a mobile species through territorial markings.  We use the example of gangs of human beings reacting to the graffiti of other gangs, but the model could equally well be applied to other mobile species.  In \cite{MLC1999,MLC2006}, Moorcroft et al. modeled how different packs of animals like coyotes and wolves base their movement on scent marking. It was discovered that both coyotes and wolves use scent marking to tag territories \cite{PM1975}. Once wolves or coyotes encounter foreign scents, they in turn mark their territory with their scent and usually head back to their own home territory. This dynamic was studied in detail in \cite{LWM1997}, where it was found that different packs of wolves can live in the same region without having contact with other packs, but each has its own territory. 

Researchers studying gang dynamics based the gang movement dynamics on existing ecological models of animal species that exhibit territorial behavior. Smith et al. \cite{SBBTV2012} combined the ecological model in \cite{MLC2006} with Hegemann et al.'s network model \cite{GSBB2011} to produce a model for gang territoriality. The new model was then solved numerically, and their results were compared to real data about gang territories in Los Angeles.  In \cite{BCD2013}, Barbaro et al. used a statistical mechanics approach to study how gang territories could be formed based on graffiti. This also drew on ideas from coyote and wolf scent-marking. The authors chose to use a spin system, which is similar to the Ising model \cite{ising} that simulates ferromagnetism. A two-dimensional lattice was used, with an agent and a graffiti spin at each site, and there were only indirect interactions between the agent spins.  The authors showed that their model exhibits a phase transition in which gangs cluster together to form territory.  

Other work on modeling criminal behavior has also been done.  This work is tangential to the work presented here, but is presented for the interested reader. Clustering methods have been used to study gang affiliations between gang members in Los Angeles \cite{GHAELHRVTBB2013}. Network models are also used to study gangs. In \cite{GSBB2011}, the authors present an agent-based model, which is coupled with a rivalry network to explore how gang rivalries are formed. In \cite{SDPTBBC2008}, a model for burglary was developed, and a continuum system consisting of a coupled reaction-diffusion equations was derived. This and similar systems were analyzed in \cite{rodriguez2013global, rodriguez2010local}, and \cite{berestycki2013traveling}.  Modifications of the model were explored in \cite{JBC2010, zipkin2014cops}. The reaction diffusion-diffusion equations were analyzed further in \cite{MW2020}, which showed that they exhibit similar behavior to chemotactic systems with cross-diffusion. Recently, in \cite{WZBS2021}, Wang et al. extended the burglary model by including independent Poison clocks in the time steps; a martingale with both a deterministic and a stochastic part was derived and analyzed. Work on riots and social segregation have also followed from this line of research \cite{berestyki2016analysis, rodriguez2016exploring}. For a more in-depth review of the crime modeling literature, the reader is referred to \cite{d2015statistical}.

Our paper is based on the work of \cite{alsenafi2018convection}, wherein the authors performed a bottom-up approach similar to the Moorcroft model \cite{MLC2006} to produce a discrete system to describe gang territorial development and formally derive from it the following system of convection-diffusion equations:
\begin{equation}
\begin{cases}
\displaystyle \frac{\partial \xi_A}{\partial t}(x,y,t) = \gamma \rho_A(x,y,t) - \lambda \xi_A(x,y,t) \\
\displaystyle \frac{\partial \xi_B}{\partial t}(x,y,t) = \gamma \rho_B(x,y,t) - \lambda \xi_B(x,y,t) \\
\displaystyle \frac{\partial \rho_A}{\partial t}(x,y,t) =  \frac{D}{4} \nabla \cdot \left[ \nabla \rho_A(x,y,t)  + 2  \beta \left(\rho_A(x,y,t) \nabla \xi_B(x, y,t) \right) \right] \\
\displaystyle \frac{\partial \rho_B}{\partial t}(x,y,t) =  \frac{D}{4} \nabla \cdot \left[ \nabla \rho_B(x,y,t)  + 2  \beta \left(\rho_B(x,y,t) \nabla \xi_A(x, y,t) \right) \right],
\end{cases}
\label{E:final_continuum_equations_intro} 
\end{equation}
where $\xi_i$ is the graffiti density of gang $i$ and $\rho_i$ is the agent density from gang $i$. The model undergoes a phase transition from no territorial development to distinct territorial formation as the parameter $\beta$ is changed. This phase transition was found both in the discrete and continuum level. The authors there only considered the case of two gangs. In contrast, in the present work, we will generalize the model and results of \cite{alsenafi2018convection} to consider any finite number of gangs. A modified version of the convection-diffusion system in \cite{alsenafi2018convection} was analyzed in \cite{BRYZ2020}, where they proved a weak stability result and identified equilibrium solutions; interestingly, though, they did not find segregated solutions in this modified system.

This paper's outline is as follows: In Section \ref{S:discrete}, we introduce our extension of the original two-gang agent-based model \cite{alsenafi2018convection}.  The rest of the article is based upon this extension.  We next define an order parameter in Section \ref{S:phases} that will be used to analyze our system's different states and  characterize phase transitions. In Section \ref{S:simulations}, we present the results of a special case of our discrete model numerical simulations as well as show our analysis of the systems' phase transitions. In Sections \ref{section:graffiti_continuum} and \ref{section:agent_continuum}, we will derive the general continuum limit from the discrete model. In Sections \ref{section:steady-state} and \ref{section:LSA}, we will derive a steady-state solution for our continuum model and perform linear stability analysis to determine whenever the well-mixed solution becomes unstable. In Section \ref{section:changing_beta}, we introduce and study two variations on the model, where parameter $\beta$ is made gang-dependent. Finally, in Section \ref{section:conclusion}, we conclude with a discussion of the results and open problems.

\subsection{\label{S:discrete}Discrete Model}

In this paper, we extend and generalize the interacting particle model in \cite{alsenafi2018convection} to now include $K$ gangs as opposed to only two, keeping all other dynamics similar. We shall use a square lattice $S$ of size $L \times L$, with periodic boundary conditions and area $1$. We assume that  we have $K$ gangs, $1, 2, \dots K$, and the number of agents belonging in each gang $j$ is denoted by $N_j$. The systems' total number of agents is denoted by $N$:
\begin{equation*}
N= \sum_{\substack{i=1 \\ i\neq j}}^K N_i
\end{equation*}
These agents are distributed over the lattice. Our model allows multiple gang agents regardless of their gang affiliation to be on the same site. We denote the number of agents of gang $j$ at site $(x,y)$ at time $t$ by $n_{j}(x,y,t)$ and their densities are $\rho_{j}(x,y,t) = \frac{n_{j}(x,y,t)}{l^2}$, where  $l=\frac{1}{L^2}$ is  the lattice spacing. The amount of graffiti belonging to gang $j$  at site $(x,y)$ on time $t$ is denoted by $g_{j}(x,y,t)$. We denote the graffiti density of gang $j$ by  $\xi_{j}(x,y,t) = \frac{g_{j}(x,y,t)}{l^2}$.

Our model assumes that every agent has to move at every time step to one of their four neighboring sites, which are the sites up, down, to the left, and to the right of it. That is, an agent currently occupying site $(x,y)$ would move to an element of the set of sites $\{(x+l,y),(x-l,y),(x,y+l),(x,y-l)\}$. The neighboring sites of $(x,y)$ will be denoted by $(\tilde x, \tilde y) \sim (x,y)$. In our model, each agent performs a biased random walk, trying to avoid the opposing gangs graffiti. Following \cite{alsenafi2018convection}, our model assumes that every agent puts down its own gang's graffiti on the lattice and this graffiti discourages the movement of agents from a different gang onto that lattice site.  However, now that we are considering more than two gangs, each gang must avoid the graffiti of all other gangs, leading us to  define the \emph{opposition sum} for gang $j$ at site $(x,y)$ at time $t$:
\begin{equation}
\psi_j(x,y,t) := \sum_{\substack{i=1 \\ i\neq j}}^K \xi_i(x, y,t).
\label{E:graffiti_complement} 
\end{equation}

We use this opposition sum to inform the movement dynamics of the agents. The probability of an agent from gang $j$  to move from site $s_1=(x_1,y_1) \in S$ to one of the neighboring sites $s_2=(x_2, y_2) \in S$ is defined to be
\begin{equation}
M_{j}(x_1 \rightarrow x_2, y_1 \rightarrow y_2, t) = \frac{e^{-\beta \psi_j(x_2, y_2,t)}}{\sum \limits_{(\tilde x, \tilde y) \sim(x_1,y_1)}e^{-\beta \psi_j(\tilde x, \tilde y, t)}}, \label{E:probability_agent_moves} 
\end{equation}
again defined analogously to \cite{alsenafi2018convection}. Here the parameter $\beta$ encodes the strength of the avoidance of other gangs' graffiti. As our model assumes that all of the agents must move at each time step, it is easily seen that 
\begin{equation}
\sum \limits_{(\tilde x, \tilde y)\sim (x,y)}   M_{j}(x \rightarrow \tilde x, y \rightarrow \tilde y, t)  = 1.
\end{equation}

\noindent The expected density of gang $j$ is therefore
\begin{align}
\rho_{j}(x,y,t + \delta t) = & \rho_{j}(x,y,t) + \sum_{ (\tilde x,\tilde y) \sim ( x, y)}  \rho_{A}(\tilde x,\tilde y, t) M_{j}( \tilde x \rightarrow x,  \tilde y \rightarrow y, t) \notag \\
& -  \rho_{j}(x, y, t)\sum_{ (\tilde x,\tilde y) \sim (x,y)}  M_{j}(x  \rightarrow  \tilde x, y  \rightarrow  \tilde y, t) \notag \\
= & \sum_{ (\tilde x,\tilde y) \sim ( x, y)}  \rho_{A}(\tilde x,\tilde y, t) M_{j}( \tilde x \rightarrow x,  \tilde y \rightarrow y, t).  \label{E:discrete_agents} 
\end{align}

For the graffiti density update rules, each agent adds graffiti at its current  site with probability $\gamma$. It is also assumed that the graffiti decays at every site with a rate of $\lambda$. Both the graffiti addition and decay are scaled by the time step $\delta t$.  Therefore, the graffiti density at site $(x, y) \in S$ at time $t + \delta t$ is
\begin{equation}
\xi_{j}(x,y,t+\delta t)=\xi_{j}(x,y,t) - \left( \delta t \cdot \lambda \right) \xi_{j}(x,y,t)  + \left( \delta t \cdot \gamma \right) \rho_{j}(x,y,t).
 \label{E:discrete_graffiti}
\end{equation}

In all of our simulations, we initially randomly distribute the agents' locations using the multivariate uniform distribution on the lattice $S$.  We also assume that the lattice is initially empty of graffiti.

\subsection{\label{S:phases}Phases and an Order Parameter}
In our simulations, we shall observe two phases, the well-mixed phase and the segregated phase.  These phases, we will see, are determined by parameter $\beta$, introduced in \eqref{E:probability_agent_moves}. In the well-mixed phase, the agents are distributed randomly throughout the lattice and their movement approximates a random walk. However, for the segregated phase, the agents' movement is a biased random walk, and the agents form territories by clustering together. In this section, we shall 
define an order parameter and use it to quantify these different phases.

\subsubsection{\label{S:Expected_density}Expected Agent Density}

We first compute the expected agent density for the well-mixed state at site $(x,y)$ for each gang $j$. In this phase, the agents from each gang are uniformly spread over the whole lattice $S$.  Thus, the expected agent density for gang $j$ at any given site is:
\begin{align}
E\left(\rho_j\right) &= \sum_{(x, y) \in S}  \rho_j(x,y)\times \frac{1}{L^2} \notag \notag \\   
   &= \sum_{(x, y) \in S}  \frac{n_j(x,y)}{l^2} \times \frac{1}{L^2}\notag \\
   &= \sum_{(x, y) \in S}  n_j(x,y) \notag \\
&= N_j.  \label{E:well-mixed_energy_approximation}
\end{align}

For the segregated phase, the agents are entirely separated into different territories, each occupied by a distinct gang. To determine the expected agent density for the segregated phase, we will need the following definitions and assumptions. We define the territory $S_j$ to be the set of all sites that are dominated by gang $j$ agents, i.e. all sites which have more gang $j$ agents than agents of another type. We also note that $S_j$ need not be connected. We also define the area of sublattice $S_j$ by $R_j$, which is the area dominated by gang $j$.  We assume that the agents from each gang are uniformly distributed in their territory, and that all sites are occupied by agents; hence, every site is assumed to contain agents from exactly one gang. These assumptions are validated in our simulation in Section \ref{S:simulations}. Accordingly, $\{S_j\}_{j=1}^{K}$ form a partition for the lattice $S$ for $j=1,2,\dots,K$.

Under these assumptions, we now calculate the expected agent density for the agent density  for gang $j$ in the segregated state by splitting the lattice $S$ into two disjoint territories, $S_j$ and its complement ${S_j}^c$. Calculating the expected agent density within the complement of gang $j$'s territory easily finds:
\begin{align*}
E\left(\rho_j\right)  &=  \sum_{(x,y)\in {S_j}^c} \frac{\rho_j(x,y)}{{R_j}^c} = 0 
\end{align*}
This last equality follows because all agents for gang $j$ are assumed to be in $S_j$ in the (perfectly) segregated phase, hence none are in ${S_j}^c$. Next, calculating the expected agent density within $S_j$ gives us:
\begin{align*} 
E\left(\rho_j\right) &= \sum_{(x,y)\in S_j} \frac{\rho_j(x,y)}{R_j}\\
&= \sum_{(x,y)\in S_j} \frac{n_j(x,y)}{l^2 \times R_j} \\
&= \frac{N_j}{l^2 R_j} 
\end{align*}
Thus, the expected agents density within $S_j$ is
\begin{align}
E\left(\rho_j(x,y)\right)&= \begin{cases} 
      \frac{N_j}{l^2 R_j}, & (x,y) \in S_j \\
      0, & (x,y) \in {S_j}^c.
\end{cases}
\label{E:segregated_energy_approximation}
\end{align}
If we further assume that the areas dominated by each gang are almost equal and that there are an equal number of agents in each gang, we deduce that $R_j=\frac{1}{K}$, where $K$ is the number of gangs. Under these assumptions, the expected density in a segregated state for an agent from gang $j$ is 
\begin{align*}
E \left( \rho_j(x,y) \right) &= \begin{cases} 
      \frac{N_j}{l^2 R_j}, & (x,y) \in S_j \\
      0, & (x,y) \in {S_j}^c
 	\end{cases}\\
      &= \begin{cases}
      \frac{K N_j}{l^2}, & (x,y) \in S_j \\
      0, & (x,y) \in {S_j}^c. 
\end{cases}
\end{align*}

\subsubsection{\label{S:order_parameter}An Order Parameter}
To investigate the phase transition, we define the following order parameter:
\begin{align} 
\mathcal{E}(t) =  ~\frac{1}{4(K-1)}\left(\frac{1}{LN}\right)^2\sum_{j=1}^K ~~\sum_{i>j}^K \sum_{(x, y) \in S} \sum_{ (\tilde x,\tilde y) \sim ( x, y)} & [ \left(\rho_j(x,y,t) - \rho_i(x,y,t)\right) \times \notag \\
&\left(\rho_j(\tilde x, \tilde y, t) - \rho_i(\tilde x, \tilde y, t) \right)].\label{E:Energy_Equation}
\end{align}  
This order parameter is modeled after the order parameter in \cite{alsenafi2018convection} and the Hamiltonian function for the Ising Model \cite{baxter2007exactly, ising}. In terms of our model, our order parameter becomes more positive if a site and its neighbors are dominated by the same gang and becomes more negative when a site and its neighbors are dominated by different gangs.  It is approximately zero if none of the gangs are dominating territory. 

This is due to the fact that in the segregated, phase the agents cluster together and this leads to there being only one gang present at site $(x,y)$. This makes the term inside the sum in equation \eqref{E:Energy_Equation} to have a large magnitude; if the same is true at the neighboring site, the second term inside the sum is identical and once we multiply the two together, the resulting value would be a positive number. However, whenever agents from all gangs become uniformly distributed throughout the lattice, this results that the two sets of parenthesis tend to be very small, and sometimes positive and sometimes negative. Thus, after summing over the whole lattice and all the gangs, the order parameter ends up near zero.

We now calculate an approximation for the order parameter when the phases are well-mixed and when they are segregated. For simplicity, in this subsection and all of our simulations, we consider the special case of three gangs and we assume that the number of agents in each gang is equal, so that $N_j = \frac{N}{3}$ for $j$ in $\{1, 2, 3\}$. The order parameter for this special case is
\begin{align}
\mathcal{E} = \frac{1}{8}\left(\frac{1}{LN}\right)^2 \sum_{(x, y) \in S} \sum_{ (\tilde x,\tilde y) \sim ( x, y)}   \biggr[ & \left(\rho_1(x,y) -\rho_2(x,y)\right) \left(\rho_1(\tilde x, \tilde y) - \rho_2(\tilde x, \tilde y) \right) \notag \\ 
+& \left(\rho_1(x,y) - \rho_3(x,y)\right) \left(\rho_1(\tilde x, \tilde y) - \rho_3(\tilde x, \tilde y) \right) \notag \\
+& \left(\rho_2(x,y) - \rho_3(x,y)\right) \left(\rho_2(\tilde x, \tilde y) - \rho_3(\tilde x, \tilde y) \right) \biggr]. \label{E:Energy_Equation_3gangs}
\end{align}

In the well-mixed state, based on equation \eqref{E:well-mixed_energy_approximation} and on our assumptions that the agents from all gangs are uniformly distributed and that each lattice site has four neighbors, our equation simplifies to
\begin{align*}
\mathcal{E} = \frac{1}{8}\left(\frac{1}{LN}\right)^2 \sum_{(x, y) \in S}   \biggr[ & 4\left(N_1 -N_2\right) \left(N_1 - N_2 \right) +4\left(N_1 - N_3\right) \left(N_1 - N_3 \right) \\
 +&4\left(N_2 - N_3\right) \left(N_2 - N_3 \right) \biggr] 
\end{align*} 
Simplifying the terms in the brackets  yields that
\begin{align*}
\mathcal{E} & = \frac{4}{8}\left(\frac{1}{LN}\right)^2 \sum_{(x, y) \in S}   \biggr[\left(N_1 -N_2\right)^2 +\left(N_1 - N_3\right)^2  +\left(N_2 - N_3\right)^2 \biggr] \\
&= \frac{1}{2}\left(\frac{1}{N}\right)^2   \biggr[\left(N_1 -N_2\right)^2 +\left(N_1 - N_3\right)^2  +\left(N_2 - N_3\right)^2 \biggr]. 
\end{align*}
However, since we assumed that the number of agents from each gang $N_1,N_2,$ and $N_3$ are equal, it follows easily that the order parameter for the agents in a well-mixed phase is
\begin{equation}
\mathcal{E} \approx 0.
\label{E:Energy_approx_well-mixed}
\end{equation}

We will next calculate the order parameter for the segregated phase. Here, we assume a perfectly segregated phase and split the lattice $S$ into the three regions $S_1$, $S_2$ and $S_3$ belonging to each gang, which gives us:
\begin{align*}
\mathcal{E} = \frac{1}{8}\left(\frac{1}{LN}\right)^2 \Biggr[ \sum_{(x, y) \in S_1} \sum_{ (\tilde x,\tilde y) \sim ( x, y)}   \biggr[ & \left(\rho_1(x,y) -\rho_2(x,y)\right) \left(\rho_1(\tilde x, \tilde y) - \rho_2(\tilde x, \tilde y) \right) \\ 
+& \left(\rho_1(x,y) - \rho_3(x,y)\right) \left(\rho_1(\tilde x, \tilde y) - \rho_3(\tilde x, \tilde y) \right) \\
+& \left(\rho_2(x,y) - \rho_3(x,y)\right) \left(\rho_2(\tilde x, \tilde y) - \rho_3(\tilde x, \tilde y) \right) \biggr].
\end{align*}
\begin{align*}
+ \sum_{(x, y) \in S_2} \sum_{ (\tilde x,\tilde y) \sim ( x, y)}   \biggr[ & \left(\rho_1(x,y) -\rho_2(x,y)\right) \left(\rho_1(\tilde x, \tilde y) - \rho_2(\tilde x, \tilde y) \right) \\ 
+& \left(\rho_1(x,y) - \rho_3(x,y)\right) \left(\rho_1(\tilde x, \tilde y) - \rho_3(\tilde x, \tilde y) \right) \\
+& \left(\rho_2(x,y) - \rho_3(x,y)\right) \left(\rho_2(\tilde x, \tilde y) - \rho_3(\tilde x, \tilde y) \right) \biggr]
\end{align*}
\begin{align*}
+ \sum_{(x, y) \in S_3} \sum_{ (\tilde x,\tilde y) \sim ( x, y)}   \biggr[ & \left(\rho_1(x,y) -\rho_2(x,y)\right) \left(\rho_1(\tilde x, \tilde y) - \rho_2(\tilde x, \tilde y) \right) \\ 
+& \left(\rho_1(x,y) - \rho_3(x,y)\right) \left(\rho_1(\tilde x, \tilde y) - \rho_3(\tilde x, \tilde y) \right) \\
+& \left(\rho_2(x,y) - \rho_3(x,y)\right) \left(\rho_2(\tilde x, \tilde y) - \rho_3(\tilde x, \tilde y) \right) \biggr] \Biggr]
\end{align*}
Using equation \eqref{E:segregated_energy_approximation}, we substitute the expectation of each $\rho_i$ for each region, which gives us the following approximation:
\begin{align*}
\mathcal{E} \approx \frac{1}{8}\left(\frac{1}{LN}\right)^2 \Biggr[ & \sum_{(x, y) \in S_1}  \biggr[  4\left(\frac{N_1}{l^2 R_1}\right)^2 + 4\left(\frac{N_1}{l^2 R_1}\right)^2 \biggr] \\ 
+& \sum_{(x, y) \in S_2}  \biggr[ 4\left(\frac{-N_2}{l^2 R_2}\right)^2 + 4\left(\frac{N_2}{l^2 R_2}\right)^2 \biggr]  \\
+& \sum_{(x, y) \in S_3}  \biggr[  4\left(\frac{-N_3}{l^2 R_3}\right)^2  + 4\left(\frac{-N_3}{l^2 R_3}\right)^2 \biggr] \Biggr].
\end{align*}
Further simplifying yields:
\begin{align*}
\mathcal{E} & \approx \frac{1}{8}\left(\frac{1}{LN}\right)^2 \left(\frac{8}{l^4} \right)\left[  \sum_{(x, y) \in S_1} \left(\frac{N_1}{ R_1}\right)^2  + \sum_{(x, y) \in S_2} \left(\frac{N_2}{ R_2}\right)^2  + \sum_{(x, y) \in S_3} \left(\frac{N_3}{ R_3}\right)^2   \right] \\
&= \left(\frac{1}{LN}\right)^2 \left(\frac{1}{l^4} \right) \left[  R_1 \left(\frac{N_1}{ R_1}\right)^2  + R_2 \left(\frac{N_2}{ R_2}\right)^2  + R_3 \left(\frac{N_3}{R_3}\right)^2 \right] \\
&= \left(\frac{L}{N}\right)^2 \left[ \frac{N_1^2}{R_1} + \frac{N_2^2}{R_2} + \frac{N_3^2}{R_3} \right].
\end{align*}
By further assuming that all gangs have the same number of agents $N_i=N/3$ and that the regions have the same area size $R_i=\frac{L^2}{3}$, the previous equation can be further simplified to
\begin{equation*}
\mathcal{E}  \approx \left(\frac{L}{N}\right)^2 \left( \frac{\left( N/3 \right)^2}{L^2/3} +  \frac{\left( N/3 \right)^2}{L^2/3} +\frac{\left( N/3 \right)^2}{L^2/3} \right) 
\end{equation*}
By further simplifying the equation, we easily see that
\begin{equation}
\mathcal{E}  \approx 1. \label{E:Segregated_energy_approx}
\end{equation}
Therefore, the order parameter for the perfectly segregated system is approximately equal to one.


\section{\label{S:simulations}Simulations of the Discrete Model} 
We now will present the results of the simulations of our discrete model. For simplicity, unless otherwise in our simulations we assume we only have $three$ gangs $1, 2$ and $3$, and that all gangs are assumed to have $50,000$ agents.  We shall also assume that the lattice size $L\times L$ is $100 \times 100$ with lattice spacing	2 $l=1$, and we will use $100,000$ time steps with each step size $\delta t =1$.

\subsection{\label{S:well-mixed}Well-Mixed State} 
\begin{figure}[!htb]
        \begin{subfigure}[b]{0.249\linewidth}
               \includegraphics[width=2cm,,keepaspectratio]{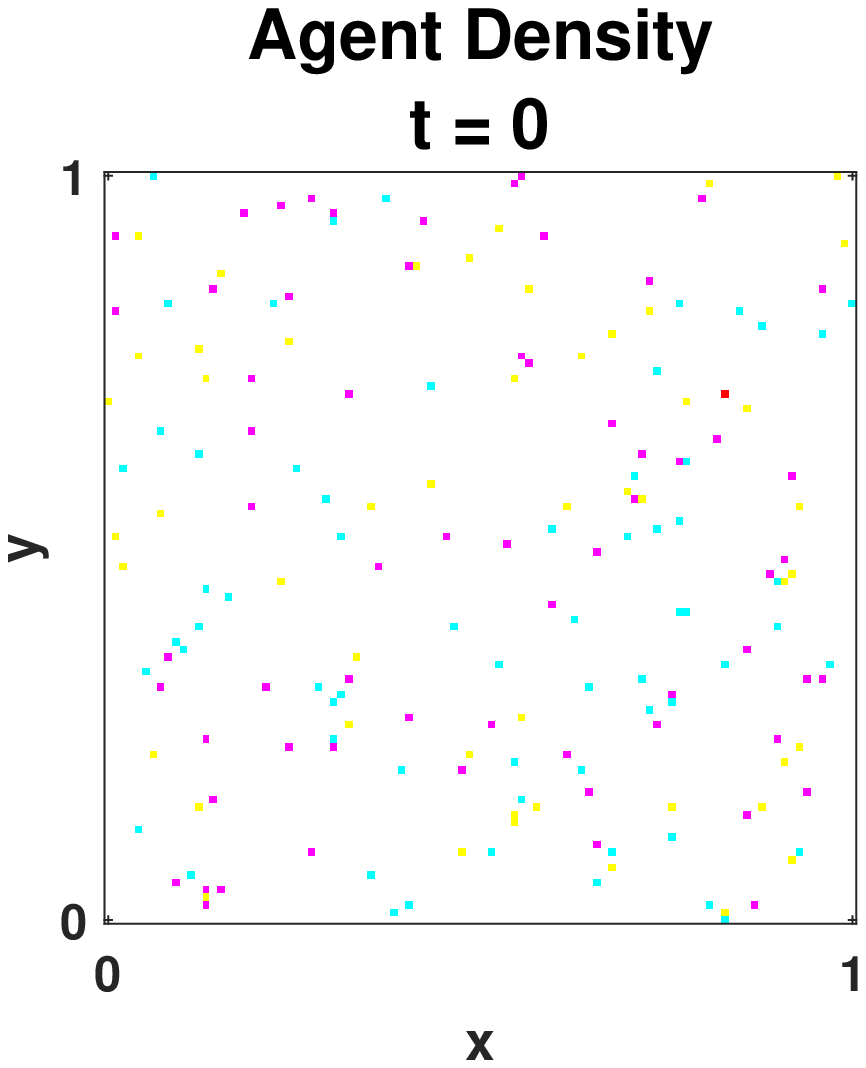}
        \end{subfigure}%
        \begin{subfigure}[b]{0.249\linewidth}
                \includegraphics[width=2cm,,keepaspectratio]{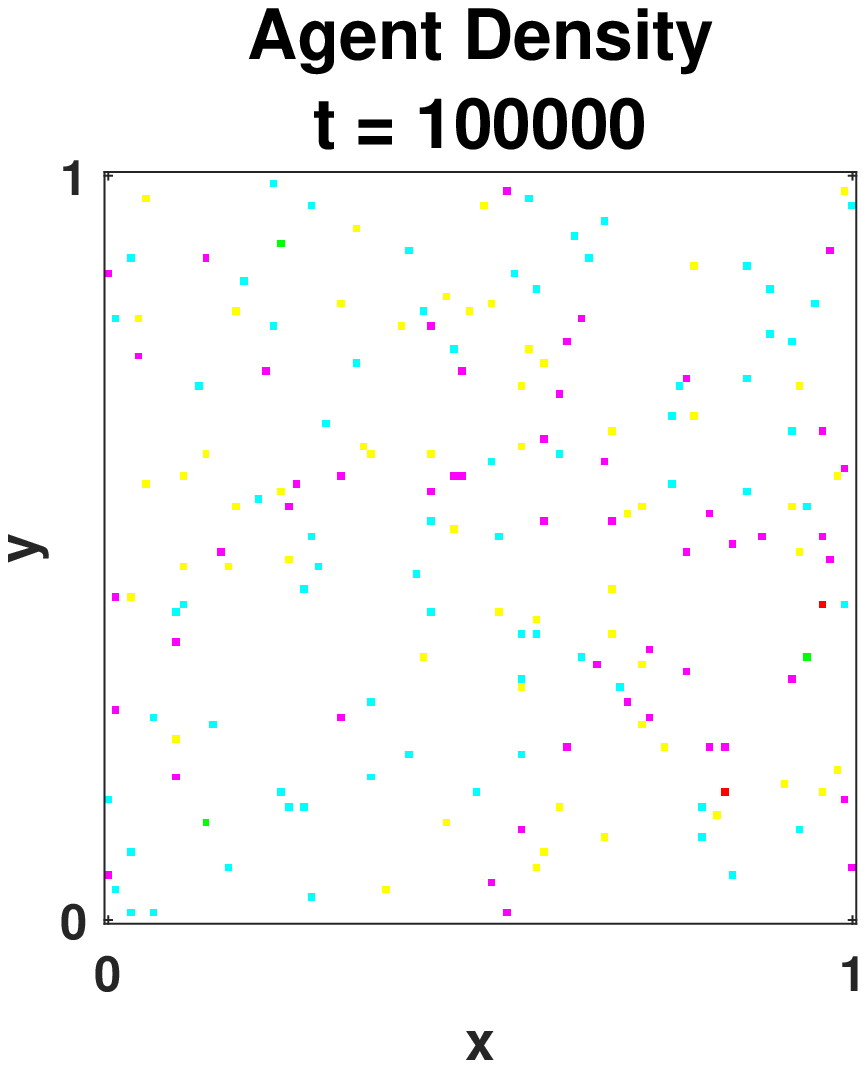}
        \end{subfigure}%
        \begin{subfigure}[b]{0.249\linewidth}
              \includegraphics[width=2cm,,keepaspectratio]{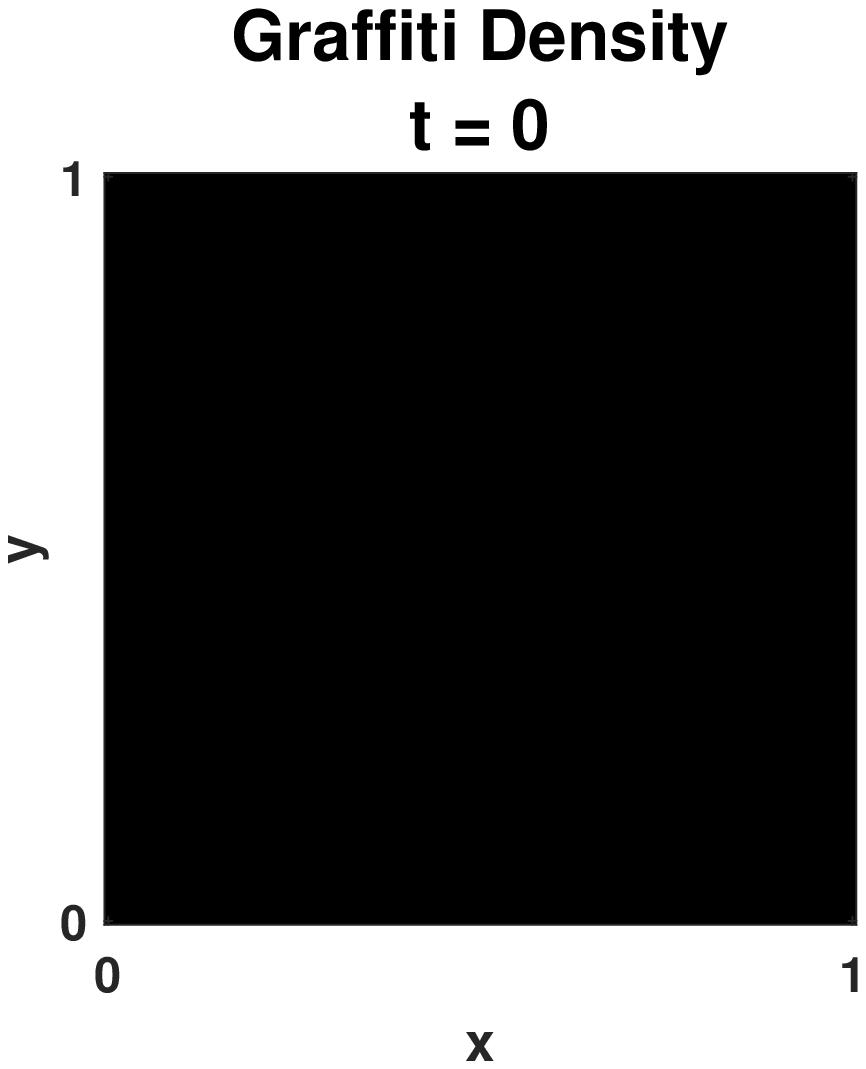}
        \end{subfigure}%
        \begin{subfigure}[b]{0.249\linewidth}
               \includegraphics[ width=2cm,keepaspectratio]{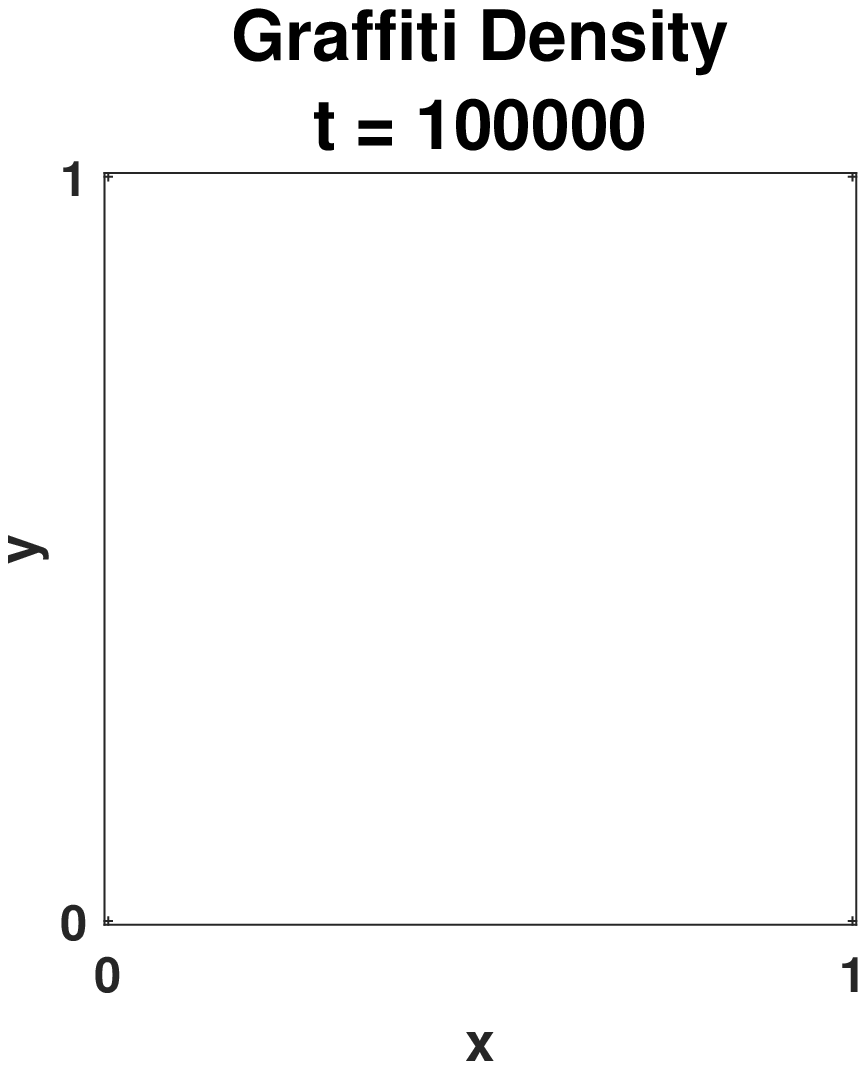}
        \end{subfigure}
                \caption[Temporal evolution of the agent and graffiti densities lattice for a well-mixed state.]{Agent (left two) and graffiti (right two) densities'  temporal evolution for a well-mixed state. Here we have $N_1 = N_2 = N_3 =50,000$, with $\lambda = \gamma =0.5$, $\beta = 5 \times 10^{-6}$, $\delta t = 1$ and the lattice size is $100 \times 100$. Note that the initial graffiti lattice appears black because it is empty.  The final graffiti lattice appears white because all sites have (almost) the same graffiti densities from all three gangs. It is clear from this figure that the agents remain well mixed over time.}
        \label{fig:Well-Mixed_Simulations}
\end{figure}

We start our simulations with $\beta = 5 \times 10^{-6}$ and the resulting lattice simulations are visualized in Figure \ref{fig:Well-Mixed_Simulations}.
The first two lattices in Figure \ref{fig:Well-Mixed_Simulations} represent the time evolution of agent density, whereas the last two lattices represent the graffiti density over time. We assign the colors red, green and blue for gangs $1$, $2$ and $3$ respectively. The color white is used if there are the same number of agents or the same amount of graffiti from all gangs at a site.  The colors cyan, magenta and yellow are used if a site has two gangs (blue and green, blue and red, or red and green, respectively). Finally, if the site is empty, then it will be assigned the color black.

From Figure \ref{fig:Well-Mixed_Simulations}, we clearly see that the gangs remain well mixed over time for  $\beta = 5 \times 10^{-6}$. We do not see any patterns being formed for the graffiti, with the initial graffiti lattice black and the final graffiti lattice white, and the gang agents' movement is in essence a two-dimensional random walk, hardly taking the opposing gangs graffiti into consideration due to the low $\beta$ value. This is due to the way the agents are allowed to move in equation \eqref{E:probability_agent_moves}, where a very small $\beta$ values give the agent a probability of nearly $0.25$ to move to one of the four neighboring sites. 

\subsection{\label{S:Segregated}Segregated State} 
\begin{figure}[!htb]
        \begin{subfigure}[b]{0.33\linewidth}
               \includegraphics[width=2.5cm,,keepaspectratio]{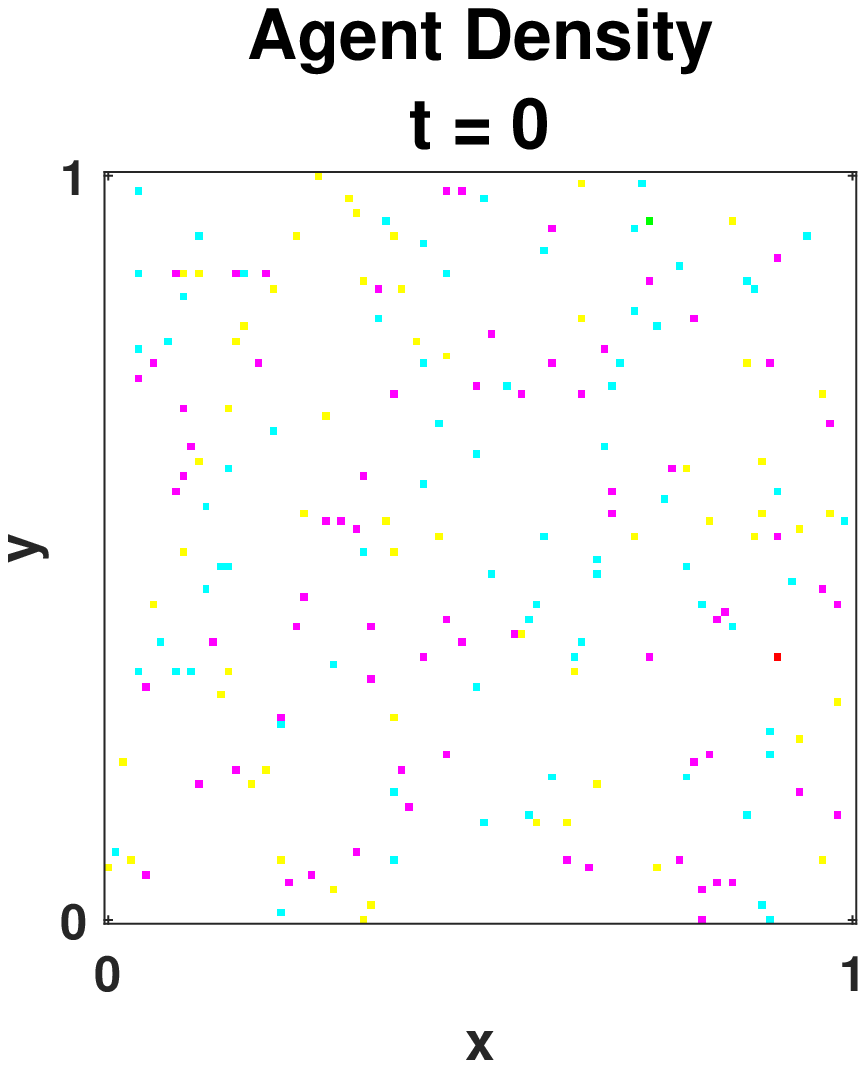}
        \end{subfigure}%
        \begin{subfigure}[b]{0.33\linewidth}
                \includegraphics[width=2.5cm,,keepaspectratio]{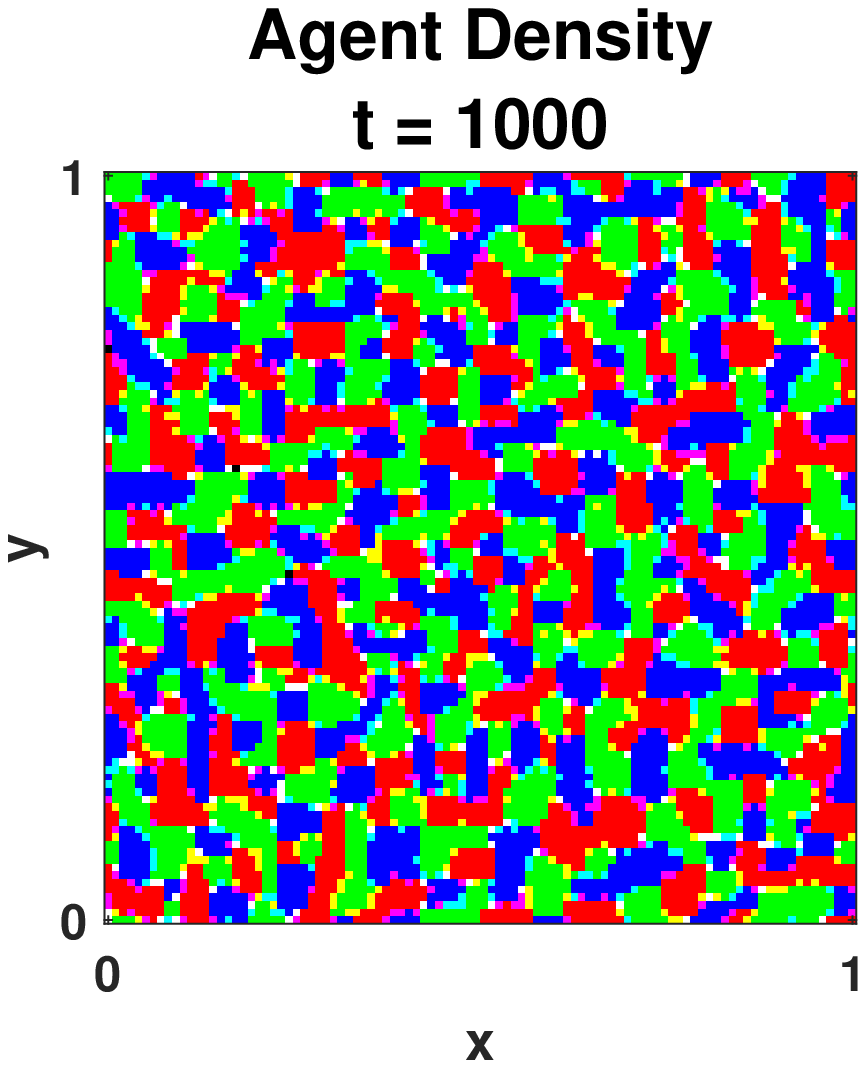}
        \end{subfigure}%
        \begin{subfigure}[b]{0.33\linewidth}
            \includegraphics[ width=2.5cm,keepaspectratio]{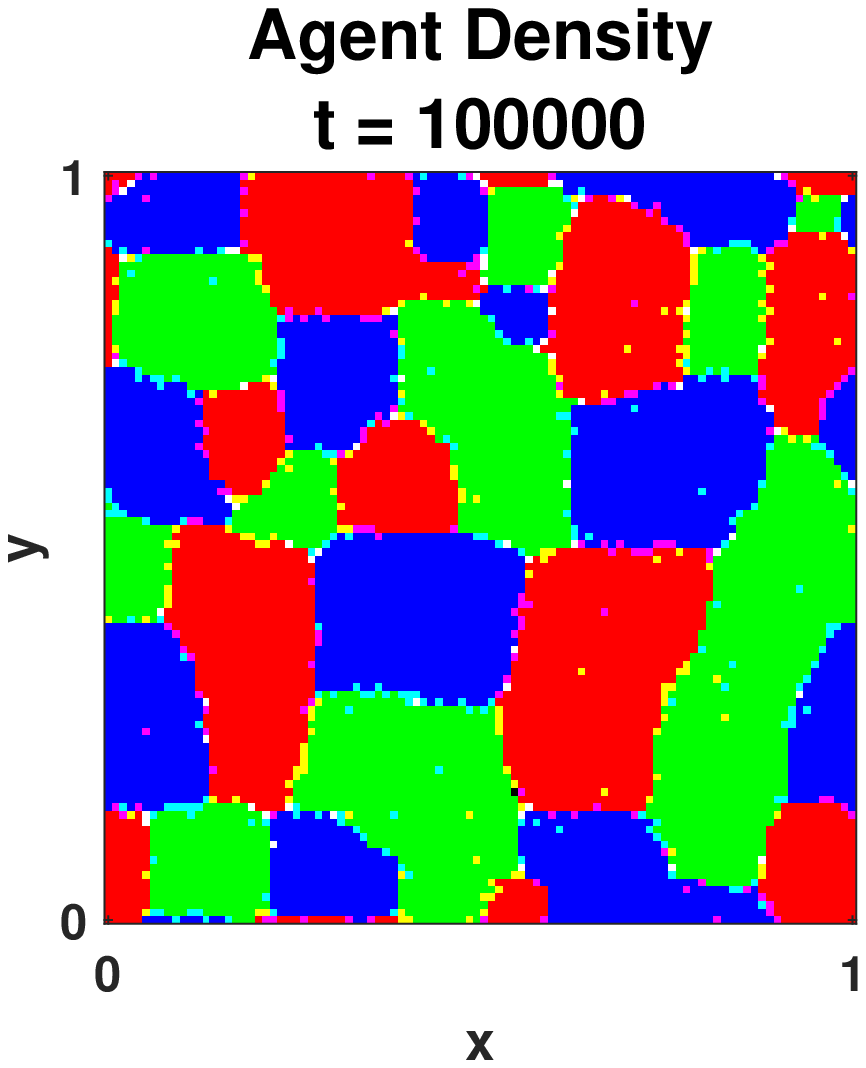}
        \end{subfigure}  
        \begin{subfigure}[b]{0.33\linewidth}
              \includegraphics[width=2.5cm,,keepaspectratio]{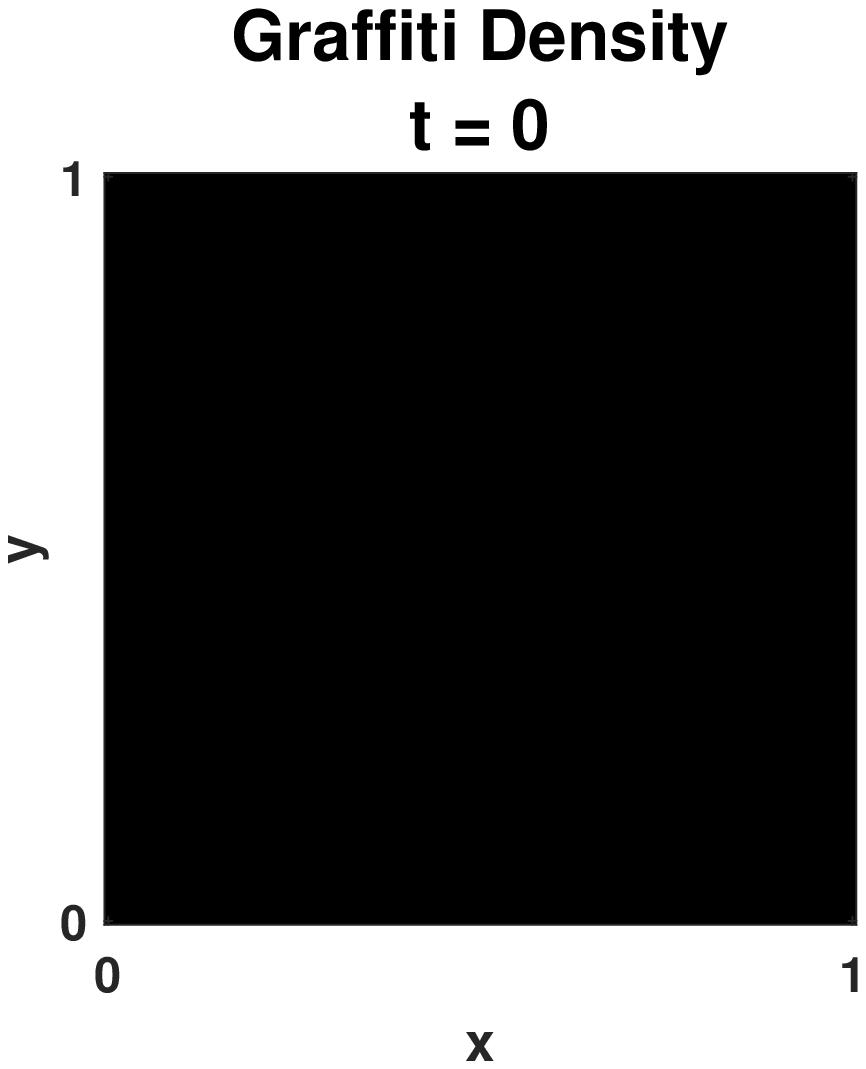}
        \end{subfigure}%
        \begin{subfigure}[b]{0.33\linewidth}
               \includegraphics[ width=2.5cm,keepaspectratio]{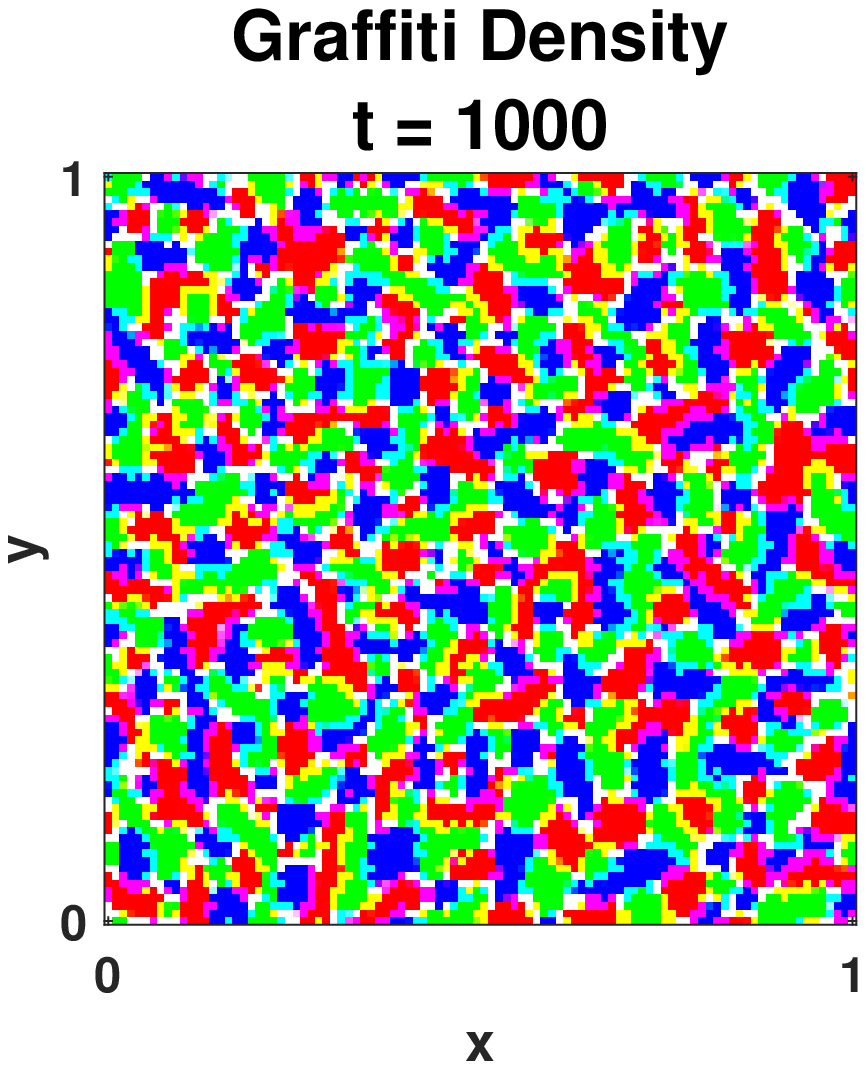}
        \end{subfigure}
        \begin{subfigure}[b]{0.33\linewidth}
                \includegraphics[ width=2.5cm,keepaspectratio]{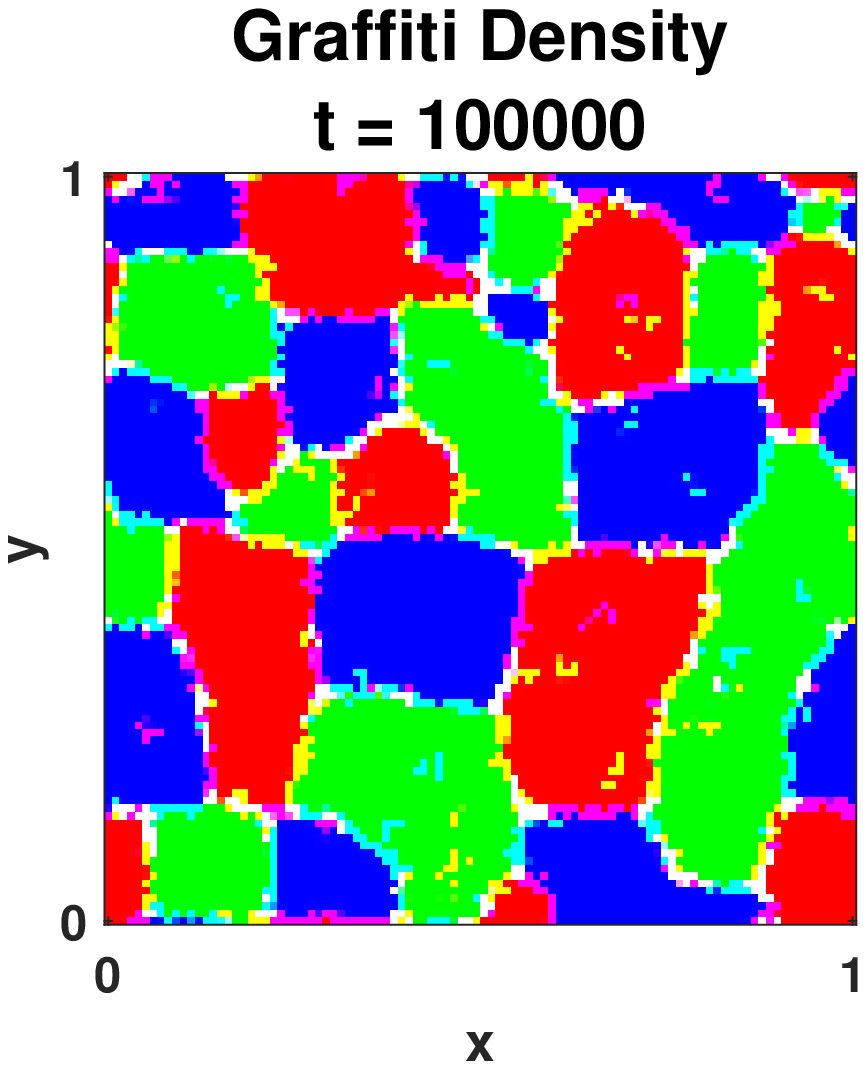}
        \end{subfigure}   
                \caption[Temporal evolution of the agent and graffiti densities lattice for a segregated state.]{Agent (top) and graffiti (bottom) densities temporal evolution  for a segregated state. Here ,we have $N_1 = N_2 = N_3 =50,000$, with $\lambda = \gamma =0.5$, $\beta = 3 \times 10^{-5}$, $\delta t = 1$ and the lattice size is $100 \times 100$. We see that the agents segregate into distinct territories, coarsening over time.}
        \label{fig:Segregated_Simulations}
\end{figure}

The value of $\beta$ is now increased so that it is equal to $3 \times 10^{-5}$ and we keep all other parameters the same. The resulting lattice is visualized in Figure \ref{fig:Segregated_Simulations}. The top row illustrates the time evolution of agent density, whereas the bottom row shows the graffiti density over time.
From Figure \ref{fig:Segregated_Simulations}, we see that initially the agents are well-mixed. However, as time evolves, we see that agents from each gang cluster together to form all-red, all-green and all-blue territories. As time increases, the patterns in both the agent and graffiti densities coarsen. From the same figure, we clearly see that the graffiti density is similar to the agents density and the agents movements in the state is based on the other gangs graffiti. That is, in this state, the $\beta$ value is large enough that the agents are reacting to the opposing gang graffiti and the agents movement is no longer an unbiased random walk.  We can also observe from this figure that the areas with more than one gang's graffiti lie at the boundaries of the territories dominated by each gang. Similarly, this is where we observe the agents overlapping, though to a lesser extent.  Presumably, this overlap enables the coarsening see in the figure.

\subsection{\label{S:Phase Transitions}System Parameters and the Discrete Phase Transition} 
\subsubsection{Effects of $\beta$} 
\begin{figure}[!htb]
        \begin{subfigure}[b]{0.495\linewidth}
               \includegraphics[width=5.0cm,,keepaspectratio]{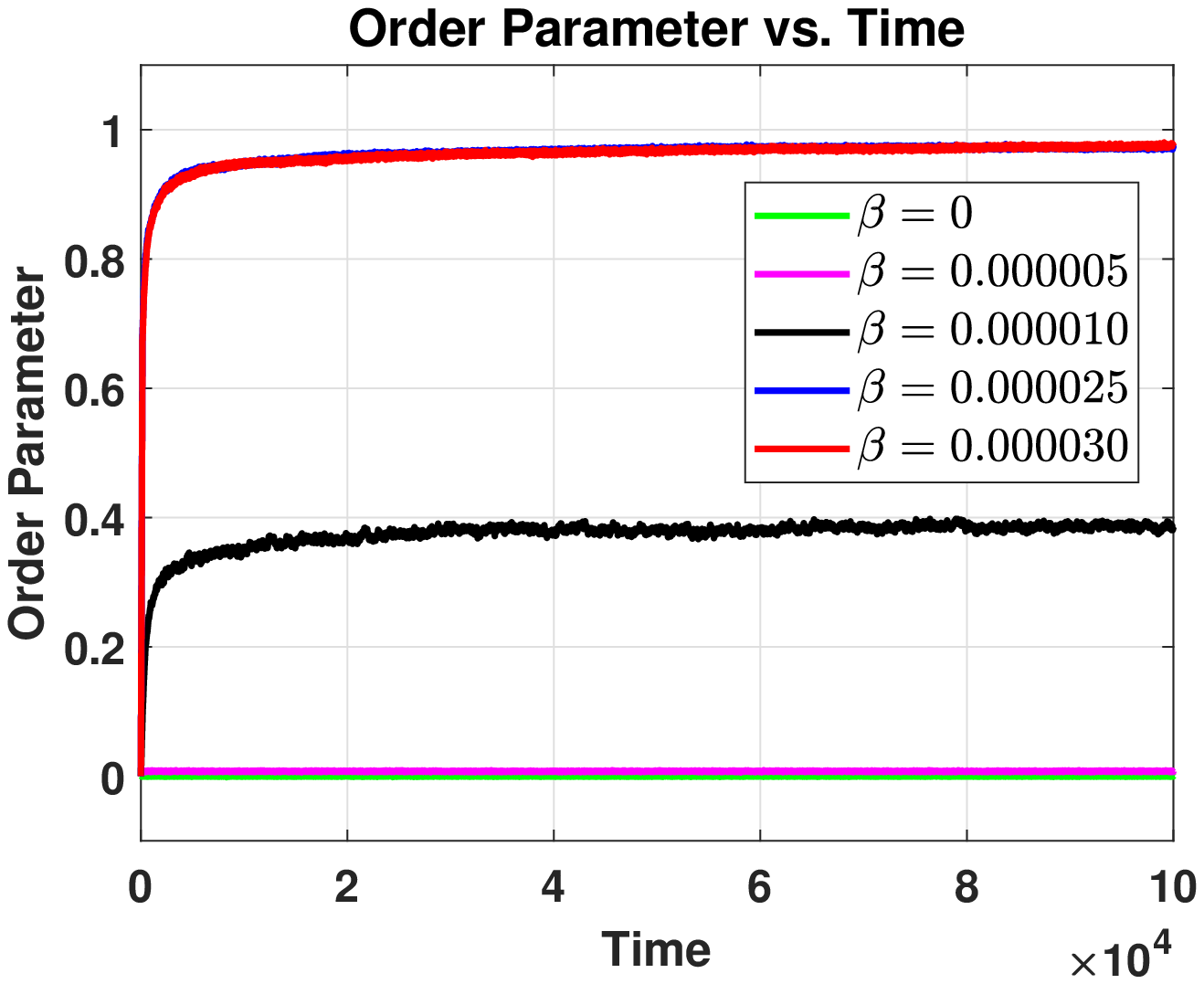}
        \end{subfigure}%
        \begin{subfigure}[b]{0.495\linewidth}
                \includegraphics[width=5.0cm,,keepaspectratio]{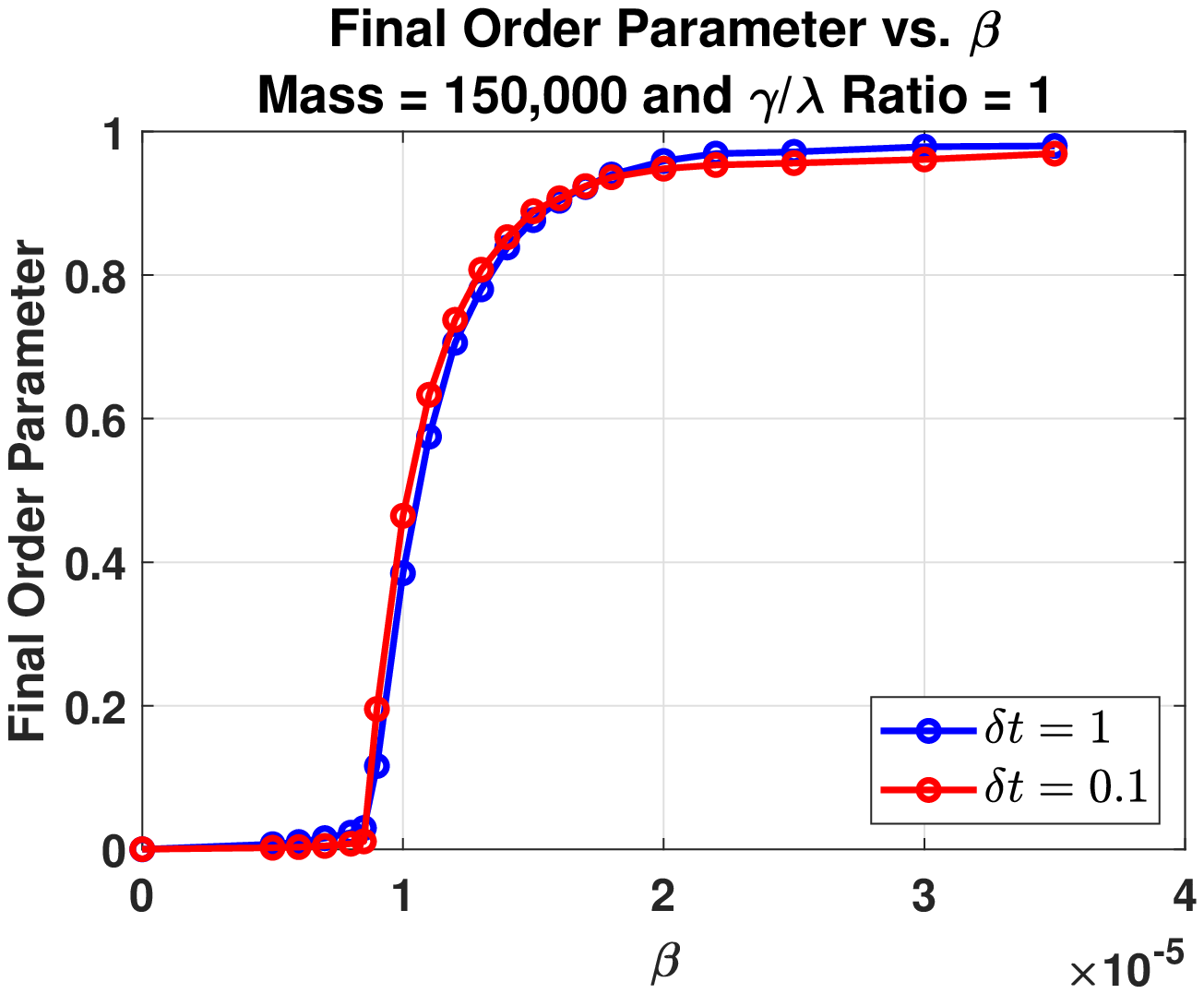}
        \end{subfigure}%
                \caption[Order Parameter.]{How changing the $\beta$ parameter affects the system. Here we have $N_1 = N_2 = N_3 =50,000$, with $\lambda = \gamma =0.5$ and the lattice size is $100 \times 100$. (Left) It is seen that for a small $\beta$ value the system remains well-mixed and the order parameter is almost zero over all time steps. For larger $\beta$ values, we see that the order parameter increases quickly as the system segregates and levels off to around one. (Right) At the final time step, we take the order parameter value for different $\beta$ values. We clearly see that as the $\beta$ value increases there is a critical $\beta$ value at which a phase transition occurs.}
        \label{fig:order_parameter}
\end{figure}

In Sections \ref{S:well-mixed} and \ref{S:Segregated}, we saw that changing the value of the parameter $\beta$ could lead to a phase transition. In order for us to study the phase transitions, we use the concept of order parameter that we introduced in Section \ref{S:order_parameter}. In equation \eqref{E:Energy_Equation_3gangs}, we defined an order parameter for a system of three gangs to be 
\begin{align}
\mathcal{E} = \frac{1}{8}\left(\frac{1}{LN}\right)^2 \sum_{(x, y) \in S} \sum_{ (\tilde x,\tilde y) \sim ( x, y)}   \biggr[ & \left(\rho_1(x,y) -\rho_2(x,y)\right) \left(\rho_1(\tilde x, \tilde y) - \rho_2(\tilde x, \tilde y) \right) \notag \\ 
+& \left(\rho_1(x,y) - \rho_3(x,y)\right) \left(\rho_1(\tilde x, \tilde y) - \rho_3(\tilde x, \tilde y) \right) \notag \\
+& \left(\rho_2(x,y) - \rho_3(x,y)\right) \left(\rho_2(\tilde x, \tilde y) - \rho_3(\tilde x, \tilde y) \right) \biggr]. \label{E:Energy_Equation_3gang}
\end{align}
This order parameter is defined to have a low value for a well-mixed phase and high value for a segregated phase; in Section \ref{S:order_parameter}, we saw that the order parameter $\mathcal{E} \approx 0$ for a well mixed state and $\mathcal{E} \approx 1$ for a fully segregated state. For our simulations, we graphed the order parameter over the course of the simulation for different values of $\beta$, visualizing the output in Figure \ref{fig:order_parameter}.

We see in the left plot in Figure \ref{fig:order_parameter}, the time evolution of the order parameter for different $\beta$ values. Here, we easily see that given enough time steps, the order parameter levels off to a certain value, presumably its asymptotic value. We see that for $\beta=0$ and $\beta=0.000005$, the order parameter remained approximately zero throughout all time steps. This is expected as the system remains well-mixed for these relatively small $\beta$ values. However, we see that once we increase the values of $\beta$, then the order parameter starts to increase. For instance, if $\beta = 0.000025$ or $\beta = 0.000030$, then the order parameter increases fairly quickly in the first $10,000$ time steps before leveling off to just under the fully-segregated value of $1$ for the remaining time steps. This shows us that for these relatively large $\beta$ values, the system segregates fairly quickly and remains segregated throughout the simulation. Finally, we also see that if we choose $\beta = 0.00001$ then the order parameter does increase and the system does exhibit some segregation, but this is not perfect segregation as the order parameter levels off to around $0.4$.

It is evident from the left plot of Figure \ref{fig:order_parameter} that there is a critical $\beta$ in which the system undergoes a phase transition. We define the critical $\beta$ to be the value where the order parameter is equal to $0.01$, and denote it by $\beta^*$. To find the value of $\beta^*$, we take the final value of the order parameter and plot it against different $\beta$ values. The output is then visualized on the right plot of Figure \ref{fig:order_parameter}. From that plot, we can see that the phase transition occurs when $\beta^* \in (0.000005, 0.000006)$.

\subsubsection{Effects of Other Parameters}

In order to investigate how other parameters such as system mass, time step, lattice size, graffiti rate and decay rate affect the system phase transition, we vary one parameter at a time while keeping all other parameters fixed. This is important since in the derivation of the continuum equations for our system, we will assume that both the time step $\delta t$ and the lattice spacing $l$ approach zero. It is therefore essential to know if a finer grid affects our discrete model as opposed to a coarser grid. We also would like to know if taking smaller or bigger time steps might affect the rate of segregation and if it has any effect on the phase transition.

We begin by studying how the time step might affect the system. To do that, we keep all our system parameters constant and decrease the time step $\delta t$ from $1$ to $0.1$; we then plot the final order parameter value for different $\beta$ values. The results are visualized on the right plot in Figure \ref{fig:order_parameter}. In the plot, it is clear that the smaller time step does not affect the rate of segregation, nor does it affect where the phase transition occurs.

\begin{figure}[!htb]
        \begin{subfigure}[b]{0.33\linewidth}
               \includegraphics[width=3.0cm,,keepaspectratio]{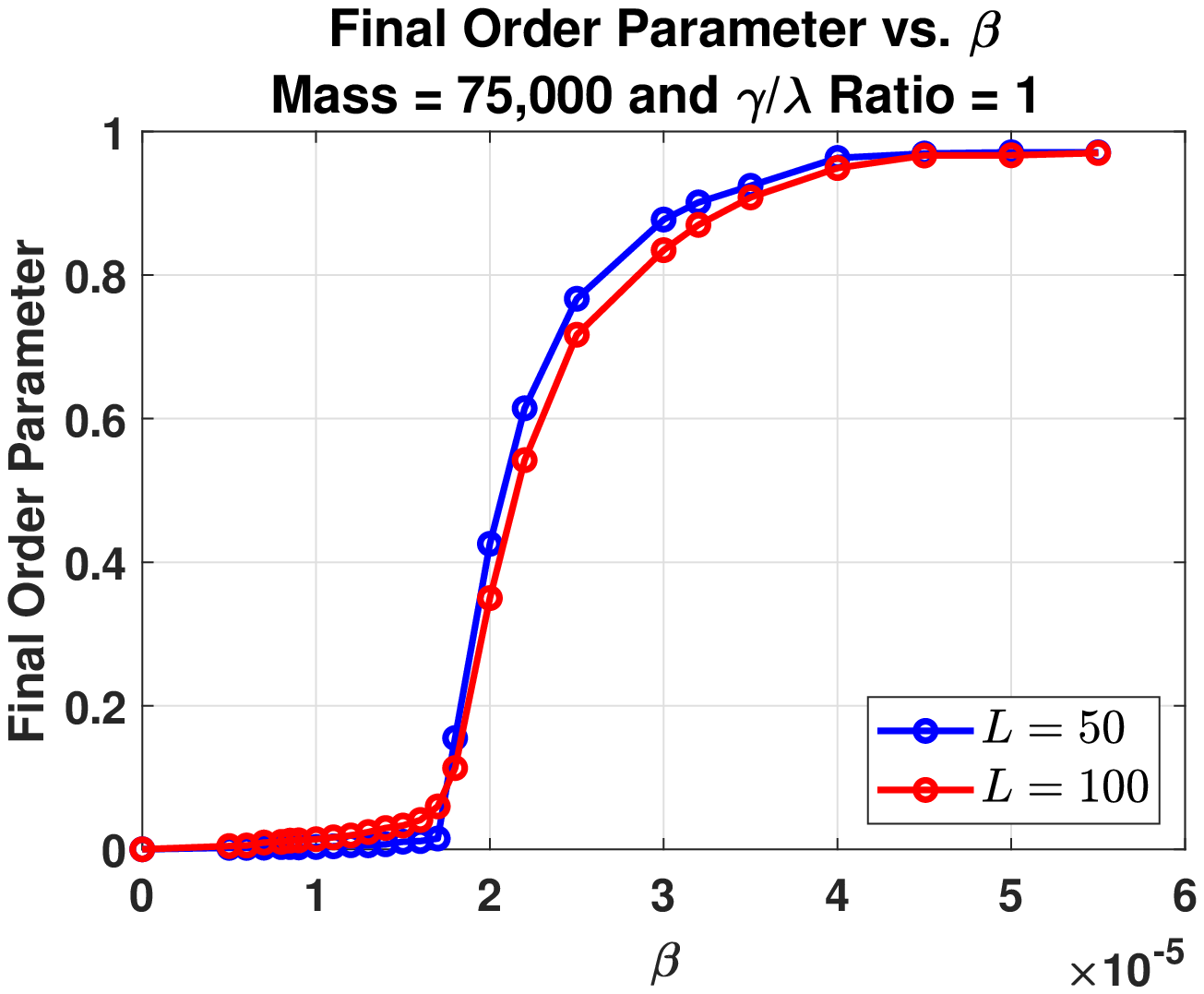}
        \end{subfigure}%
        \begin{subfigure}[b]{0.33\linewidth}
                \includegraphics[width=3.0cm,,keepaspectratio]{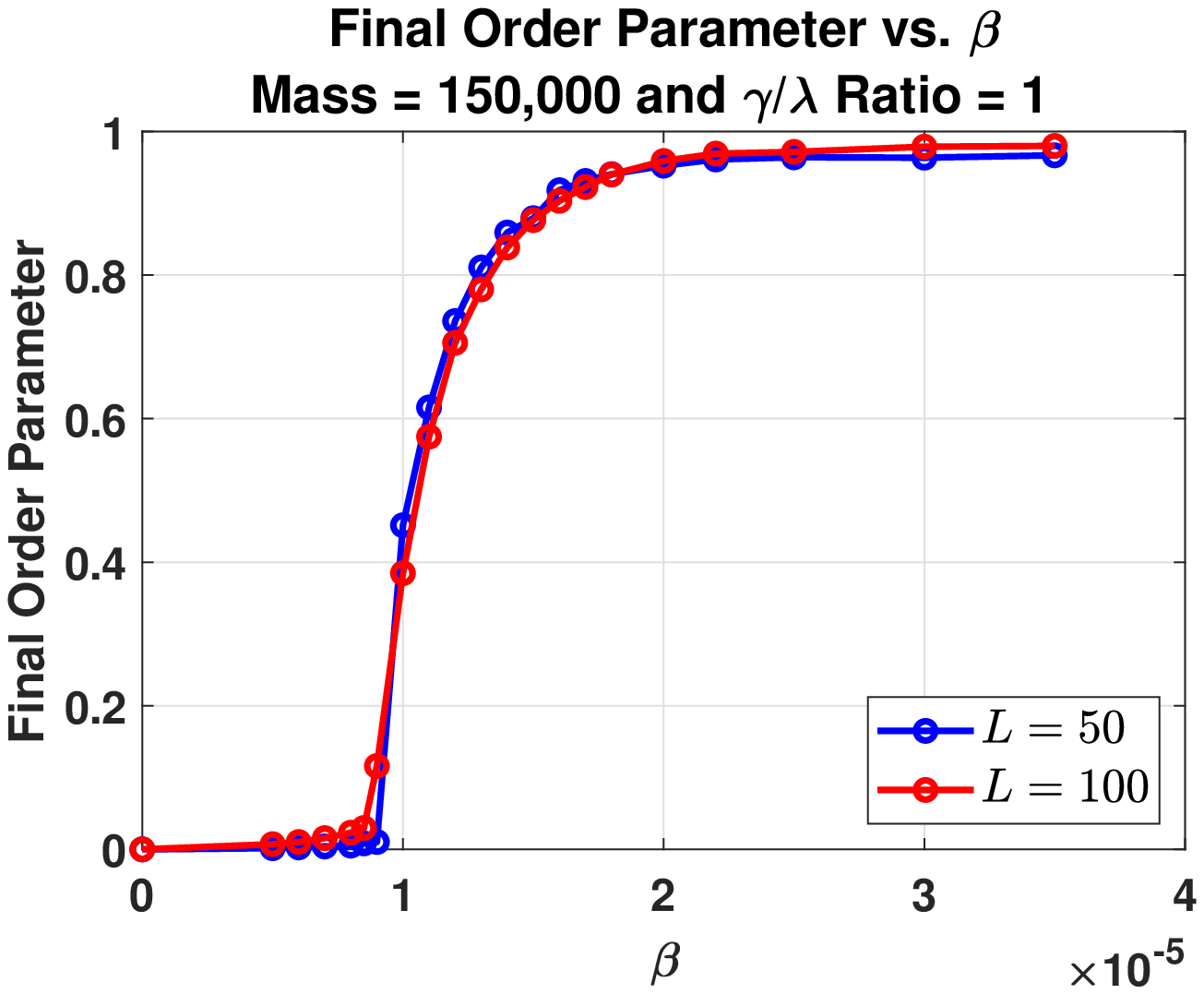}
        \end{subfigure}%
        \begin{subfigure}[b]{0.33\linewidth}
                \includegraphics[width=3.0cm,,keepaspectratio]{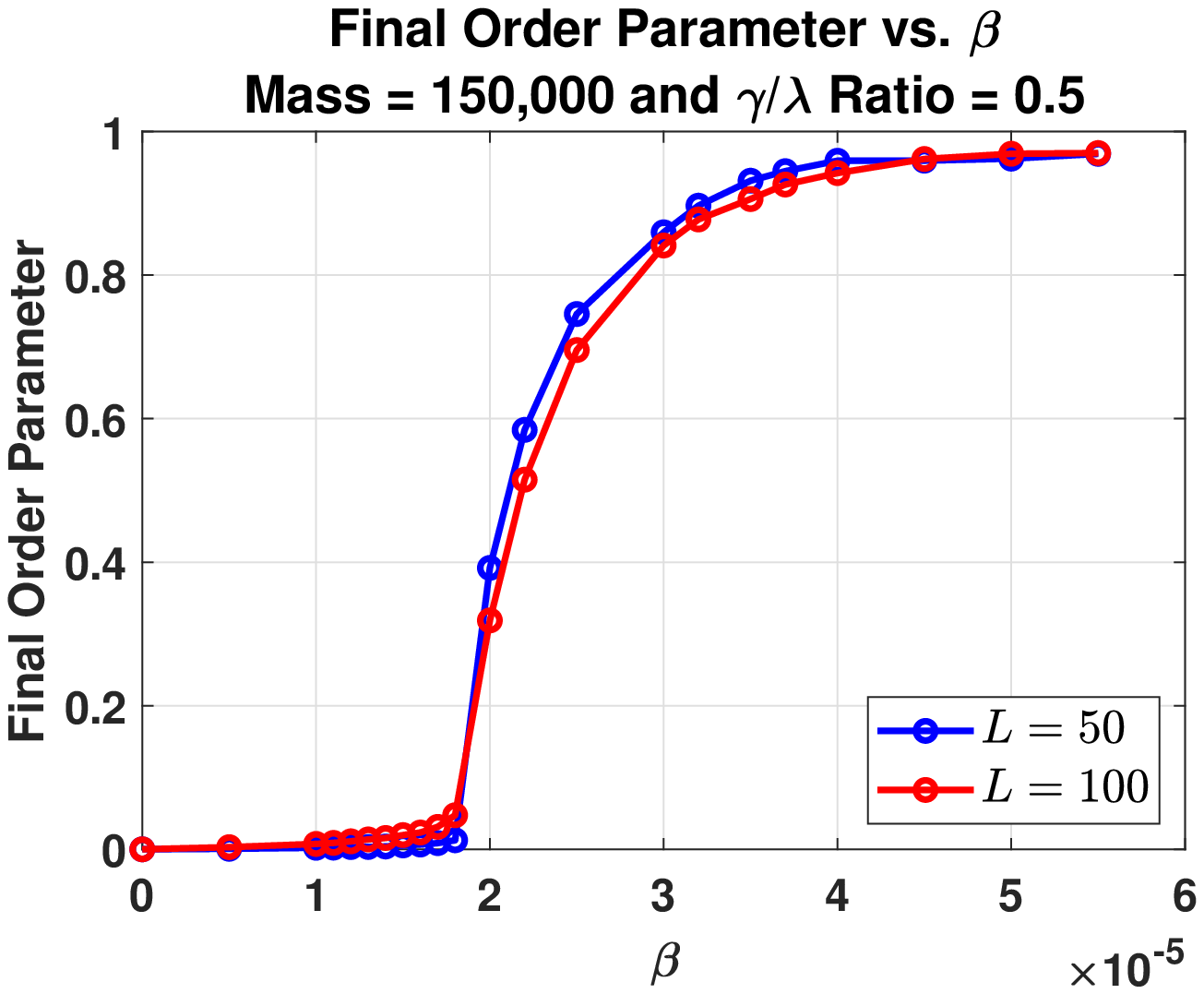}
        \end{subfigure}%
                \caption[Order Parameter.]{The Effect of Parameters on the Phase Transition. Here we have $N_1 = N_2 = N_3 =50,000$, with $\lambda = \gamma =0.5, \delta t = 1$ and the lattice size is $100 \times 100$. In all three plots, it is seen that for small $\beta$ values, the system remains well-mixed and the resulting order parameter values is approximately zero over time. However, for larger $\beta$ values we see that the order parameter increases quickly as the system segregates and levels off to around one.The order parameter value is taken at the final time step for different $\beta$ values. We clearly see that as the $\beta$ value increases there is a critical $\beta$ value in which a phase transition occurs. The three plots show the effects of changing the mass and the ratio $\gamma/\lambda$.}
        \label{fig:order_parameter2}
\end{figure}

We were also interested in how the mass might affect our system. In Figure \ref{fig:order_parameter2}, we see in the first plot that when the mass is $75,000$ the critical $\beta$ at which the phase transition occurs is about $1.8 \times 10^{-5}$. However, in the middle plot, the mass is increased to $150,000$ and this time the phase transition occurs around $0.9 \times 10^{-5}$. Thus, we notice that as the systems' mass increases, the resulting phase transition happens at a smaller $\beta$ value. Physically, this makes sense, since having a larger number of agents implies that there will be more graffiti being added at each site and thus a smaller $\beta$ value should be sufficient for the agents to react to the graffiti field. 

We also investigated how the ratio $\frac{\gamma}{\lambda}$ might change where the phase transition occurs. Again, we kept all other parameters fixed and changed the value of the ratio by altering the decay rate $\lambda$. Having a higher decay rate means that the graffiti is decaying more quickly and thus each site would have less graffiti. We found that by decreasing the $\frac{\gamma}{\lambda}$ ratio, a higher $\beta$ value is needed for segregation. This is evident in  the middle and right plots in Figure \ref{fig:order_parameter2}. There, we clearly see that when the $\frac{\gamma}{\lambda} = 1$, the critical $\beta$ is around $0.9 \times 10^{-5}$, whereas when the $\frac{\gamma}{\lambda}$ is decreased to $0.5$, the critical $\beta$ is around $1.8 \times 10^{-5}$. Physically, this is due to the fact that less graffiti on a site means that a larger $\beta$ value is necessary for the agents to react to it.

We were also interested in investigating how changing the grid size might affect the segregation when we alter the other parameters. In Figures \ref{fig:order_parameter} and \ref{fig:order_parameter2}, we see that increasing the number of sites from $L=50$ to $L=100$ does not have any noticeable effect on our discrete model.

\section{Deriving the Convection-Diffusion System} \label{section:contiuum_background}

In this section, we will formally derive the continuum equations of our system and will prove that the limiting system of convection-diffusion equations is
\begin{equation}
\begin{cases}
\displaystyle \frac{\partial \xi_j}{\partial t}(x,y,t) = \gamma \rho_j(x,y,t) - \lambda \xi_j(x,y,t) \\
\displaystyle \frac{\partial \rho_j}{\partial t}(x,y,t) =  \frac{D}{4} \nabla \cdot \left[ \nabla \rho_j(x,y,t)  + 2  \beta \left(\rho_j(x,y,t) \nabla \left( \sum_{\substack{i=1 \\ i\neq j}}^K \xi_i(x, y,t) \right) \right) \right] 
\end{cases}
\label{E:final_continuum_equations} 
\end{equation}
\noindent with periodic boundary conditions, where $j \in \{1, 2, \dots, K\}$.
Since our discrete model is a multiple-gang extension of the two-species model in \cite{alsenafi2018convection}, we proceed with finding the continuum equations by following the steps of the derivation of the continuum model therein. With minor modifications, the same derivation goes through for this multiple-gang case.

Deriving the continuum limits from discrete models is of great interest to the mathematical community; for example, even just from the crime modeling literature, we can refer you to papers \cite{dolak2005kinetic, JBC2010, SBB2010, SDPTBBC2008}. These continuum equations are often formally derived by assuming appropriate smoothness of the gang density and graffiti density and taking both the grid spacing and time step to zero, as we will do here. The continuum partial differential equations give us more tools for understanding the macroscopic behavior of the model.

\subsection{Continuum Graffiti Density} \label{section:graffiti_continuum}
We start by formally deriving the continuum equations for graffiti densities, recalling that for $j \in \{ 1, 2, \dots, K\}$, the discrete model \eqref{E:discrete_graffiti}:
\begin{equation*}
\xi_j(x,y,t+\delta t)=\xi_j(x,y,t) - \delta t \cdot \lambda \cdot \xi_j(x,y,t)  + \delta t \cdot \gamma \cdot \rho_j(x,y,t).
\end{equation*}
Rearranging the equation and dividing by $\delta t$ gives us:
\begin{equation*}
\frac{\xi_j(x,y,t+\delta t) - \xi_j(x,y,t) }{\delta t} = \gamma \cdot \rho_j(x,y,t) - \lambda \cdot \xi_j(x,y,t).
\end{equation*}
This is now in the form of a difference equation.  Assuming sufficient smoothness of the agent and graffiti densities $\rho_j$ and $\xi_j$, we take $\delta t \rightarrow 0$.  This gives us the final form of the graffiti continuum equation for gang $j$:
\begin{equation}
\frac{\partial \xi_j}{\partial t}(x,y,t) = \gamma \rho_j(x,y,t) - \lambda \xi_j(x,y,t). \label{E:graffiti_continuum_eqn_1}
\end{equation}

\subsection{Continuum Agent Density} \label{section:agent_continuum}
\subsubsection{Tools for the Derivation}Deriving the continuum equations for the agent densities is more complex, so before we begin, we define several quantities that will be useful in the derivation. We will need equation \eqref{E:graffiti_complement}, which we recall here:
\begin{equation*}
\psi_j(x,y,t) := \sum_{\substack{i=1 \\ i\neq j}}^K \xi_i(x, y,t).
\end{equation*}
Employing this notation, the first quantity we define is
\begin{equation}
T_j(x,y,t) := \frac{e^{\beta \psi_j(x,y,t)}}{4 + l^2 \left(  \beta^2 \left( \nabla \psi_j(x,y,t)\right)^2  - \beta \Delta \psi_j(x,y,t) \right) }. \label{D:T_A} 
\end{equation}
We will use $T_j$ to account for the influences of the neighbors and neighbor's neighbors in the discrete model.

Next, we derive approximations to $\nabla T_j$ and $\Delta T_j$, which we will use later in this section. For simplicity, the notation $(x,y,t)$ will be dropped as there will be no neighbors $(\tilde x, \tilde y, t) $ in the derivation of these quantities. We start by simplifying $T_j$ using Taylor series approximations. Recall the Taylor expansion
\begin{equation*}
\frac{1}{x+h} = \frac{1}{x} - \frac{h}{x^2} + \mathcal{O}(h^2).
\end{equation*}
We apply Taylor this expansion  to $T_j$ with $x=4$ and $h=l^2 \left(  \beta^2 \left(\nabla \psi_j\right)^2 - \beta \Delta \psi_j \right)$:
\begin{align}
T_j &= \frac{e^{\beta \psi_j}}{4}\left( 1 - \frac{l^2}{4}\left(\beta^2 \left(\nabla \psi_j\right)^2  - \beta \Delta \psi_j \right)  \right) + \mathcal{O}(l^4). \label{D:T_A_approximation}
\end{align}
Note that here, we are depending on the smoothness of $\psi_j$. Then, by taking the gradient of \eqref{D:T_A_approximation}, we find that
\begin{align}
\nabla T_j &= \frac{e^{\beta \psi_j}}{4} \left( \beta \nabla \psi_j - \frac{l^2}{4}\left(\beta^3 (\nabla \psi_j)^3 + \beta^2 \nabla \psi_j \Delta \psi_j - \beta \nabla^3 \psi_j \right) \right) +\mathcal{O}(l^4). \label{D:T_A_gradient}
\end{align}

\begin{align}
\intertext{We also can find $\Delta T_j$:}
\Delta T_j &= \nabla \cdot (\nabla T_j) \notag \\
\Rightarrow \Delta T_j &= \frac{e^{\beta \psi_j}}{4} \bigg( \left(\beta^2 (\nabla \psi_j)^2 + \beta \Delta \psi_j\right) - \frac{l^2}{4}\Big( 4 \beta^3 (\nabla \psi_j)^2 \Delta \psi_j  \notag \\ \notag
&+ \beta^2(\Delta \psi_j)^2  +\beta^4 (\nabla \psi_j)^4 - \beta \nabla^4 \psi_j   \Big) \bigg)  + \mathcal{O}(l^4). \label{D:T_A_laplacian} \\
\end{align}


We now focus our attention on the movement probability. Starting from definition \eqref{E:probability_agent_moves}, recall that the probability that an agent from gang $j$ moves from site $(x,y)$ to a neighboring site $(x_1,y_1)$ is
\begin{equation*}
M_j(x \rightarrow x_1, y \rightarrow y_1, t)  = \frac{e^{-\beta \psi_j(x_1, y_1,t)}}{\sum \limits_{(\tilde x, \tilde y) \sim(x,y)}e^{-\beta \psi_j(\tilde x, \tilde y, t)}},
\end{equation*}
where $(\tilde x, \tilde y)$ are the neighbors of site $(x,y)$.
We now slightly modify the above definition  so that we evaluate the probability that an agent at a neighboring site  $(\tilde x, \tilde y)$ moves to site  $(x,y)$:
\begin{equation}
M_j(\tilde x \rightarrow  x, \tilde y \rightarrow  y, t)  = \frac{e^{-\beta \psi_j(x, y,t)}}{\sum \limits_{(\tilde{\tilde x}, \tilde{\tilde y}) \sim (\tilde x,\tilde y)}e^{-\beta \psi_j(\tilde{\tilde x}, \tilde{\tilde y}, t)}},\label{L:movement_from_neighbors_1}
\end{equation} 
where $(\tilde{\tilde x}, \tilde{\tilde y})$ are the four neighbors of site $(\tilde x,\tilde y)$.

To remove the presence of the neighbors' neighbors $(\tilde{\tilde x}, \tilde{\tilde y})$ from the denominator, we apply the discrete Laplacian to find thats
\begin{equation}
\sum \limits_{(\tilde{\tilde x}, \tilde{\tilde y}) \sim (\tilde x,\tilde y)}e^{-\beta \psi_j(\tilde {\tilde x},\tilde {\tilde y},t)} = 4e^{-\beta \psi_j(\tilde x,\tilde y,t)} + l^2 \Delta \left( e^{-\beta \psi_j(\tilde x,\tilde y,t)}\right) + \mathcal{O}(l^4). 
\label{L:movement_from_neighbors_2}
\end{equation}
Noting that 
\begin{align}
 \Delta e^{-\beta \psi_j(\tilde x,\tilde y,t)} &= \nabla \cdot \nabla \left(e^{-\beta \psi_j(\tilde x,\tilde y,t)}\right) \notag \\ 
&= \nabla \cdot \left( -\beta \nabla \psi_j(\tilde x,\tilde y,t) e^{-\beta \psi_j(\tilde x,\tilde y,t)}\right)  \notag \\
&= \left[ \beta^2 \left(\nabla \psi_j(\tilde x,\tilde y,t) \right)^2 - \beta \Delta \psi_j(\tilde x,\tilde y,t) \right]e^{-\beta \psi_j(\tilde x,\tilde y,t)}, \label{L:movement_from_neighbors_3}
\end{align}
Combining equations (\ref{L:movement_from_neighbors_2}) and (\ref{L:movement_from_neighbors_3}) gives us
\begin{align*}
\sum \limits_{(\tilde{\tilde x}, \tilde{\tilde y}) \sim (\tilde x,\tilde y)} e^{-\beta \psi_j(\tilde {\tilde x},\tilde {\tilde y},t)} =& e^{-\beta \psi_j(\tilde x,\tilde y,t)} \left( 4 + l^2\left( \beta^2 \left( \nabla \psi_j(\tilde x,\tilde y,t) \right)^2 - \beta \Delta \psi_j(\tilde x,\tilde y,t) \right) \right)\\ 
&+\mathcal{O}(l^4). 
\end{align*}
Substituting it back into equation (\ref{L:movement_from_neighbors_1}), and replacing the denominator gives us 
\begin{align*}
M_j(\tilde x \rightarrow  x, \tilde y \rightarrow  y, t) &=  \frac{e^{-\beta \psi_j( x, y,t)}}{\left[4 + l^2 \left( \beta^2 \left(  \nabla \psi_j(\tilde x,\tilde y,t) \right)^2 -\beta \Delta \psi_j(\tilde x,\tilde y,t) \right) \right]e^{-\beta \psi_j(\tilde x,\tilde y,t)} + \mathcal{O}(l^4)}\\
 &\approx e^{-\beta \psi_j(x, y,t)} \left[ \frac{e^{\beta \psi_j(\tilde x,\tilde y,t)}}{4 + l^2 \left( \beta^2\left( \nabla \psi_j(\tilde x,\tilde y,t) \right)^2 -\beta \Delta \psi_j(\tilde x,\tilde y,t) \right) }  \right].  \\  
\end{align*}
The term inside the large brackets in the equation above takes the form of \eqref{D:T_A} where  $(x, y, t)$ is replaced with $(\tilde x,\tilde y,t)$, yielding the following approximation:
\begin{equation}
M_j(\tilde x \rightarrow  x, \tilde y \rightarrow  y, t) \approx e^{-\beta \psi_j(x, y,t)} T_j(\tilde x,\tilde y,t).
\label{L:movement_from_neighbors}
\end{equation}

\subsubsection{The Derivation} We now have all the tools needed to formally derive the agent density continuum equations. We will be using the discrete Laplacian approximation in order to approximate the influence of the neighbors of site $(x,y)$. We will also be using equation (\ref{L:movement_from_neighbors}) to simplify the discrete model.

Starting from the discrete model, we recall equation (\ref{E:discrete_agents}):
\begin{equation*}
\begin{split}
\rho_j(x,y,t + \delta t) = &\rho_j(x,y,t) + \sum_{ (\tilde x,\tilde y) \sim ( x, y)}  \rho_j(\tilde x,\tilde y, t) M_j( \tilde x \rightarrow x,  \tilde y \rightarrow y, t) \\
&-  \rho_j(x, y, t)\sum_{ (\tilde x,\tilde y) \sim (x,y)}  M_j(x  \rightarrow  \tilde x, y  \rightarrow  \tilde y, t).
\end{split}
\end{equation*}
 Rearranging the equation and dividing both sides by $\delta t$ gives us
\begin{equation*}
\begin{split}
 \frac{\rho_j(x,y,t + \delta t) - \rho_j(x,y,t)}{\delta t} = \frac{1}{\delta t} \left[ \sum_{ (\tilde x,\tilde y) \sim (x,y)}   \rho_j( \tilde x,  \tilde y, t)  M_j( \tilde x \rightarrow x,  \tilde y \rightarrow y, t) \right. \\ 
 \left. - \rho_j(x, y, t) \sum_{ (\tilde x,\tilde y) \sim (x,y)} M_j(x  \rightarrow  \tilde x, y  \rightarrow  \tilde y, t)\right].
\end{split} 
\end{equation*}
By equation (\ref{L:movement_from_neighbors}), and noting that each agent has to move to one of the neighboring sites,
\begin{align}
\frac{\rho_j(x,y,t + \delta t) - \rho_j(x,y,t)}{\delta t} = \frac{1}{\delta t} \Bigg[ &e^{-\beta \psi_j(x,y,t)} \sum_{ (\tilde x,\tilde y) \sim (x,y)}   \rho_j( \tilde x,  \tilde y, t)  T_j( \tilde x,  \tilde y, t)  \notag \\ 
 -& \rho_j(x, y, t) +\mathcal{O}(l^4) \Bigg].  \label{T:agents_continuum_eqn_1}
 \end{align}
Using the discrete Laplacian technique, we can approximate the contribution of the neighboring sites, giving on the right-hand side
\begin{align*}
\frac{1}{\delta t} \Bigg[ &e^{-\beta \psi_j(x,y,t)}\bigg(4 \rho_j(x,y,t)T_j(x,y,t)+l^2 \Delta \Big(\rho_j(x,y,t)T_j(x,y,t)\Big)\bigg)  \\
&-\rho_j(x,y,t) +\mathcal{O}(l^4) \Bigg]. 
\end{align*}
The notation $(x,y,t)$ is again dropped  as there are no longer any neighbors $(\tilde x, \tilde y, t)$ remaining in this derivation. Hence, we write,
\begin{equation}
\frac{1}{\delta t} \Bigg[ e^{-\beta \psi_j}\bigg(4 \rho_j T_j+l^2 \Delta \Big(\rho_j T_j)\Big)\bigg) -\rho_j +\mathcal{O}(l^4) \Bigg]. \label{E:no_neighbours}
\end{equation}
\\From definition (\ref{D:T_A}), $T_j(x,y,t)$ is substituted back into the first term of (\ref{E:no_neighbours}):
\begin{align}
&\frac{1}{\delta t} \Bigg[ 4\rho_j e^{-\beta \psi_j}\left( \frac{e^{\beta \psi_j}}{4 + l^2 \left( \left( \beta \nabla \psi_j \right)^2 - \beta\Delta \psi_j\right)} \right)+ l^2e^{-\beta \psi_j}\Delta \Big(\rho_j T_j\Big) - \rho_j  + \mathcal{O}(l^4) \Bigg].  \notag 
\intertext{Simplifying the expression yields}
&\frac{1}{\delta t} \Bigg[ 4\rho_j\left( \frac{1}{4 + l^2 \left( \left( \beta \nabla \psi_j \right)^2 - \beta \Delta \psi_j \right)} \right) - \rho_j+ l^2e^{-\beta \psi_j}\Delta \Big(\rho_j T_j\Big) +\mathcal{O}(l^4) \Bigg]. \label{T:agents_continuum_eqn_2}
\end{align}
Using a Taylor series expansion on the first term within the brackets yields,  
\begin{equation} 
\left( \frac{1}{4 + l^2 \left( \left( \beta \nabla \psi_j \right)^2 - \beta \Delta \psi_j\right)} \right) = \frac{1}{4} - \frac{l^2 \left( \left(\beta \nabla \psi_j \right)^2 -\beta \Delta \psi_j \right) }{4^2} + \mathcal{O}(l^4).  \notag
\end{equation}
Expression \eqref{T:agents_continuum_eqn_2} thus becomes,
\begin{equation}
\frac{1}{\delta t} \left[ 4\rho_j\left(\frac{1}{4} - \frac{l^2 \left( \left (\beta \nabla \psi_j\right) ^2 -\beta \Delta \psi_j \right) }{4^2} \right) - \rho_j  + l^2e^{-\beta \psi_j}\Delta \Big(\rho_j T_j\Big)  +\mathcal{O}(l^4)\right]. \notag 
\end{equation}
Simplifying the expression yields 
\begin{equation}
\begin{split}
\frac{\rho_j(x,y,t + \delta t) - \rho_j(x,y,t)}{\delta t} =& \frac{l^2}{\delta t} \left[ -\frac{\rho_j}{4}\left( \left( \beta \nabla \psi_j \right)^2 -\beta \Delta \psi_j \right)   + e^{-\beta \psi_j}\Delta \Big(\rho_j T_j\Big)  \right] \\
&+\mathcal{O}\left( \frac{l^4}{\delta t} \right).
 \label{T:agents_continuum_eqn_3}
 \end{split}
\end{equation}
However, we can further simplify this by noting that
\begin{equation}
\Delta \Big(\rho_j T_j \Big) =   \Big(T_j \Delta \rho_j+ 2 \nabla T_j \nabla \rho_j +\rho_j \Delta T_j\Big). \notag
\end{equation}
From \eqref{D:T_A_approximation} through \eqref{D:T_A_laplacian}, we have
\begin{align*}
T_j &= \frac{e^{\beta \psi_j}}{4} + \mathcal{O}(l^2), \notag \\
\nabla T_j &= \frac{\beta e^{\beta \psi_j} }{4}\nabla \psi_j + \mathcal{O}(l^2),\\
\Delta T_j &= \frac{e^{\beta \psi_j} }{4} \left(\beta \Delta \psi_j +  (\beta \nabla \psi_j)^2 \right) + \mathcal{O}(l^2).
\end{align*}
Therefore,
\begin{align}
\Delta \Big(\rho_j T_j\Big) &= \frac{e^{\beta \psi_j}}{4} \Delta\rho_j  + \frac{2 \beta e^{\beta \psi_j} }{4} \nabla \psi_j  \nabla \rho_j +   \frac{e^{\beta \psi_j}}{4}\rho_j \left(\beta \Delta \psi_j +  (\beta \nabla \psi_j)^2 \right)  +\mathcal{O}(l^2)  \notag \\
&= \frac{e^{\beta \psi_j}}{4}  \left[\Delta \rho_j +  2 \beta \nabla \psi_j \nabla \rho_j  + \rho_j  \left( \left(\beta \nabla \psi_j\right)^2 + \beta \Delta \psi_j \right) \right] +\mathcal{O}(l^2). \label{T:agents_continuum_eqn_4}
\end{align}
\noindent Substituting (\ref{T:agents_continuum_eqn_4}) back into (\ref{T:agents_continuum_eqn_3}) gives us
\begin{align*}
\frac{\rho_j(x,y,t + \delta t) - \rho_j(x,y,t)}{\delta t} &=  \frac{l^2}{4 \delta t} \Bigg[ -\rho_j\left( (\beta \nabla \psi_j)^2 -\beta \Delta \psi_j \right) + \Delta \rho_j + 2 \beta \nabla \psi_j \nabla \rho_j  \\
  &\qquad  +  \rho_j  \left(\left(\beta \nabla \psi_j \right)^2 + \beta \Delta \psi_j \right) \Bigg] + \mathcal{O}\left(\frac{l^4}{\delta t}\right)   \\
&=\frac{l^2}{4\delta t}  \Bigg[ \Delta \rho_j  +  2\beta\nabla \rho_j \nabla \psi_j + 2\beta \rho_j \Delta \psi_j  \bigg] + \mathcal{O}\left(\frac{l^4}{\delta t}\right) \\
&=\frac{l^2}{4\delta t}  \Bigg[ \Delta \rho_j  +  2\beta \left(\nabla \rho_j \nabla \psi_j +  \rho_j \Delta \psi_j \right) \bigg] + \mathcal{O}\left(\frac{l^4}{\delta t}\right) \\
&=\frac{l^2}{4\delta t}\Bigg[ \Delta \rho_j  + 2 \beta \nabla \cdot \Big(\rho_j \nabla \psi_j \Big) \Bigg] + \mathcal{O}\left(\frac{l^4}{\delta t}\right)   \\
&= \frac{l^2}{4\delta t} \nabla \cdot \Bigg[ \nabla \rho_j  + 2\beta \Big(\rho_j \nabla \psi_j \Big) \Bigg] +\mathcal{O}\left(\frac{l^4}{\delta t}\right).
\end{align*}
Assuming that the agent density $\rho_j$ is sufficiently smooth and the following limits 
\begin{equation}
\begin{split}
l&\rightarrow 0, \\
\delta t &\rightarrow 0, \\
\frac{l^2}{\delta t} &\rightarrow D,
\end{split}
\end{equation}
gives us the final form for the continuum equations for the density of gang $j$ agents:
\begin{equation}
\frac{\partial \rho_j}{\partial t} =  \frac{D}{4} \nabla \cdot \Bigg[ \nabla \rho_j  + 2  \beta \Big(\rho_j \nabla \psi_j \Big) \Bigg]. \label{E:agents_continuum_eqn_1}
\end{equation}

Finally, from (\ref{E:graffiti_continuum_eqn_1}) and (\ref{E:agents_continuum_eqn_1}), and using equation \eqref{E:graffiti_complement} to express everything in terms of agent density and graffiti density, the limiting convection-diffusion system for our model is
\begin{equation}\label{T:continuum_eqns}
\begin{cases} 
\displaystyle \frac{\partial \xi_j}{\partial t}(x,y,t) = \gamma \rho_j(x,y,t) - \lambda \xi_j(x,y,t) \\
\displaystyle \frac{\partial \rho_j}{\partial t}(x,y,t) =  \frac{D}{4} \nabla \cdot \left[ \nabla \rho_j(x,y,t)  + 2  \beta \left(\rho_j(x,y,t) \nabla \left( \sum_{\substack{i=1 \\ i\neq j}}^K \xi_i(x, y,t) \right) \right) \right] 
\end{cases}
\end{equation}
for $j=1,2, \dots K$ with periodic boundary conditions.

\subsection{Steady-State Solutions} \label{section:steady-state}

Considering steady-state solutions for the graffiti density, we find from the evolution equations for for the graffiti density that
\begin{align}
\frac{\partial \xi_j}{\partial t}(x,y,t) &= 0 \notag \\
\Rightarrow \gamma \rho_j(x,y,t) - \lambda \xi_j(x,y,t)  &= 0\notag \\
\Rightarrow \xi_j &= \frac{\gamma}{\lambda} \rho_j. \label{E:steadystategraffitiA}
\end{align}
\noindent We now focus our attention on the steady-state solutions for the agent density of gang $j$:
\begin{align}
\frac{\partial \rho_j}{\partial t}(x,y,t) =  \frac{D}{4} \nabla \cdot \left[ \nabla \rho_j(x,y,t)  + 2  \beta \left(\rho_j(x,y,t) \nabla \left( \sum_{\substack{i=1 \\ i\neq j}}^K \xi_i(x, y,t) \right) \right) \right] \notag
&= 0, \notag
\end{align}
considering solutions of the form
\begin{align*}
\nabla \rho_j(x,y,t)  + 2  \beta \left(\rho_j(x,y,t) \nabla \left( \sum_{\substack{i=1 \\ i\neq j}}^K \xi_i(x, y,t) \right) \right) &= c.
\end{align*}
Using the steady-state graffiti density derived in equation \eqref{E:steadystategraffitiA}, we find that
\begin{align}
\nabla \rho_j(x,y,t)  + \frac{2 \beta \gamma^2}{\lambda^2} \left(\rho_j(x,y,t)\nabla \cdot \left( \sum_{\substack{i=1 \\ i\neq j}}^K  \rho_i(x,y,t)\right) \right) &= c. \label{E:steadystate_rho1.}
\end{align}
Any form of $\rho_j(x,y,t)$ satisfying the above equation  is a steady-state solution of our system. For simplicity, we suppose that $\rho_j$ is a constant for all $j$.  In this case, the steady-state solution of our problem takes the  form:
\begin{align} \label{E:steadystateequation}
\begin{cases}
\xi_j &= \frac{\gamma}{\lambda} \rho_j, \\
\rho_j &= c_j,
\end{cases}
\end{align}
for $j=1,2, \dots, K$ and $c_j$ is a positive constant. 

\begin{figure}[!htb]
        \begin{subfigure}[b]{0.495\linewidth}
               \includegraphics[width=4.0cm,,keepaspectratio]{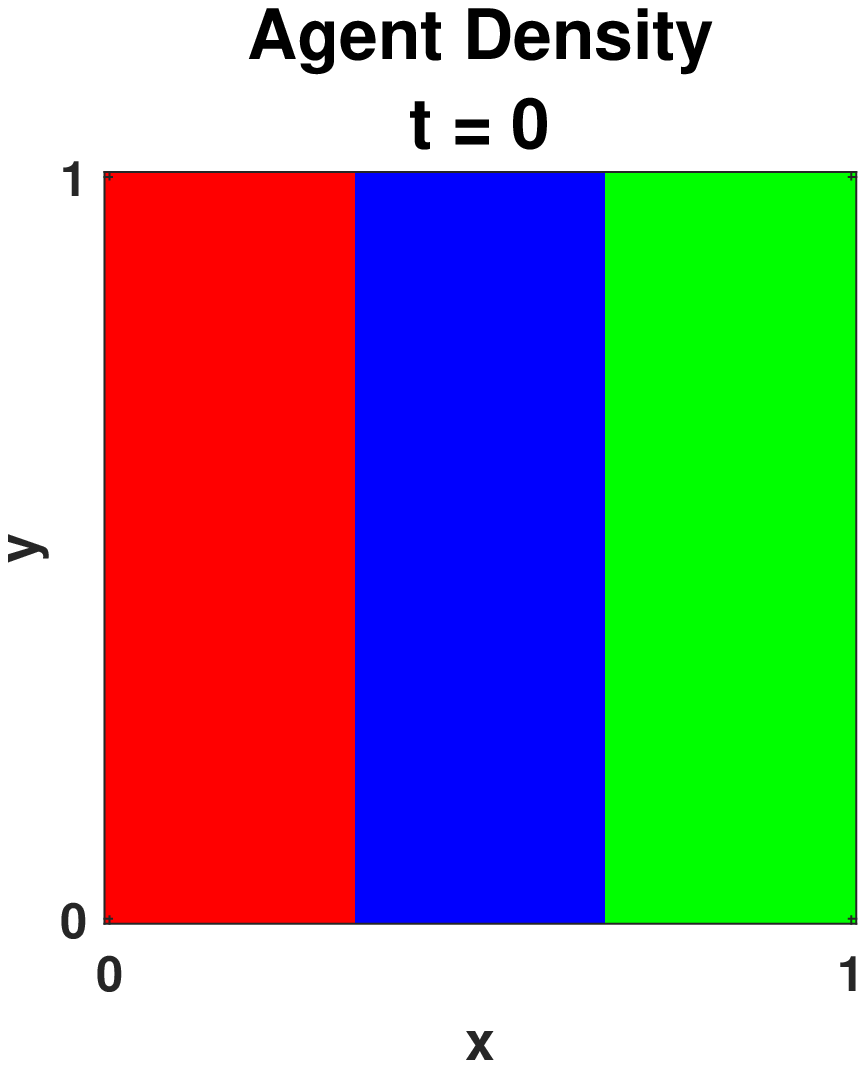}
        \end{subfigure}%
        \begin{subfigure}[b]{0.495\linewidth}
                \includegraphics[width=4.0cm,,keepaspectratio]{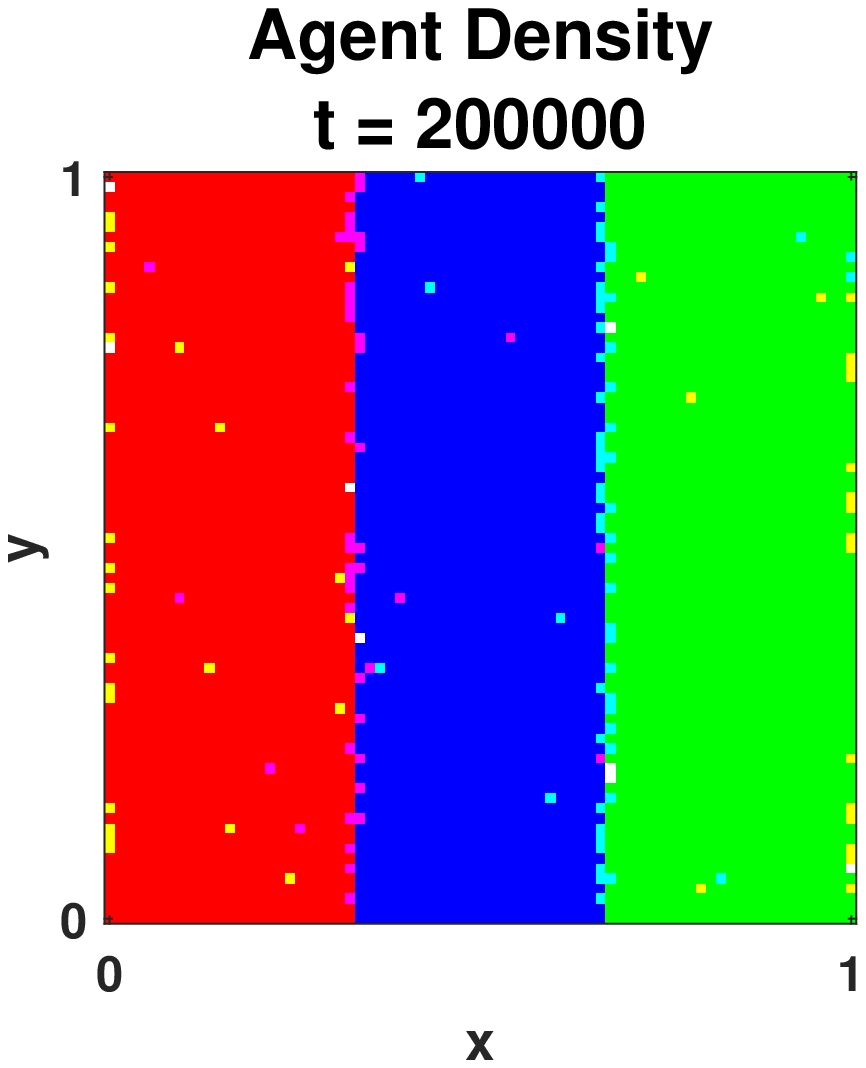}
        \end{subfigure}%
                \caption[Steady-State.]{Temporal evolution of a steady-state solution for the agent density. Here we have $N_1 = N_2 = N_3 =50,625$, with $\lambda = \gamma =0.5$, $\beta = 3 \times 10^{-4}$ and the lattice size is $75 \times 75$. It is shown that if the system started initially at a steady-state then it would remain there.}
        \label{fig:seg_state_1}
\end{figure}

To test whether these steady-state solutions of the continuum system (\ref{E:steadystateequation}) agree with the discrete model, we first start our simulations in a steady-state solution from \eqref{E:steadystateequation}. Here we will start the simulations with the agents from all three gangs are completely segregated. The results of the simulation are visualized in Figure \ref{fig:seg_state_1}. In that figure, we clearly see that the agents remain segregated and the system does not deviate from the steady-state solutions over time.

\begin{figure}[!htb]
        \begin{subfigure}[b]{0.495\linewidth}
               \includegraphics[width=4.0cm,,keepaspectratio]{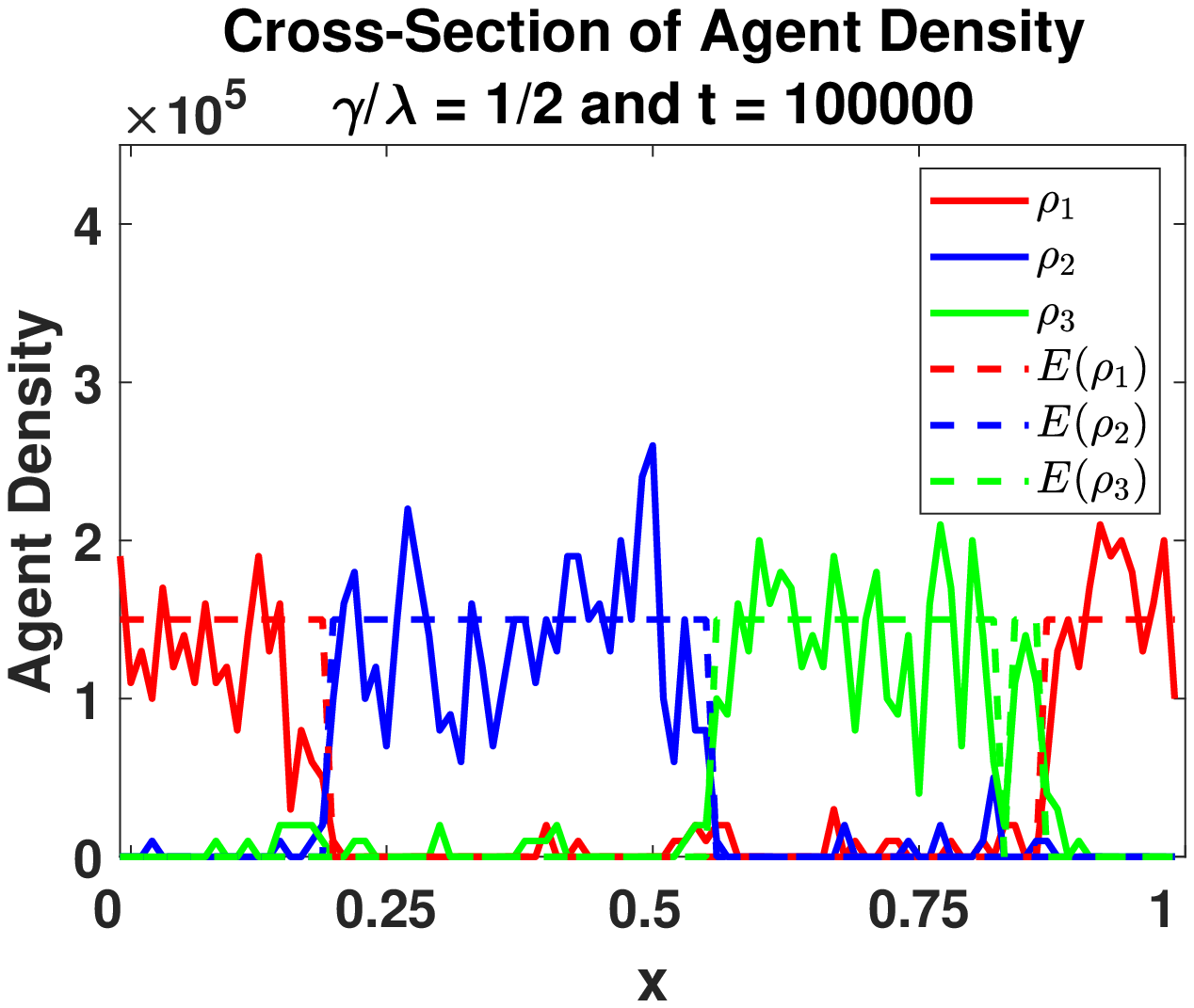}
        \end{subfigure}%
        \begin{subfigure}[b]{0.495\linewidth}
                \includegraphics[width=4.0cm,,keepaspectratio]{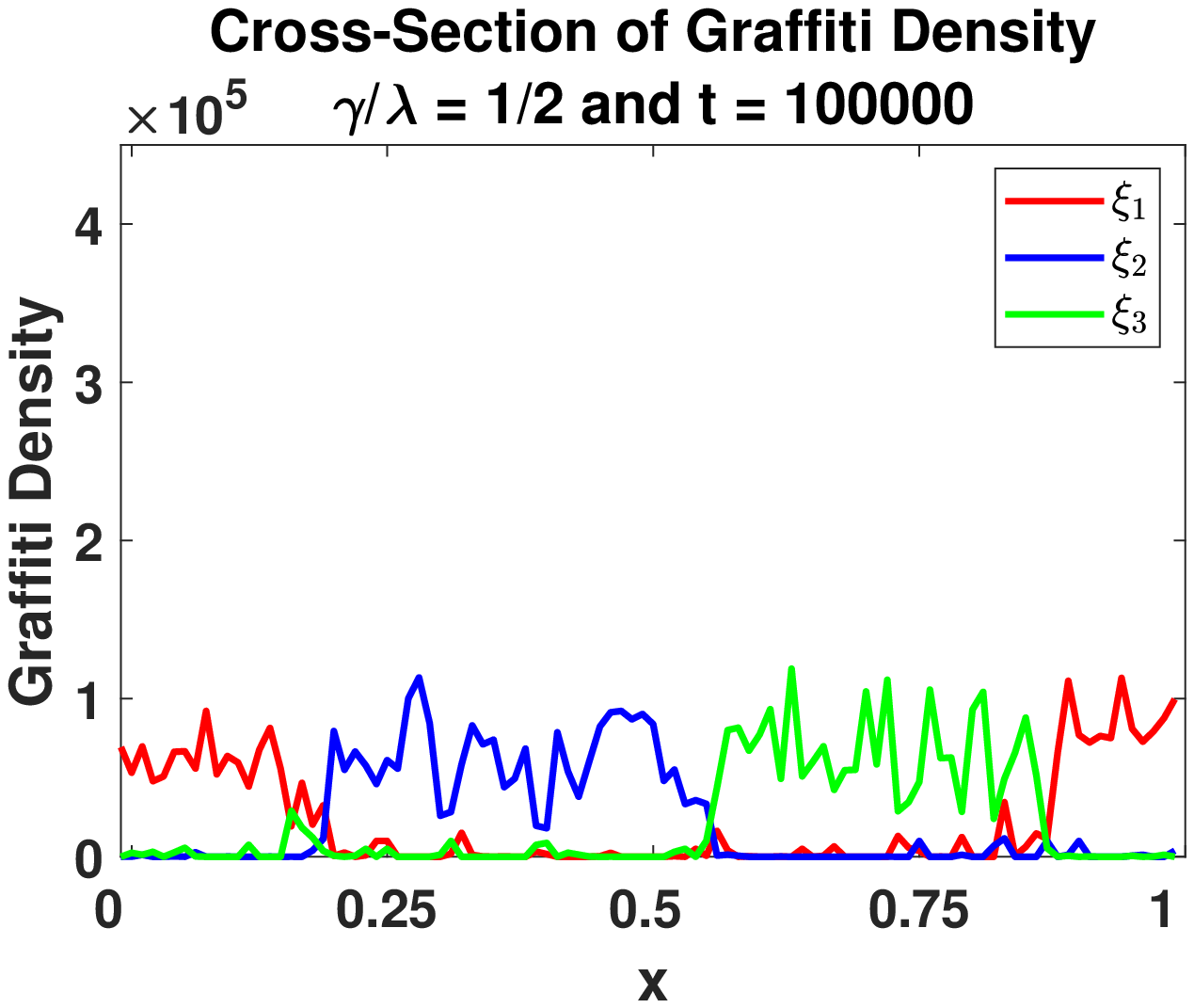}
        \end{subfigure}%
        
        \begin{subfigure}[b]{0.495\linewidth}
               \includegraphics[width=4.0cm,,keepaspectratio]{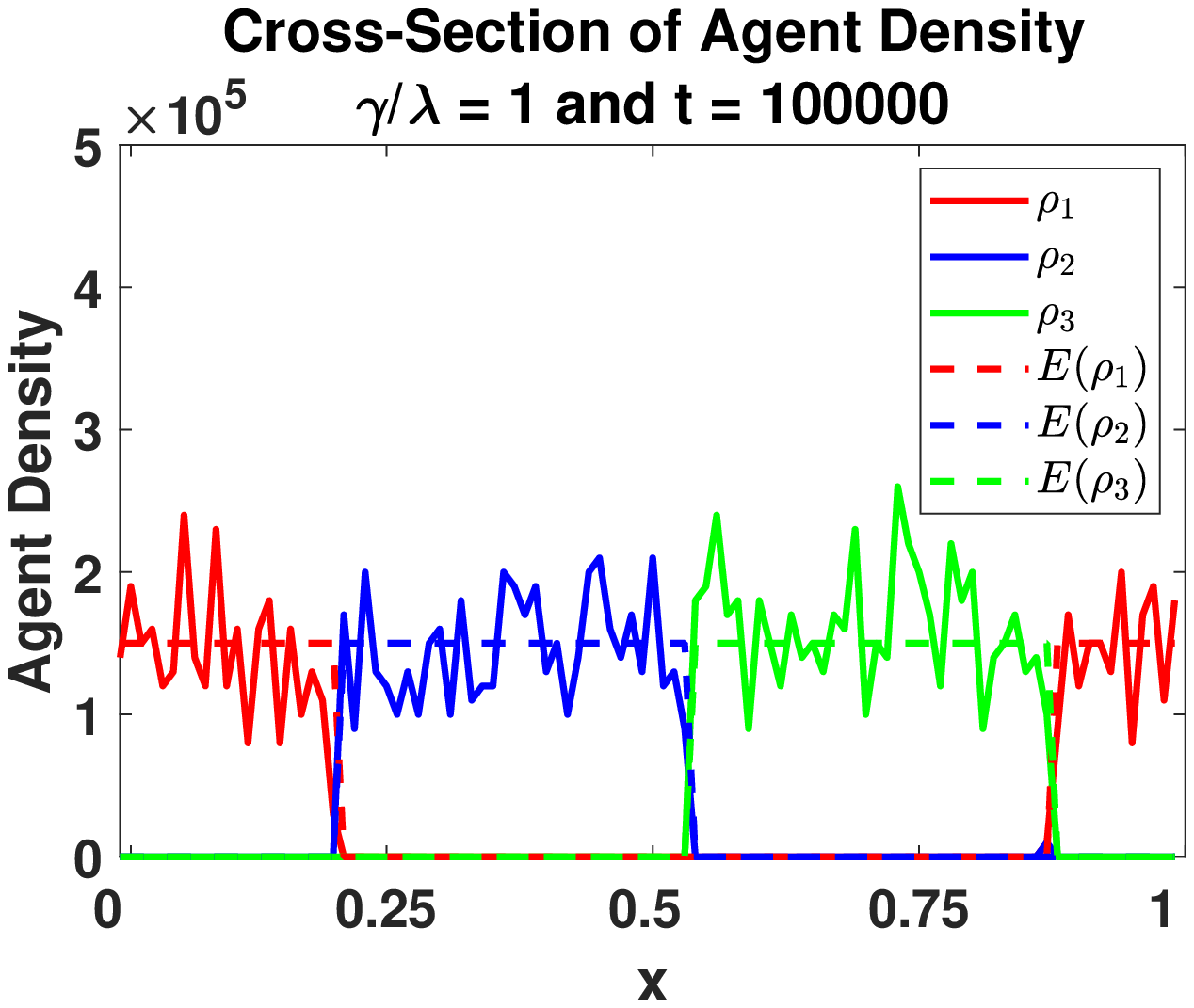}
        \end{subfigure}%
        \begin{subfigure}[b]{0.495\linewidth}
                \includegraphics[width=4.0cm,,keepaspectratio]{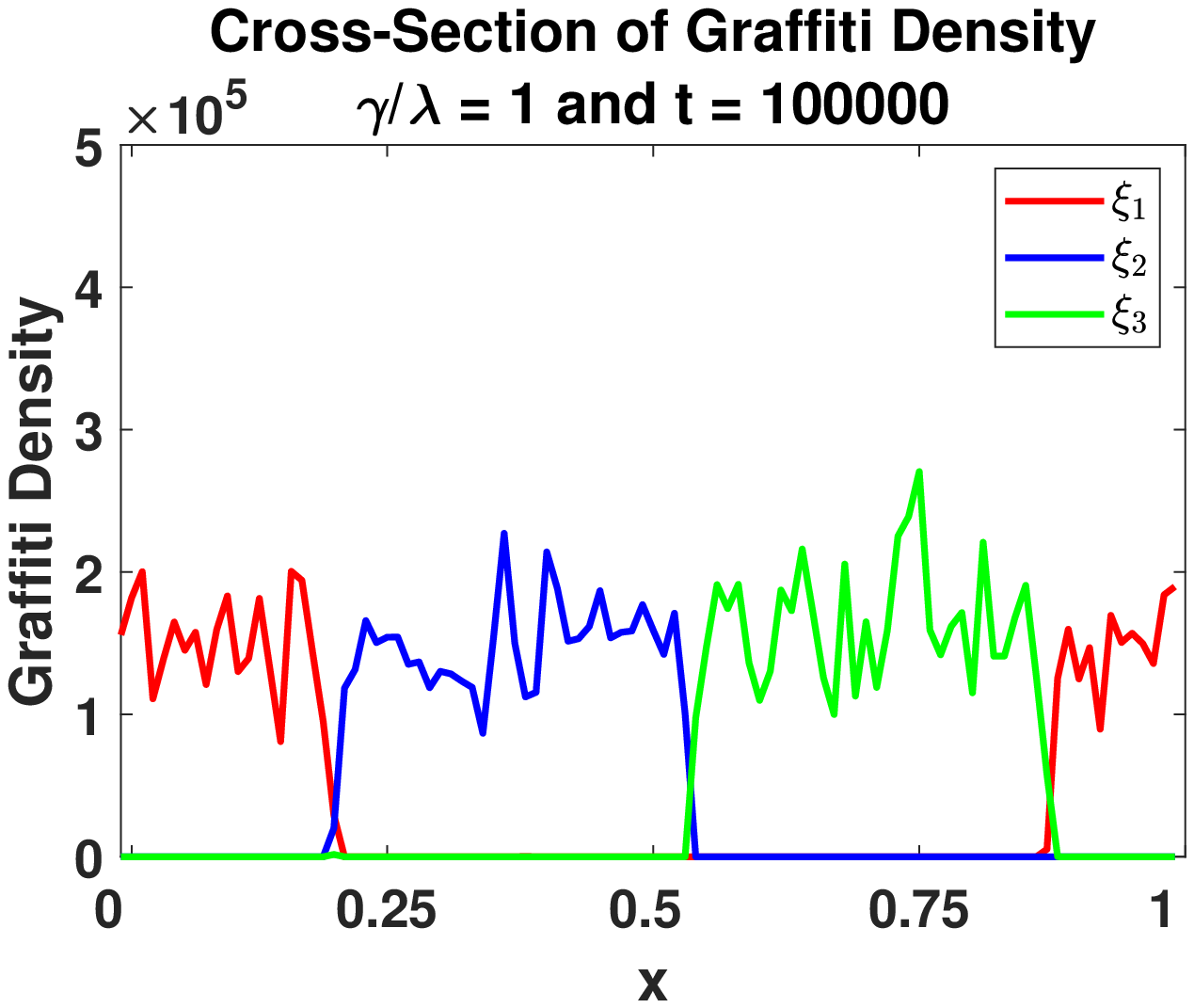}
        \end{subfigure}%
        
        \begin{subfigure}[b]{0.495\linewidth}
               \includegraphics[width=4.0cm,,keepaspectratio]{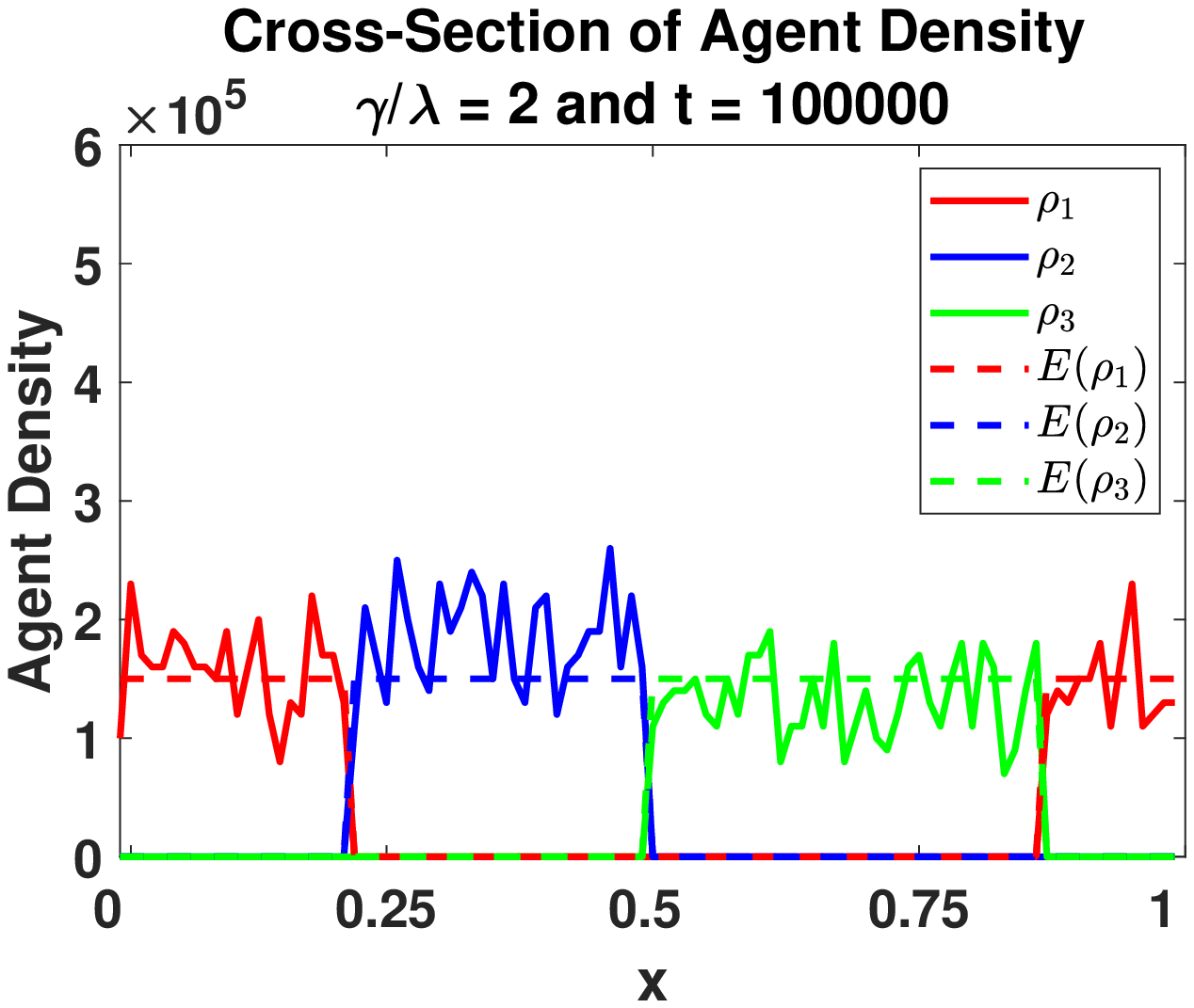}
        \end{subfigure}%
        \begin{subfigure}[b]{0.495\linewidth}
                \includegraphics[width=4.0cm,,keepaspectratio]{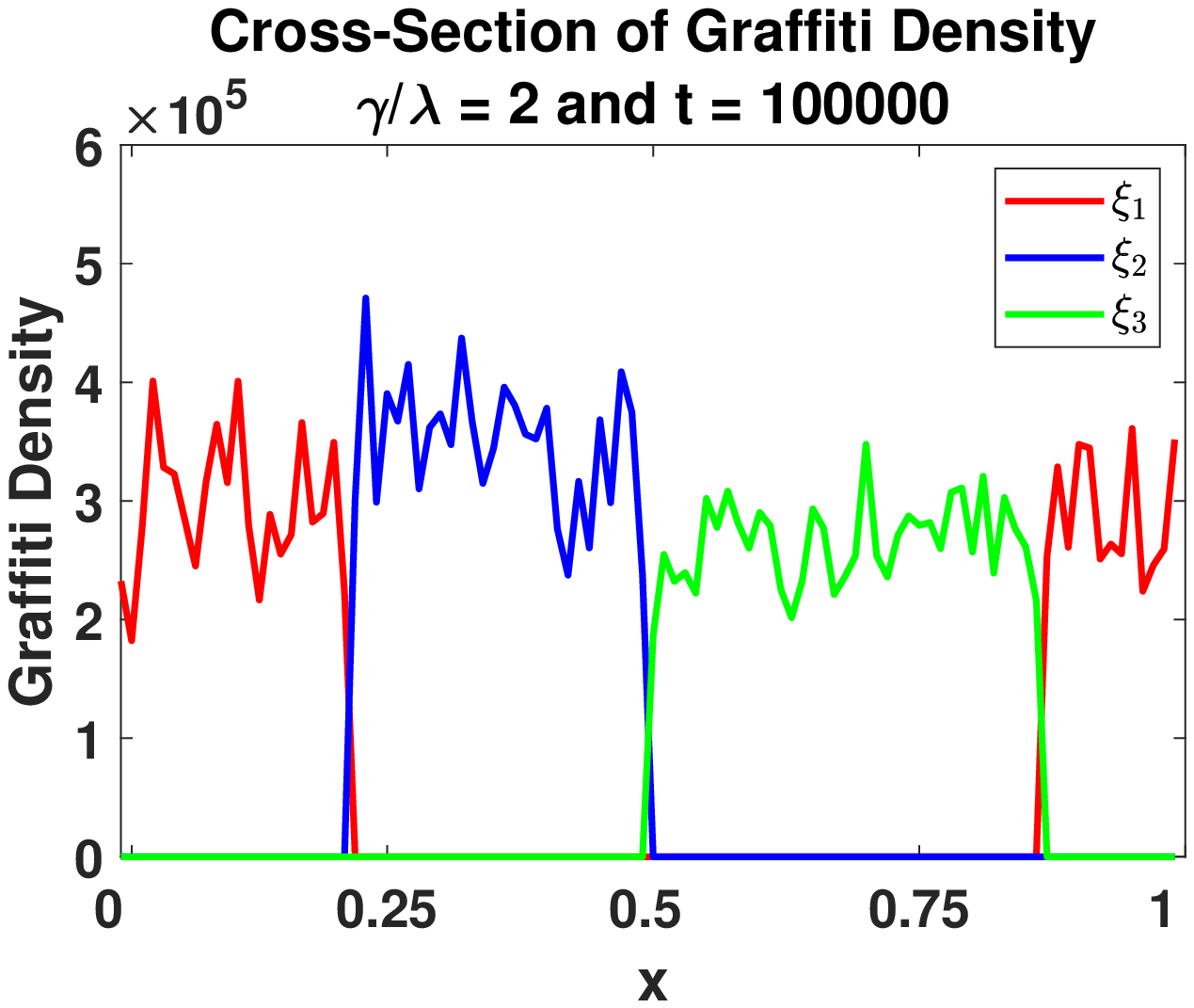}
        \end{subfigure}%
                \caption[Steady-State.]{Cross-sectional slices of the agent and graffiti densities for different $\frac{\gamma}{\lambda}$ at the final time step for a segregated state. Here we have $N_1=N_2=N=3 = 50,000$ with $\delta t = 1$, $\beta = 3 \times 10^{-4}$ and the lattice size is $100 \times 100$. In the top row we have $\gamma = 0.25$ and $\lambda = 0.5$, for the middle row $\gamma = 0.5$ and $\lambda = 0.5$ and  $\gamma = 0.5$ and $\lambda = 0.25$ for the bottom row. From looking at the cross-sectional slices we clearly that  $\xi_j \approx \frac{\gamma}{\lambda} \rho_j, j=1,2,3$ and this agrees with our steady-state solution which was $\xi_j = \frac{\gamma}{\lambda} \rho_j$.}
        \label{fig:seg_state_2}
\end{figure}

From the steady-state solutions of the continuum system in (\ref{E:steadystateequation}) we see that the graffiti density is  $\xi = \frac{\gamma}{\lambda} \rho$, which implies that the steady-state solution of the graffiti and agent densities at a site are equal up to scaled amount of $\frac{\gamma}{\lambda}.$ To check whether this generally holds for our discrete model in Section \ref{S:simulations}, we test our discrete system with several ratio values that are different. We also  take a cross-sectional slice over the lattice at the first and final time steps. This would allow us to see if the steady-state of the discrete model and the continuum system agree. We visualize the results in Figure \ref{fig:seg_state_2}.

\section{Linear Stability Analysis} \label{section:LSA}

To have a better understanding of our system we linearize our model by considering a perturbation of  the equilibrium solution (\ref{E:steadystateequation}) to the well-mixed state. We assume that our perturbations are of the form $\epsilon = \delta e^{\alpha t}e^{ikx}$, with $\delta < < 1$, and in that case our solution will take the following form:
\begin{align}
\begin{cases}
\xi_j &= \bar{\xi_j} + \delta_{\xi_j} e^{\alpha t}e^{ikx}\\
\rho_j &= \bar{\rho_j} + \delta_{\rho_j} e^{\alpha t}e^{ikx}\label{E:dimensionalpertubation}
\end{cases}
\end{align}
Here, $e^{ikx} = \cos(kx) + i\sin(kx)$, where $k$ represents the wave number of the spatial wave. In order for the equilibrium solution to be stable, $\alpha$ must be negative so that it forces the perturbations to decay as time increases. For more examples of this kind of perturbation being used to study the stability of equilibrium solutions, the interested reader is referred to \cite{BLP2002,JBC2010,SDPTBBC2008,WLM1996}.

To analyze the dynamics of these solutions, we now substitute \eqref{E:dimensionalpertubation} into the evolution equations (\ref{T:continuum_eqns}). We start with the first equation:
\begin{equation}
\frac{\partial  \xi_j}{ \partial  t} =  \gamma  \rho_j - \lambda \xi_j. \label{E:dimensionallsa1}
\end{equation}
Substituting \eqref{E:dimensionalpertubation} into (\ref{E:dimensionallsa1}) yields
\begin{align}
\frac{\partial}{\partial  t} \left( \bar{\xi_j} + \delta_{\xi_j} e^{\alpha t}e^{ikx} \right) &= \gamma (\bar{\rho_j} +\delta_{\rho_j} e^{\alpha t}e^{ikx}) -\lambda (\bar{\xi_j} +\delta_{\xi_j} e^{\alpha t}e^{ikx} ). \notag \\
\intertext{Since we assumed $\bar{\xi_j}$ to be an equilibrium solution, its derivative with respect to time is zero, }
\alpha \delta_{\xi_j} e^{\alpha t}e^{ikx} &= (\gamma \bar{\rho_j} - \lambda \bar{\xi_j} ) + (\gamma \delta_{\rho_j}-\lambda \delta_{\xi_j})e^{\alpha t}e^{ikx}, \notag \\
&=(\gamma \delta_{\rho_j}-\lambda \delta_{\xi_j})e^{\alpha t}e^{ikx}. \notag
\intertext{Since $\gamma  \bar{\rho_j} - \lambda \bar{\xi_j} =   \frac{\partial  \bar{\xi_j}}{ \partial  t} = 0$. Hence}
\alpha \delta_{\xi_j} &= (\gamma \delta_{\rho_j}-\lambda \delta_{\xi_j}), \text{ for j=1,2, \dots, K.} \label{E:dimensionalstable1}
\end{align}
Next, we substitute \eqref{E:dimensionalpertubation}  into the evolution equation
\begin{equation*}
\frac{\partial \rho_j}{\partial t} =  \frac{D}{4} \nabla \cdot \Big[ \nabla \rho_j  + 2  \beta \big(\rho_j \nabla \psi_j \big) \Big] 
\end{equation*}
\noindent giving us
\begin{align}
\frac{\partial}{\partial  t}  \left( \bar{\rho_j} + \delta_{\rho_j} e^{\alpha t}e^{ikx} \right) &=  \frac{D}{4} \Delta \left( \bar{\rho_j} + \delta_{\rho_j} e^{\alpha t}e^{ikx} \right) \notag \\
&\quad+  \frac{D\beta}{2} \nabla\cdot \left( (\bar{\rho_j} + \delta_{\rho_j} e^{\alpha t}e^{ikx} ) \nabla \left( \sum_{\substack{l=1 \\ l\neq j}}^K (\bar{\xi_l} + \delta_{\xi_l} e^{\alpha t}e^{ikx})\right) \right). \notag
\end{align}
\noindent Since we are working in one dimension and our equilibrium solution is a constant in both space and time,
\begin{align}
\alpha \delta_{\rho_j} e^{\alpha t}e^{ikx} &= \frac{-D|k|^2}{4} \delta_{\rho_j} e^{\alpha t}e^{ikx}  + \frac{D\beta}{2} \frac{d}{d x} \left( (\bar{\rho_j} + \delta_{\rho_j} e^{\alpha t}e^{ikx} ) (ik \sum_{\substack{l=1 \\ l\neq j}}^K \delta_{\xi_l} e^{\alpha t}e^{ikx} )\right) \notag \\
&=  \frac{-D|k|^2}{4} \delta_{\rho_j} e^{\alpha t}e^{ikx}  + \frac{D\beta}{2} \frac{d}{d x} \left( ik\bar{\rho_j}  \sum_{\substack{l=1 \\ l\neq j}}^K \delta_{\xi_l} e^{\alpha t}e^{ikx}  \right) +\mathcal{O}(\delta_{\rho_j}\sum_{\substack{l=1 \\ l\neq j}}^K\delta_{\xi_l}) \notag \\
&=  \frac{-D|k|^2}{4} \delta_{\rho_j} e^{\alpha t}e^{ikx}  - \frac{D\beta |k|^2}{2}\bar{\rho_j} \sum_{\substack{l=1 \\ l\neq j}}^K\delta_{\xi_l} e^{\alpha t}e^{ikx}  +\mathcal{O}(\delta_{\rho_j}\sum_{\substack{l=1 \\ l\neq j}}^K\delta_{\xi_2}) \notag \\
&= \frac{-D|k|^2}{4} \left( \delta_{\rho_j} + 2 \beta \bar{\rho_j} \sum_{\substack{l=1 \\ l\neq j}}^K\delta_{\xi_l} \right) e^{\alpha t}e^{ikx} +\mathcal{O}(\delta_{\rho_j}\sum_{\substack{l=1 \\ l\neq j}}^K\delta_{\xi_l}). \notag 
\end{align}
We can safely neglect the term  $\displaystyle \mathcal{O}(\delta_{\rho_j}\sum_{\substack{l=1 \\ l\neq j}}^K\delta_{\xi_l})$, since $|\delta_{\rho_j}|,|\delta_{\xi_l}| < < 1$; therefore,
\begin{equation}
\alpha \delta_{\rho_j} = \frac{-D|k|^2}{4} \left( \delta_{\rho_j} + 2 \beta \bar{\rho_j} \sum_{\substack{l=1 \\ l\neq j}}^K\delta_{\xi_l} \right), \text{ for j=1, 2, $\dots$, K.} \label{E:dimensionalstable3}
\end{equation}

\begin{figure}[!htb]
        \begin{subfigure}[b]{0.33\linewidth}
               \includegraphics[width=4.0cm,,keepaspectratio]{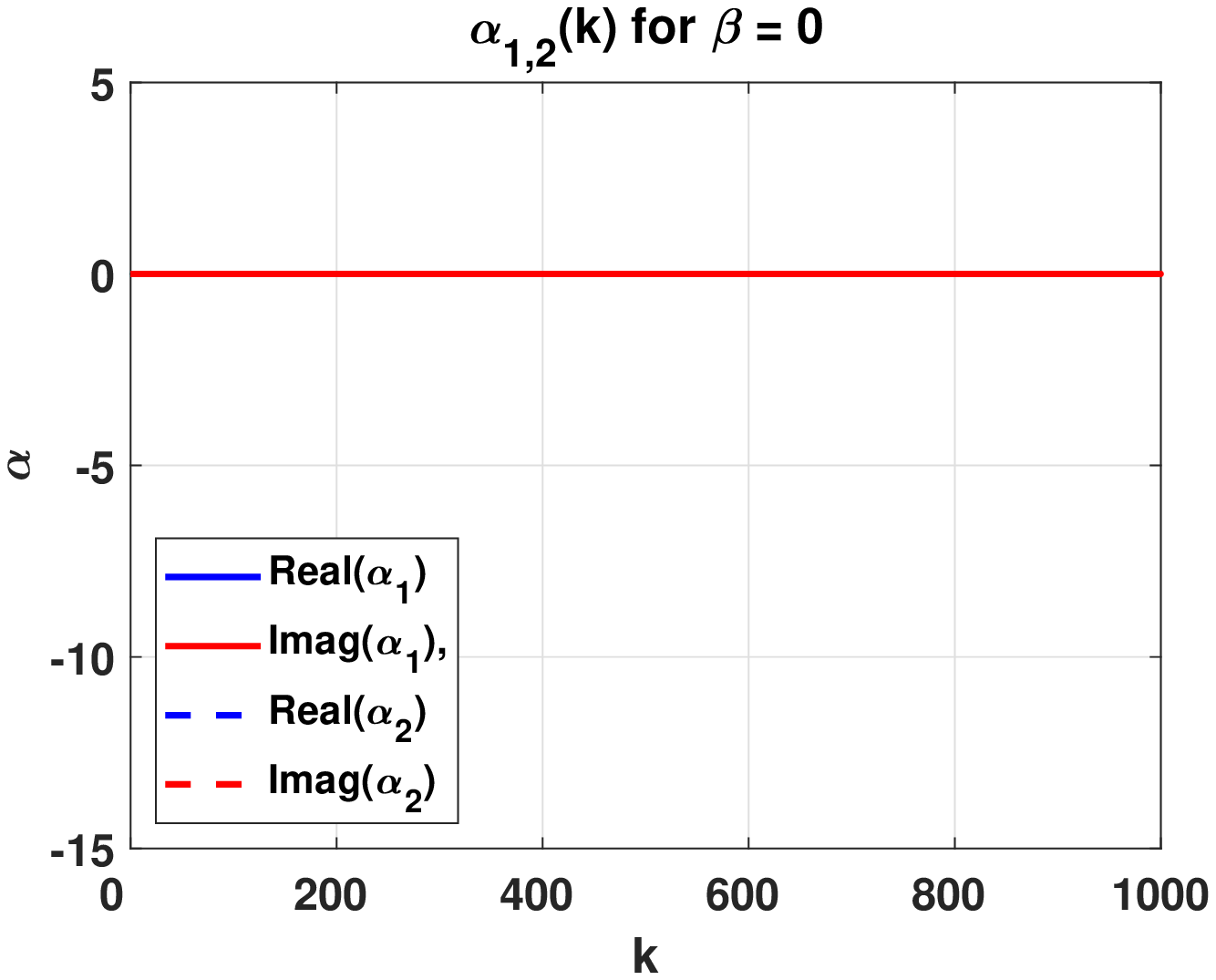}
        \end{subfigure}%
        \begin{subfigure}[b]{0.33\linewidth}
                \includegraphics[width=4.0cm,,keepaspectratio]{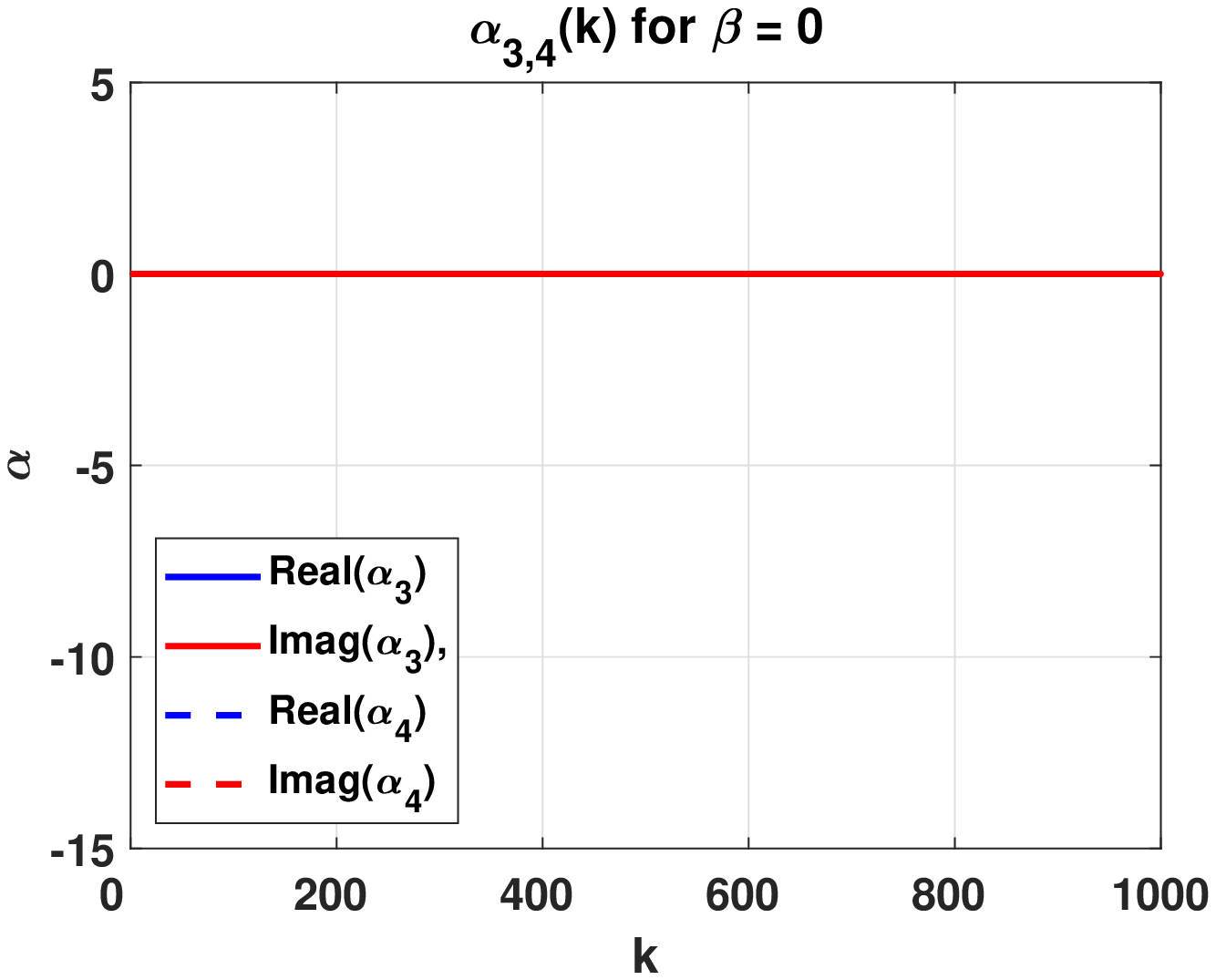}
        \end{subfigure}%
        \begin{subfigure}[b]{0.33\linewidth}
                \includegraphics[width=4.0cm,,keepaspectratio]{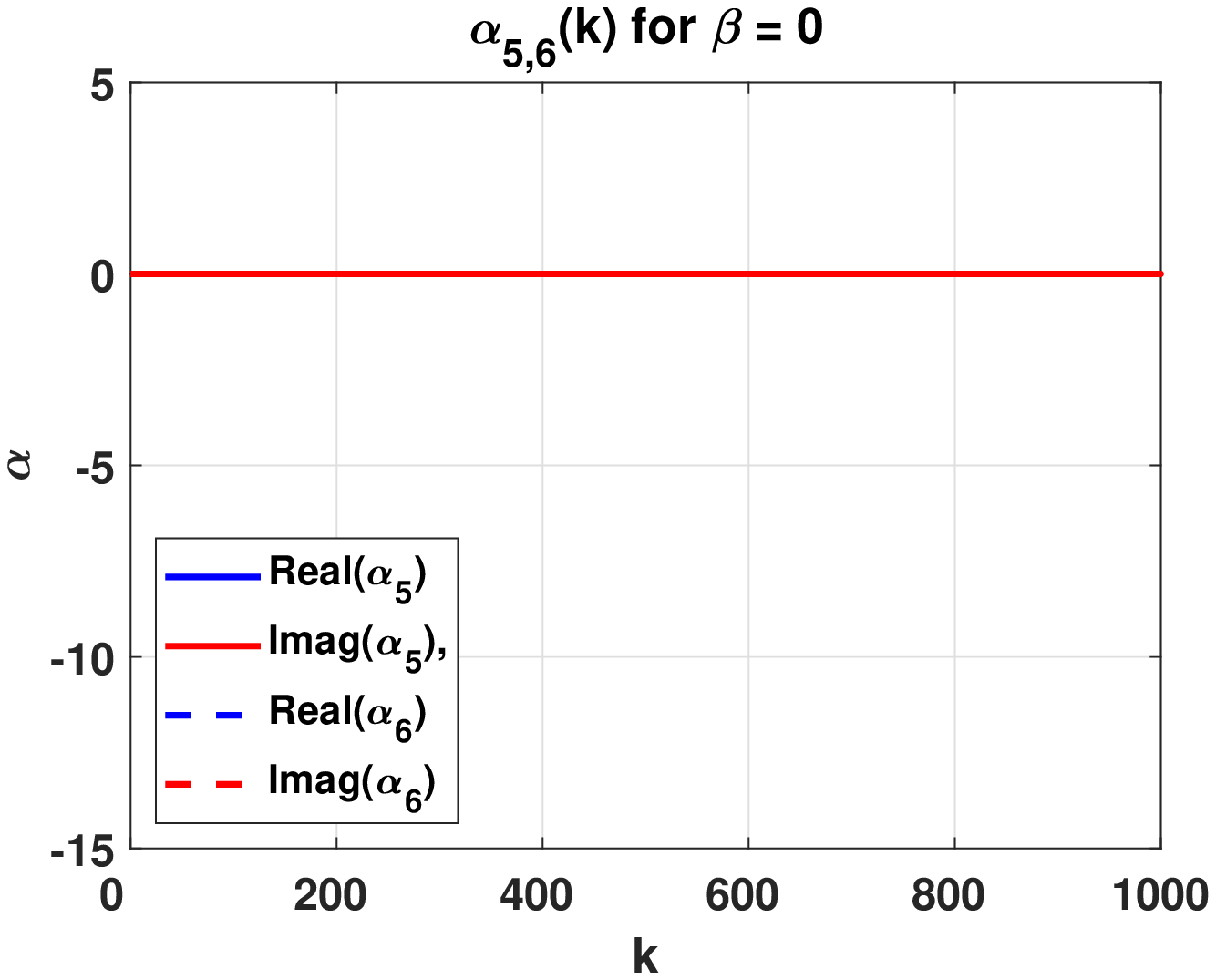}
        \end{subfigure}%
        
        \begin{subfigure}[b]{0.33\linewidth}
               \includegraphics[width=4.0cm,,keepaspectratio]{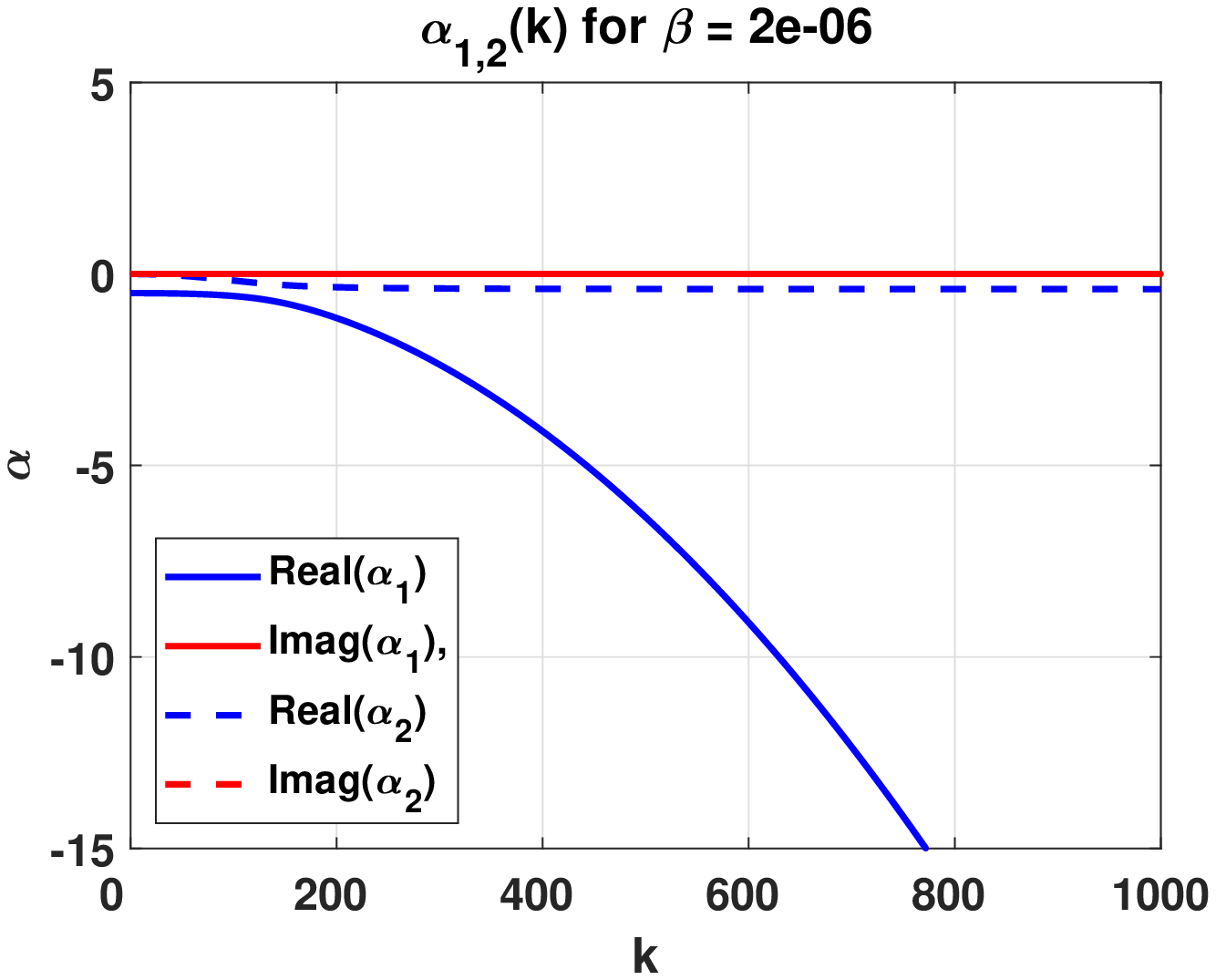}
        \end{subfigure}%
        \begin{subfigure}[b]{0.33\linewidth}
                \includegraphics[width=4.0cm,,keepaspectratio]{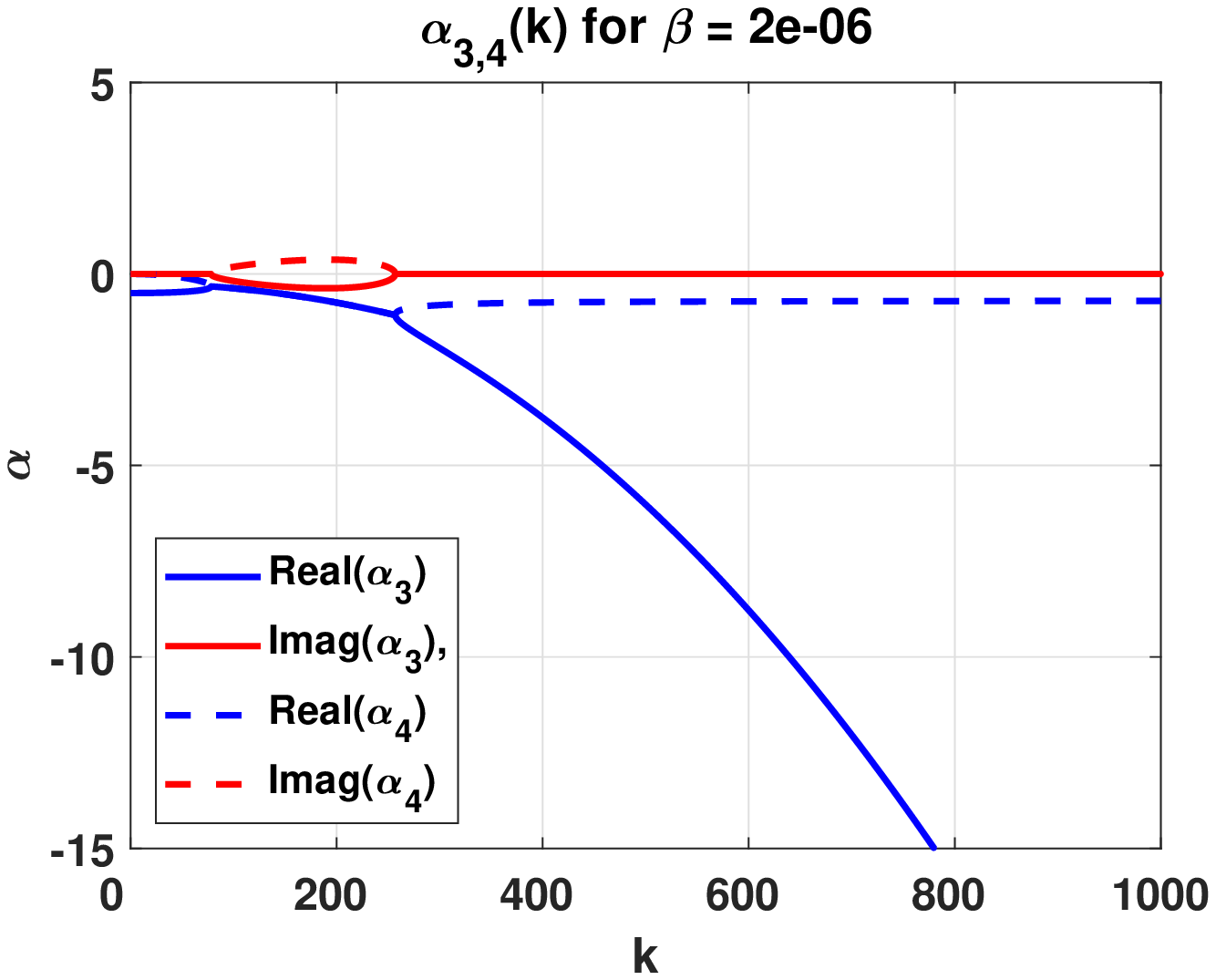}
        \end{subfigure}%
        \begin{subfigure}[b]{0.33\linewidth}
                \includegraphics[width=4.0cm,,keepaspectratio]{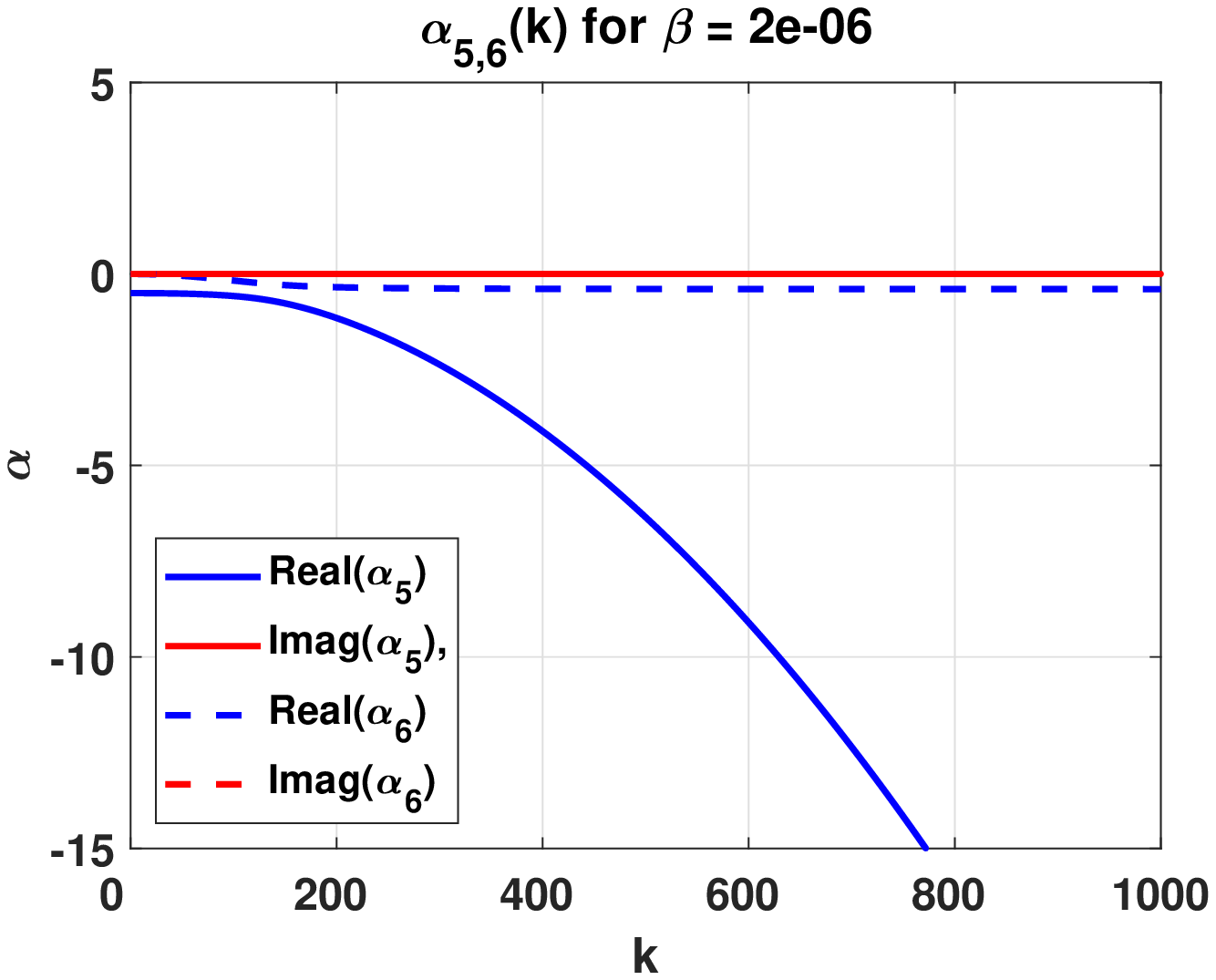}
        \end{subfigure}%
 
         \begin{subfigure}[b]{0.33\linewidth}
               \includegraphics[width=4.0cm,,keepaspectratio]{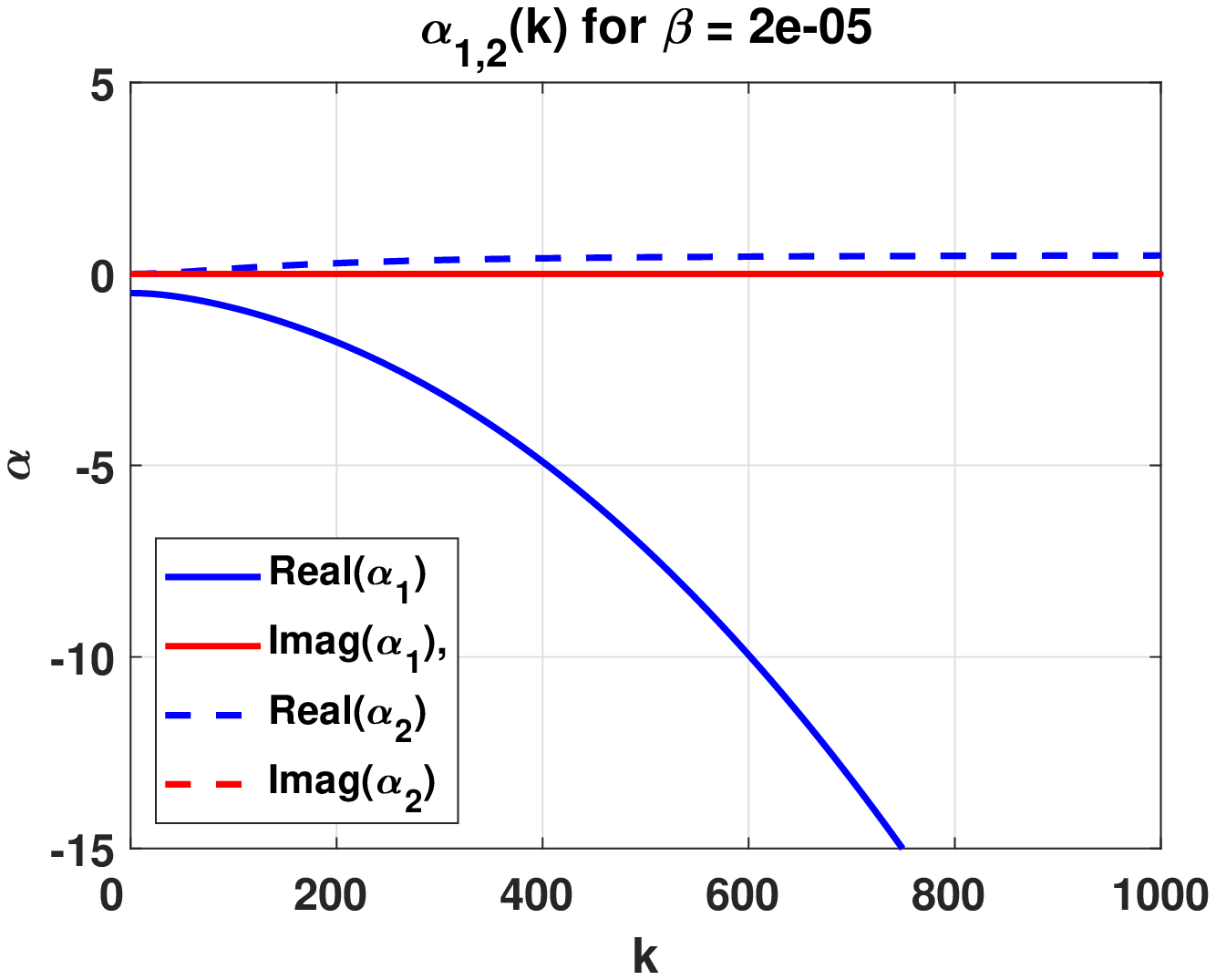}
        \end{subfigure}%
        \begin{subfigure}[b]{0.33\linewidth}
                \includegraphics[width=4.0cm,,keepaspectratio]{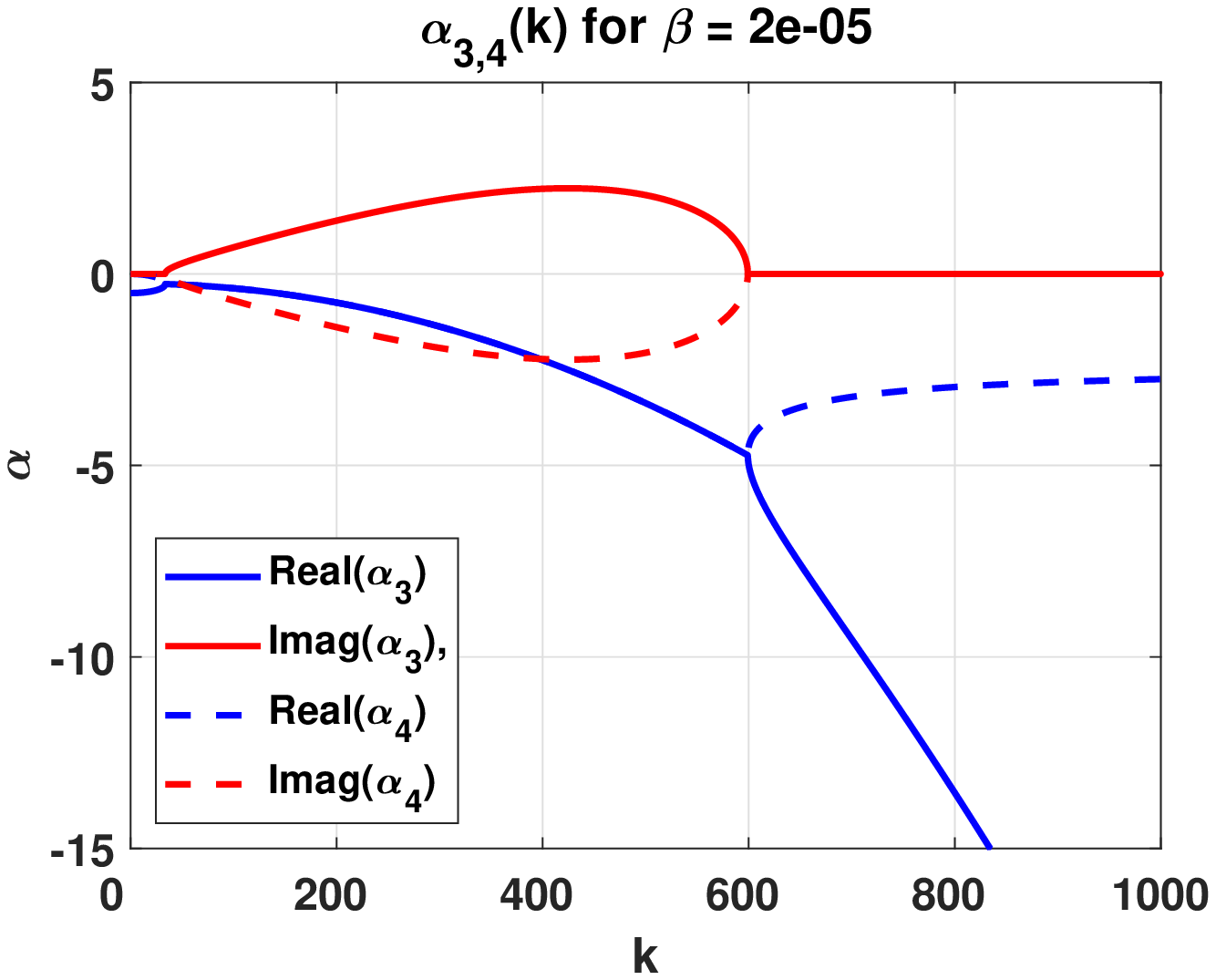}
        \end{subfigure}%
        \begin{subfigure}[b]{0.33\linewidth}
                \includegraphics[width=4.0cm,,keepaspectratio]{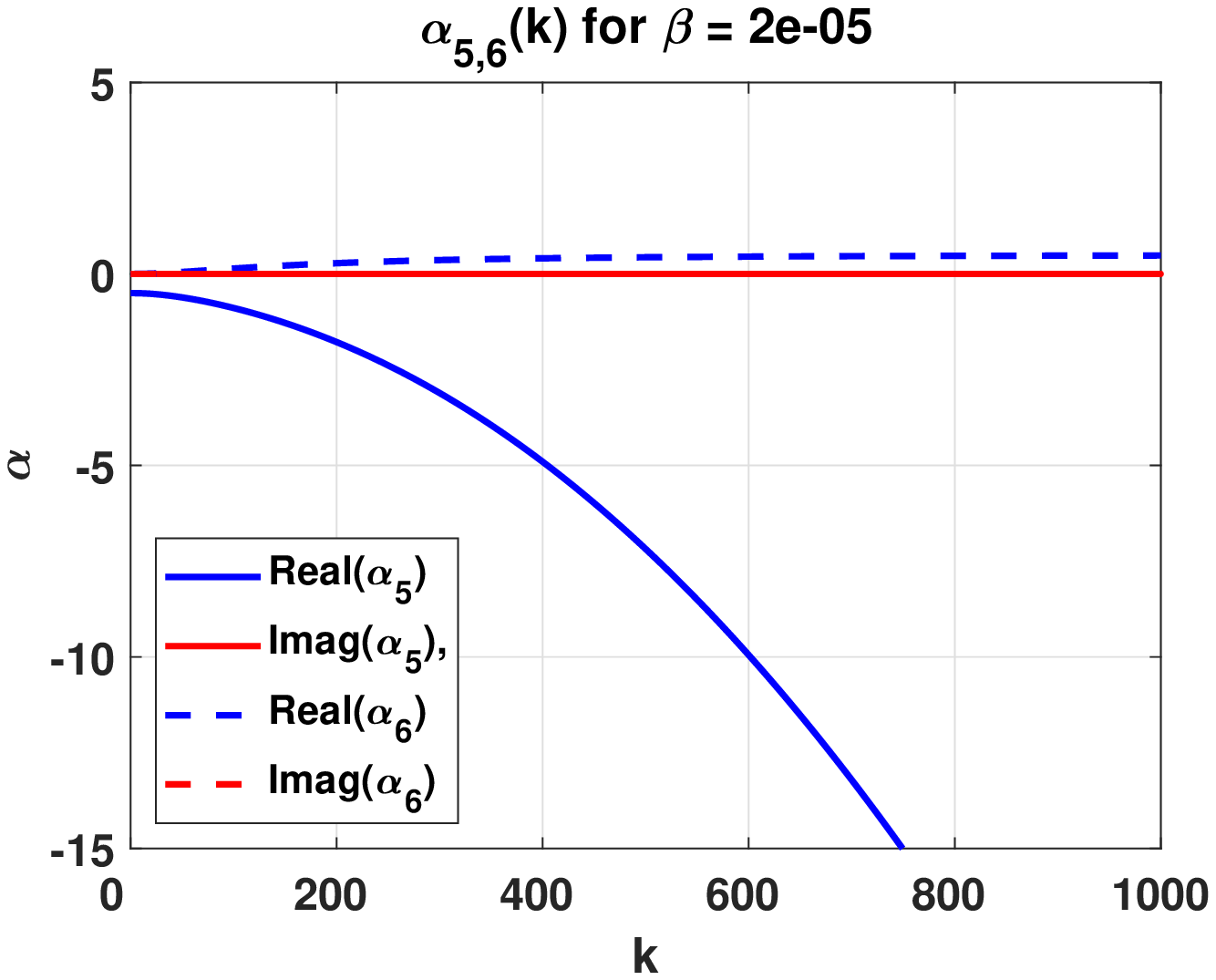}
        \end{subfigure}%
                \caption[Linear Stability Analysis.]{The six eigenvalues versus the wave number k plotted for different $\beta$ values. Here we have $D= 0.0001$, mass  $ = 150,000$ and the $\frac{\gamma}{\lambda}$ ratio $= 1$.}
        \label{fig:LSA}
\end{figure}

Next, we write the equations from \eqref{E:dimensionalstable1} and \eqref{E:dimensionalstable3} in a systems form: 
\begin{align*}
(\gamma \delta_{\rho_j}-\lambda \delta_{\xi_j}) &= \alpha \delta_{\xi_j}  \\
 \frac{-D|k|^2}{4} \left( \delta_{\rho_j} + 2 \beta \bar{\rho_j} \sum_{\substack{l=1 \\ l\neq j}}^K \delta_{\xi_l} \right) &= \alpha \delta_{\rho_j}, \text{ where }j=1, \dots, K. 
\end{align*}
For simplicity we consider the case where $K=3$, with all gangs having the same $\beta$ parameter and write the system in matrix vector format, giving us:
\begin{equation*}
\begin{bmatrix}
-\lambda & 0 & 0 & \gamma & 0 & 0 \\
0 & -\lambda & 0 & 0 &\gamma & 0\\
0 & 0 &-\lambda & 0 & 0 &\gamma\\
0 & \frac{-\beta D\bar{\rho_1} |k|^2}{2} & \frac{-\beta D\bar{\rho_1} |k|^2}{2} & \frac{-D|k|^2}{4} & 0 & 0 \\
\frac{-\beta D\bar{\rho_2} |k|^2}{2} & 0 & \frac{-\beta D\bar{\rho_2} |k|^2}{2} & 0 & \frac{-D|k|^2}{4} & 0 \\
\frac{-\beta D\bar{\rho_3} |k|^2}{2} &  \frac{-\beta D\bar{\rho_3} |k|^2}{2} & 0 &  0 & 0 & \frac{-D|k|^2}{4}  \\
\end{bmatrix} \begin{bmatrix}  \delta_{\xi_1}\\  \delta_{\xi_2}\\  \delta_{\xi_3}\\  \delta_{\rho_1}\\  \delta_{\rho_2} \\  \delta_{\rho_3}\end{bmatrix} = \alpha \begin{bmatrix}  \delta_{\xi_1}\\  \delta_{\xi_2}\\  \delta_{\xi_3}\\ \delta_{\rho_1}\\  \delta_{\rho_2} \\  \delta_{\rho_3}\end{bmatrix}.
\end{equation*}
This gives us
\begin{align}
F \vec{\delta} &= \alpha \vec{\delta} \notag \\
\iff \left(F - \alpha I_4 \right)\vec{\delta} &= 0, \notag
\end{align}
which reduces to an eigenvalue problem for matrix $F$. For the problem to have a non trivial solution (i.e. $\vec{\delta} \neq 0$), the determinant of $(F - \alpha I_4)$ must be zero. Therefore,
\begin{equation*}
\begin{vmatrix}
-(\lambda + \alpha) & 0 & 0 & \gamma & 0 & 0 \\
0 & -(\lambda + \alpha) & 0 & 0 &\gamma & 0\\
0 & 0 &-(\lambda + \alpha) & 0 & 0 &\gamma\\
0 & \frac{-\beta D\bar{\rho_1} |k|^2}{2} & \frac{-\beta D\bar{\rho_1} |k|^2}{2} & -\left( \frac{D|k|^2}{4} + \alpha \right) & 0 & 0 \\
\frac{-\beta D\bar{\rho_2} |k|^2}{2} & 0 & \frac{-\beta D\bar{\rho_2} |k|^2}{2} & 0 & -\left( \frac{D|k|^2}{4} + \alpha \right) & 0 \\
\frac{-\beta D\bar{\rho_3} |k|^2}{2} &  \frac{-\beta D\bar{\rho_3} |k|^2}{2} & 0 &  0 & 0 & -\left( \frac{D|k|^2}{4} + \alpha \right)  \\
\end{vmatrix}=0,
\end{equation*}
giving us the following characteristic polynomial
\begin{align*}
f(\alpha) &= \frac{1}{64}  \Big[4 \alpha D^2 (\alpha + \lambda) (3 \alpha^2 + 6 \alpha \lambda + 3 \lambda^2 -  4 \beta^2 \gamma^2 (\bar{\rho_1} \bar{\rho_2} + \bar{\rho_1} \bar{\rho_3} + \bar{\rho_2} \bar{\rho_3})) k^4 \notag \\ 
   & + D^3 \Big(\alpha^3 + 3 \alpha^2 \lambda + \lambda^3 + 16 \beta^3 \gamma^3 \bar{\rho_1}\bar{\rho_2}\bar{\rho_3} -   4 \beta^2 \gamma^2 \lambda(\bar{\rho_1} \bar{\rho_2} + \bar{\rho_1} \bar{\rho_3} + \bar{\rho_2} \bar{\rho_3}) \notag \\
  &+ \alpha (3 \lambda^2 - 4 \beta^2 \gamma^2 (\bar{\rho_1} \bar{\rho_2} + \bar{\rho_1} \bar{\rho_3} + \bar{\rho_2} \bar{\rho_3}))\Big) k^6  + 64 \alpha ^3 (\alpha + \lambda)^3 + 48 \alpha^2 D (\alpha + \lambda)^3 k^2 \Big] \notag \\
  &=0.
\end{align*}
Solving the characteristic polynomial gives six eigenvalues, which we solve numerically using Mathematica and plot the results for different values of $\beta$ in Figure \ref{fig:LSA}.

 To determine the stability of our system, we recall that the system becomes linearly unstable when any eigenvalue has a positive real part.  We see in the top two rows of Figure \ref{fig:LSA} that for small $\beta$ values, none of the eigenvalues had a positive real part, and thus the system remained stable. In terms of our model, this makes sense since in our discrete simulations, the system remains well-mixed  for these $\beta$ values. However, when we increase the value of $\beta$ to $0.00005$, we see in the bottom three figures that the second and sixth eigenvalues have a positive real part, and the system has thus become linearly unstable. This again agrees with our physical intuition. We note that the values $\beta$ from the discrete model matches those of the linearized system of partial differential equations.

\section{Variations of the model: varying $\beta$ by gang} \label{section:changing_beta}

In Section \ref{S:discrete} equation  \eqref{E:probability_agent_moves}, we defined the probability that an agent from gang $j$  moves from site $s_1=(x_1,y_1) \in S$ to one of the neighboring sites $s_2=(x_2, y_2) \in S$ to be
\begin{equation*}
M_{j}(x_1 \rightarrow x_2, y_1 \rightarrow y_2, t) = \frac{e^{-\beta \psi_j(x_2, y_2,t)}}{\sum \limits_{(\tilde x, \tilde y) \sim(x_1,y_1)}e^{-\beta \psi_j(\tilde x, \tilde y, t)}},
\end{equation*}
with
\begin{equation*}
\psi_j(x,y,t) := \sum_{\substack{i=1 \\ i\neq j}}^K \xi_i(x, y,t)
\end{equation*}
from equation \eqref{E:graffiti_complement}. The parameter $\beta$ then controls how strongly each gang reacts to the graffiti of the other gangs. However, it is reasonable to consider that this parameter $\beta$ might vary by gang. Here, we explore variations of the model incorporating this idea. In this section, we will make two different modifications of \eqref{E:probability_agent_moves} and explore how these modifications affect the system of PDEs and the segregation behavior of the model.

\subsection{Timidity Model (Variation 1)} 
In the first modification of the model, instead of having identical $\beta$ values for all gangs, we change it so that gang $j$ has a distinct corresponding $\beta$ value, denoted by $\beta_j$. This $\beta_j$ determines how much attention gang $j$ places on the graffiti of the other gangs. In essence, this $\beta_j$ encodes the timidity of gang $j$, with higher $\beta_j$ corresponding to higher timidity, causing gang $j$ to more strongly avoid other gangs' graffiti. Hence, the modified definition for movement becomes, 
\begin{equation}
M_{j}(x_1 \rightarrow x_2, y_1 \rightarrow y_2, t) = \frac{e^{-\beta_j \psi_j(x_2, y_2,t)}}{\sum \limits_{(\tilde x, \tilde y) \sim(x_1,y_1)}e^{-\beta_j \psi_j(\tilde x, \tilde y, t)}}. \label{E:probability_agent_moves_beta1} 
\end{equation}

In this variation of the model, gang $j$ avoids all other gangs' graffiti with rate $\beta_j$. All of the graffiti from other gangs count equally and are identically avoided.  For example, let us consider the case of three gangs $1$, $2$, and $3$ such that gang $2$ has a relatively large $\beta_2$ value, gang $3$ has a relatively small $\beta_3$ value, and gang $1$ has an intermediate $\beta_1$ value. Then gang $2$'s agents would strongly avoid areas where the other two gangs, $1$ and $3$, have tagged. Gang $3$'s agents, on the other hand, would more freely on the lattice, as the small $\beta_3$ value leads it to not place much importance on other gangs' graffiti.  Gang $1$'s agents' movement dynamics would lie somewhere in between.

If one follows the derivation of the continuum equations in Section \ref{section:contiuum_background} but replacing \eqref{E:probability_agent_moves} with \eqref{E:probability_agent_moves_beta1}, it can be easily shown that the resulting system of equations for $j=1,2, \dots, K$ are
\begin{equation}\label{T:continuum_eqns_beta1}
\begin{cases} 
\displaystyle \frac{\partial \xi_j}{\partial t}(x,y,t) = \gamma \rho_j(x,y,t) - \lambda \xi_j(x,y,t) \\
\displaystyle \frac{\partial \rho_j}{\partial t}(x,y,t) =  \frac{D}{4} \nabla \cdot \left[ \nabla \rho_j(x,y,t)  + 2  \beta_j \left(\rho_j(x,y,t) \nabla \left( \sum_{\substack{i=1 \\ i\neq j}}^K \xi_i(x, y,t) \right) \right) \right]
\end{cases}
\end{equation}
\noindent with periodic boundary conditions.  We can see that the $\beta_j$ values will then affect the balance between the diffusion and the advection terms differently depending on the gang affiliation, making diffusion relatively stronger for those gangs with lower $\beta_j$ values.

\begin{figure}[!htb]
        \begin{subfigure}[b]{0.249\linewidth}
               \includegraphics[width=1.75cm,,keepaspectratio]{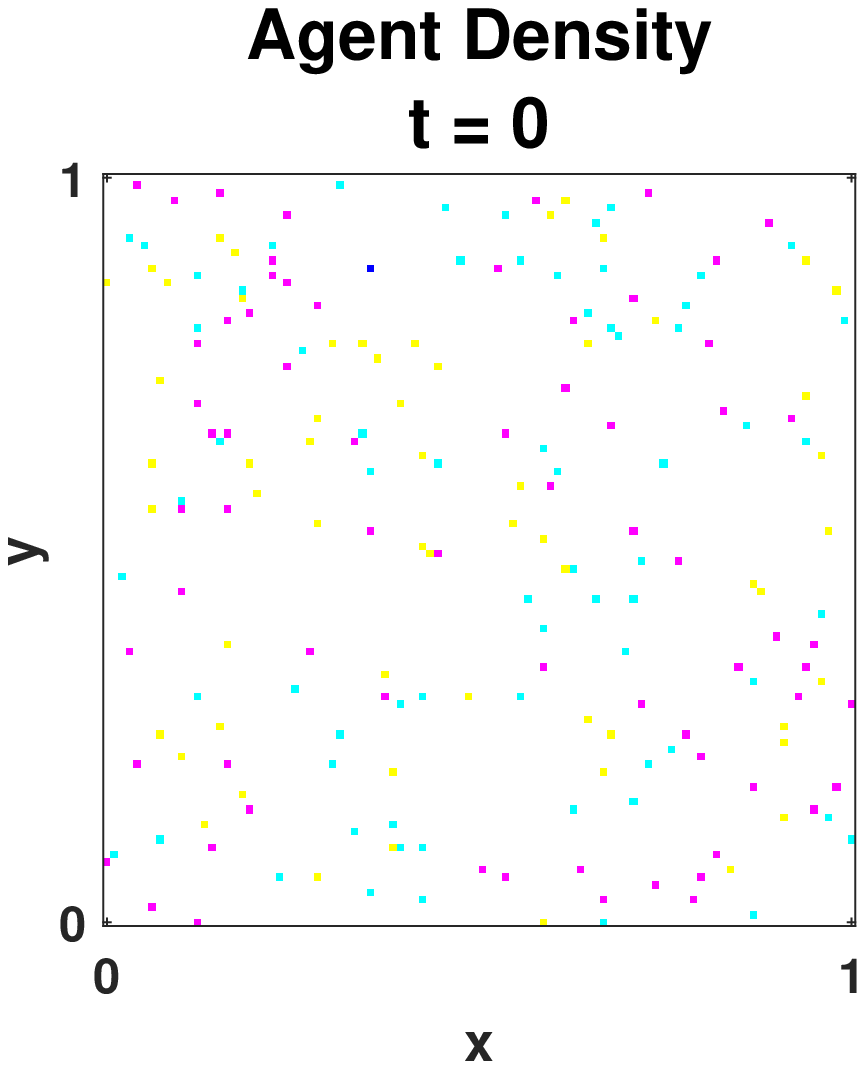}
        \end{subfigure}%
        \begin{subfigure}[b]{0.249\linewidth}
                \includegraphics[width=1.75cm,,keepaspectratio]{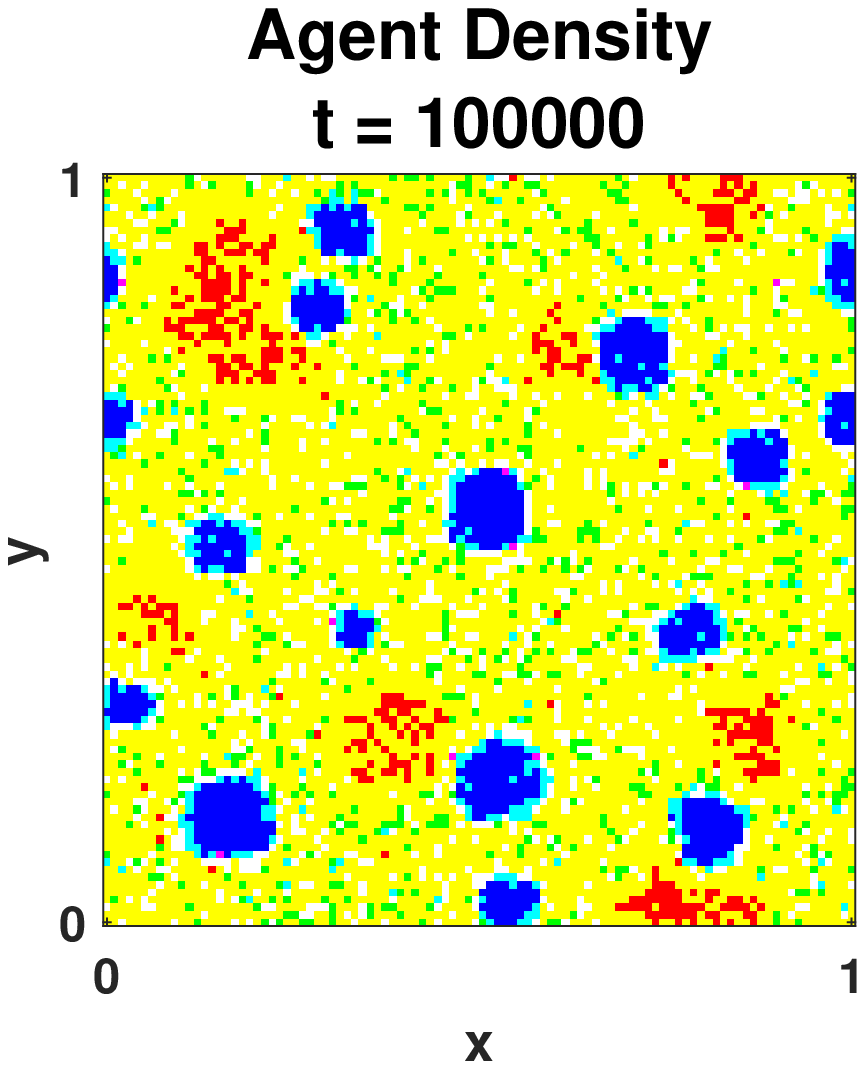}
        \end{subfigure}%
        \begin{subfigure}[b]{0.249\linewidth}
              \includegraphics[width=1.75cm,,keepaspectratio]{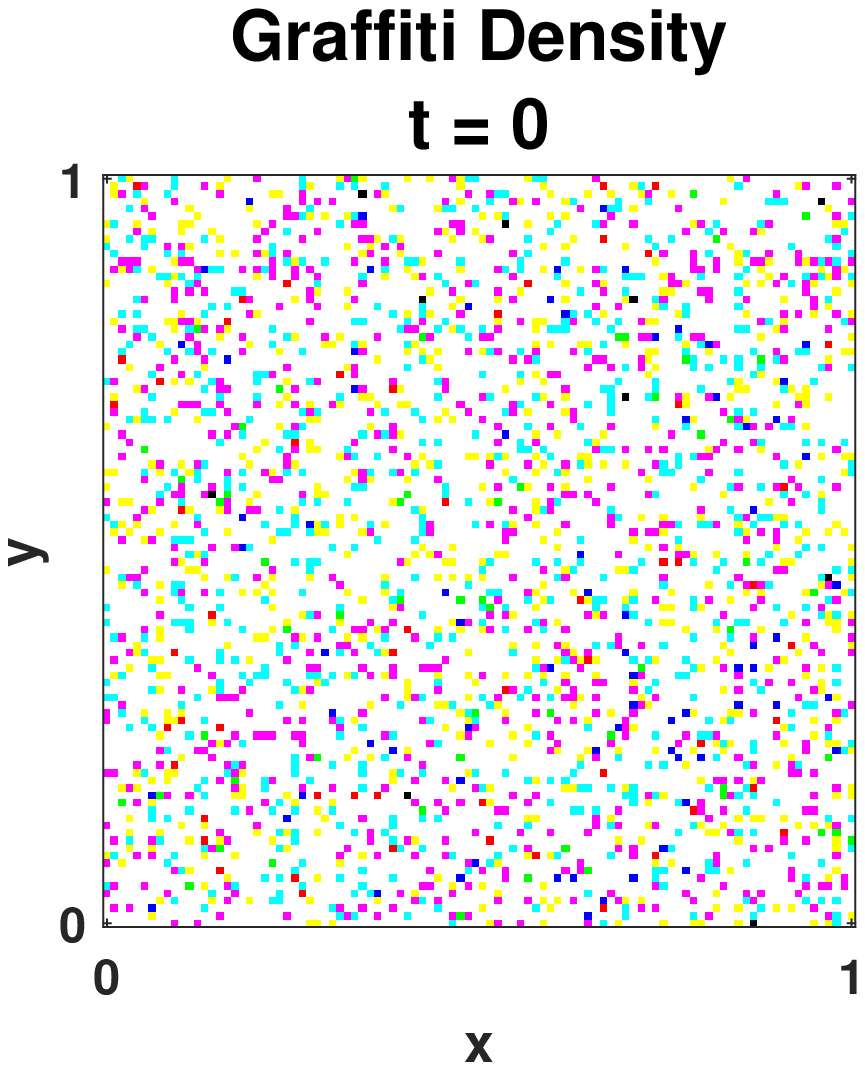}
        \end{subfigure}%
        \begin{subfigure}[b]{0.249\linewidth}
               \includegraphics[ width=1.75cm,keepaspectratio]{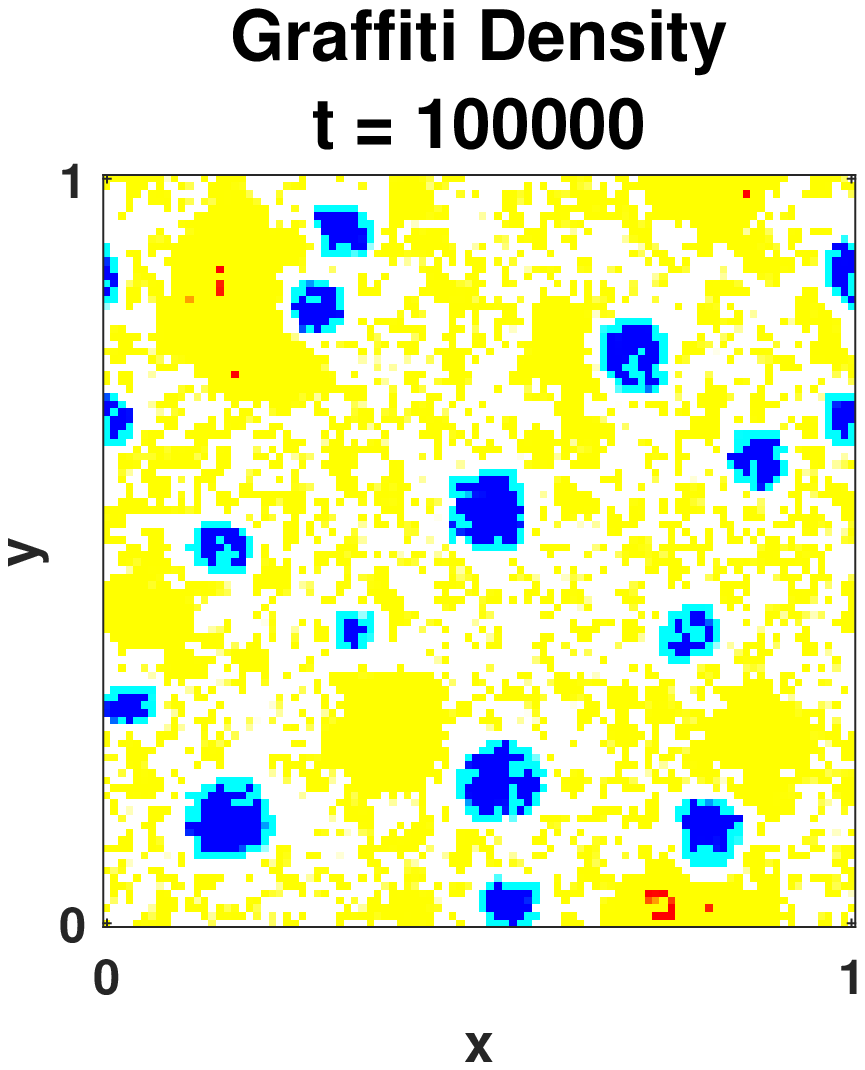}
        \end{subfigure}
        \begin{subfigure}[b]{0.33\linewidth}
               \includegraphics[width=3.0cm,,keepaspectratio]{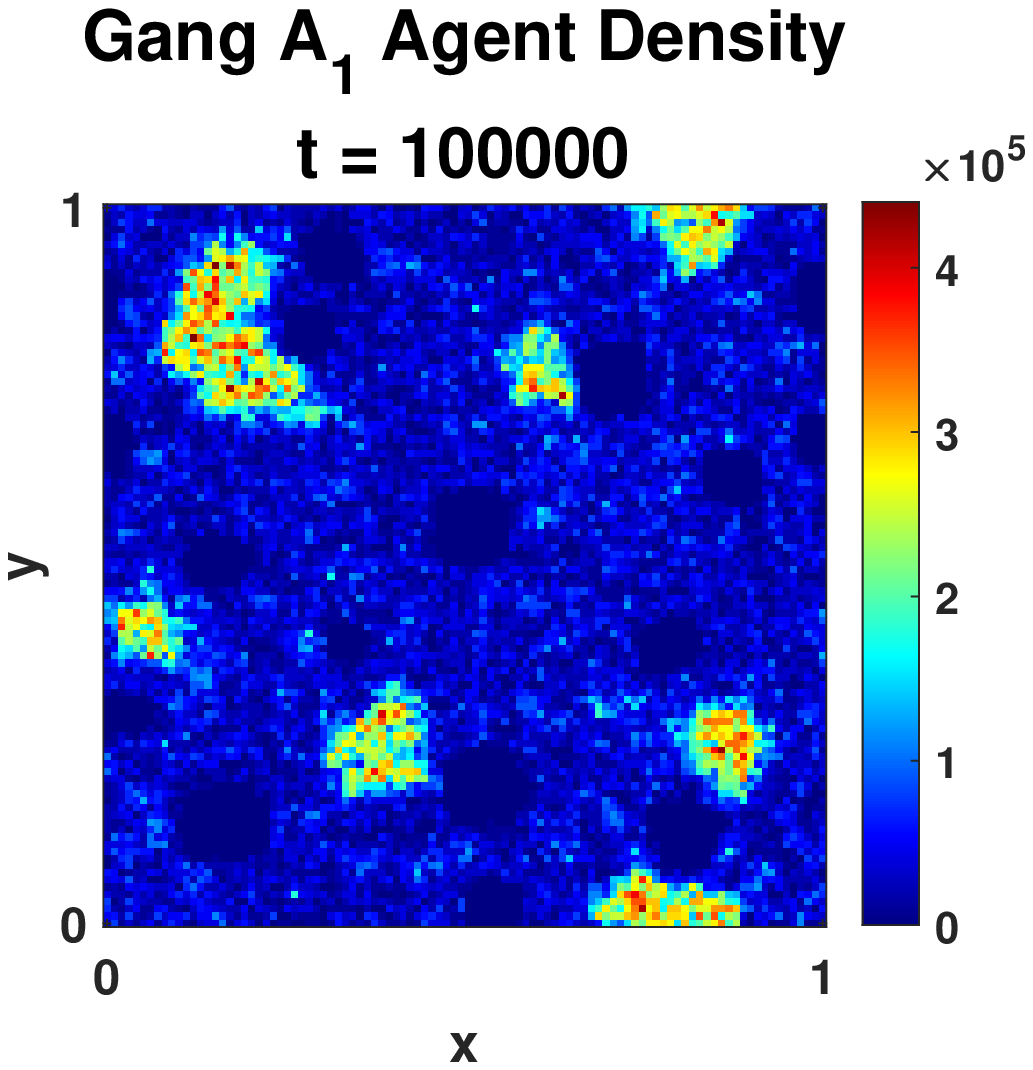}
        \end{subfigure}%
           \begin{subfigure}[b]{0.33\linewidth}
               \includegraphics[width=3.0cm,,keepaspectratio]{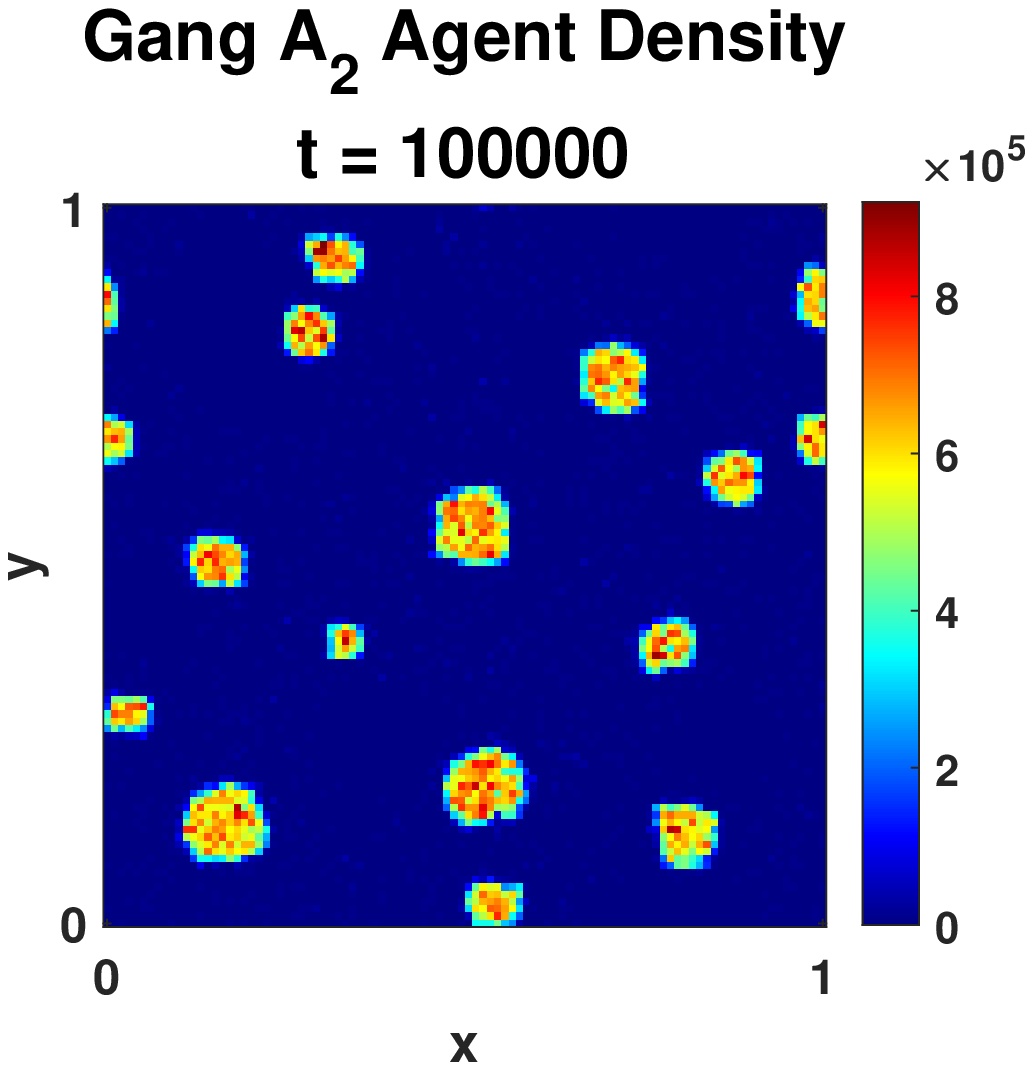}
        \end{subfigure}%
         \begin{subfigure}[b]{0.33\linewidth}
               \includegraphics[width=3.0cm,,keepaspectratio]{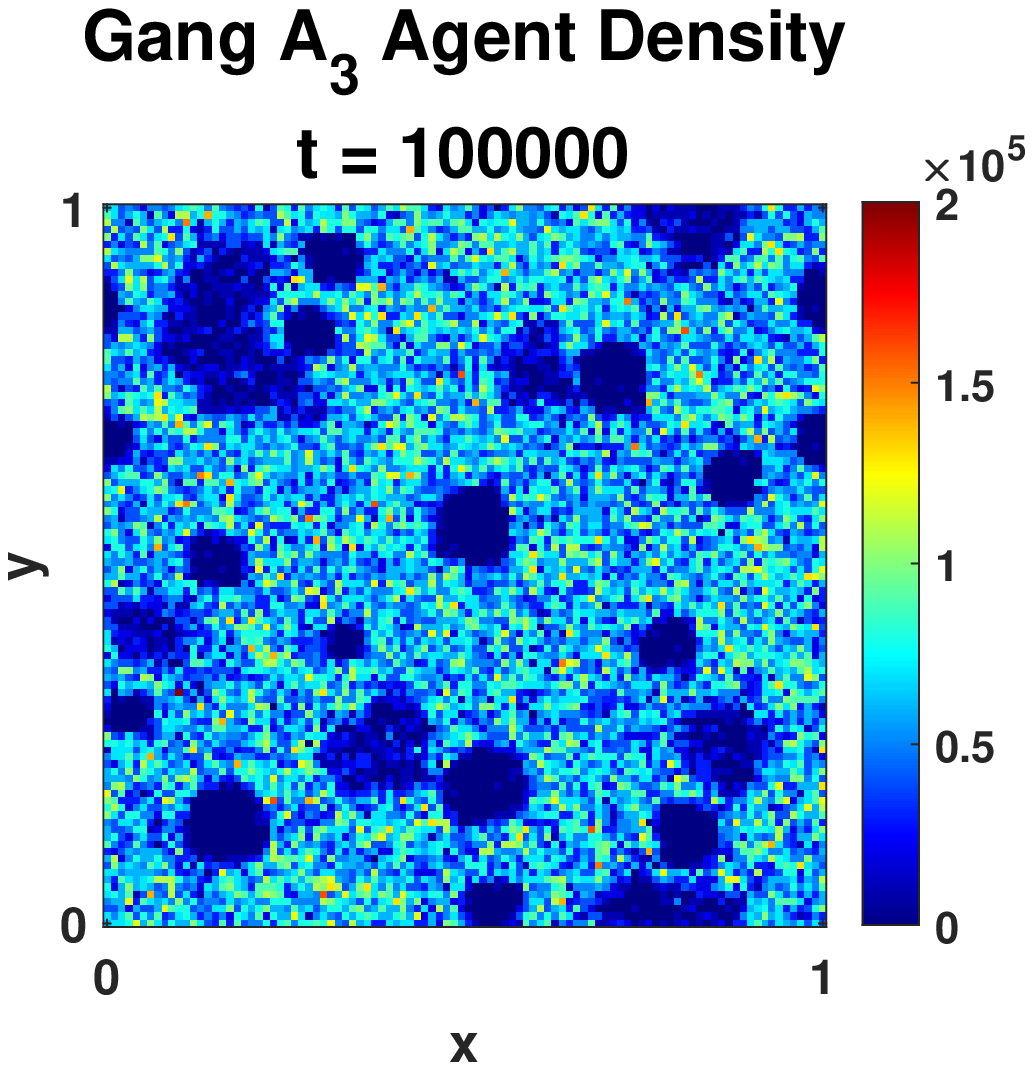}
        \end{subfigure}%
                \caption[Temporal evolution of the agent and graffiti densities lattice for a well-mixed state when $\beta$ depends on the gang.]{\textbf{Top row:} Agent and graffiti densities'  temporal evolution for the Timidity Model.  Here $\beta_{1} = 2 \times 10^{-5}, ~\beta_{2} = 3.5 \times 10^{-5}$ and $\beta_{3} = 0.5 \times 10^{-5}$. We also have $N_1 = N_2 = N_3 =50,000$, with $\lambda = \gamma =0.5$, $\delta t = 1$ and the lattice size is $100 \times 100$. It is clearly seen that the agents segregate over time.  \textbf{Bottom row:} The densities for gangs 1 (left), 2 (middle), and 3 (right) can be seen after $100,000$ time steps.}
        \label{fig:beta_changes_1_Simulations}
\end{figure}

\begin{figure}
\begin{center}
\includegraphics[width=8cm]{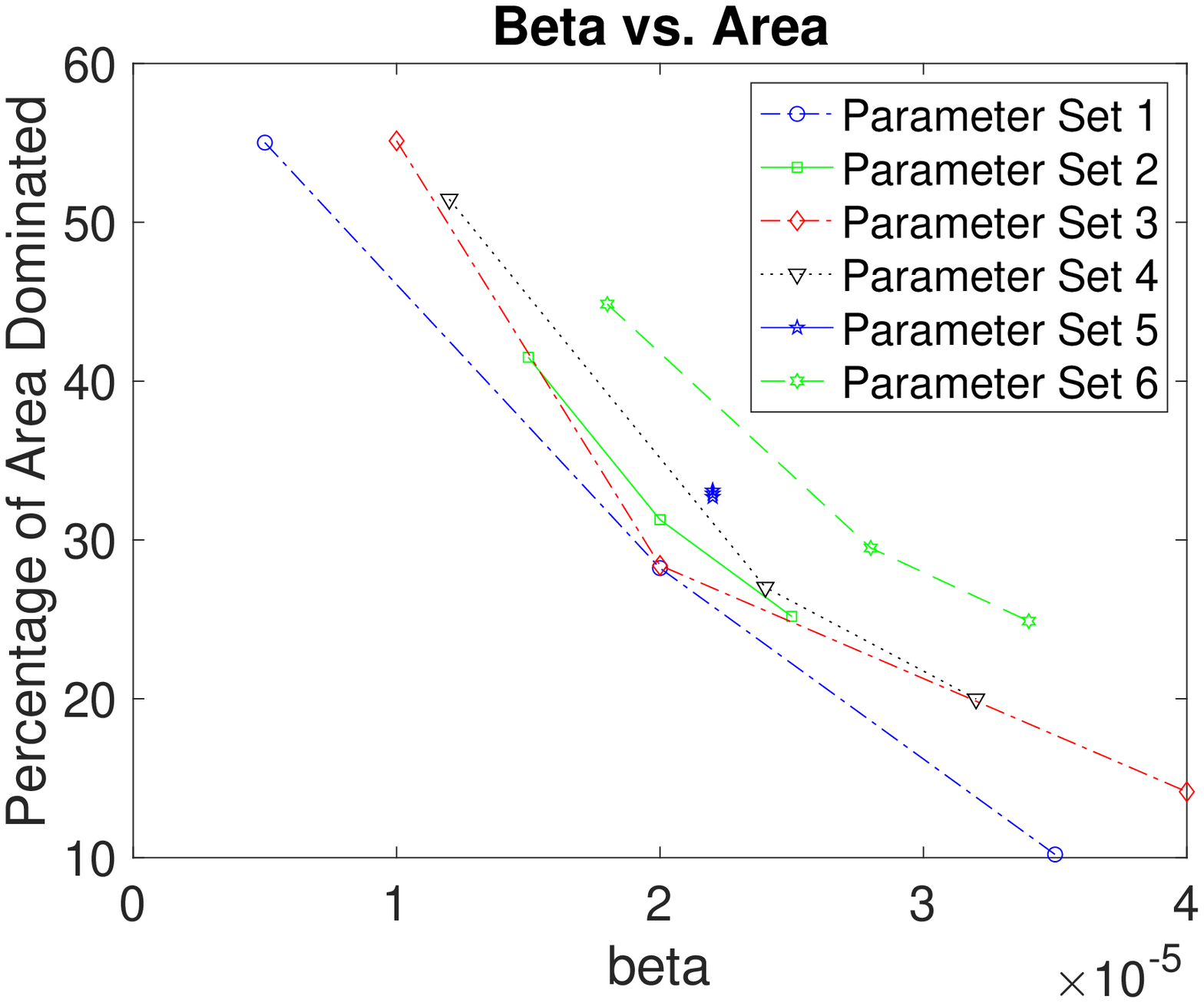}
\end{center}
\includegraphics[width=3.85cm]{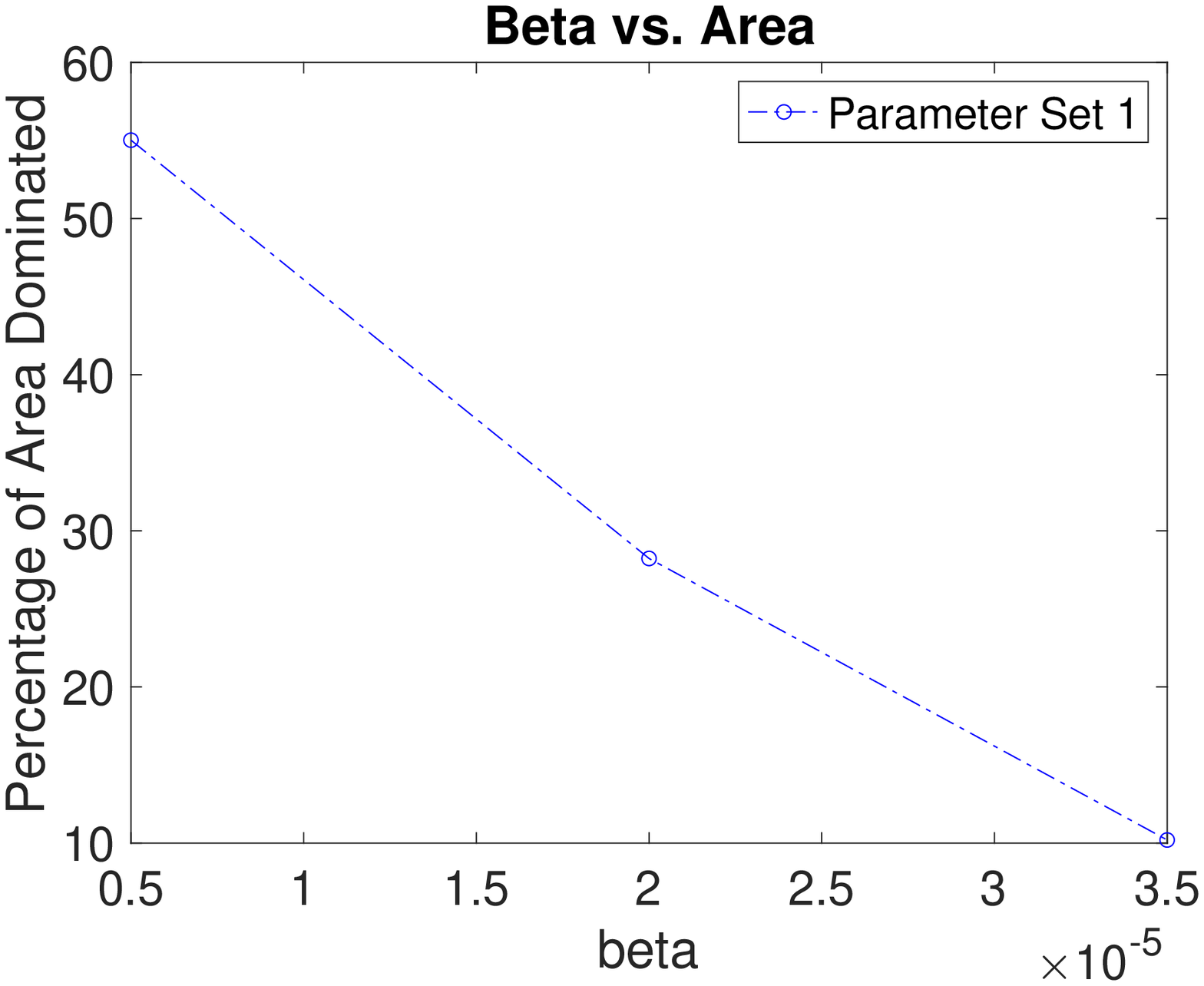}
\includegraphics[width=3.85cm]{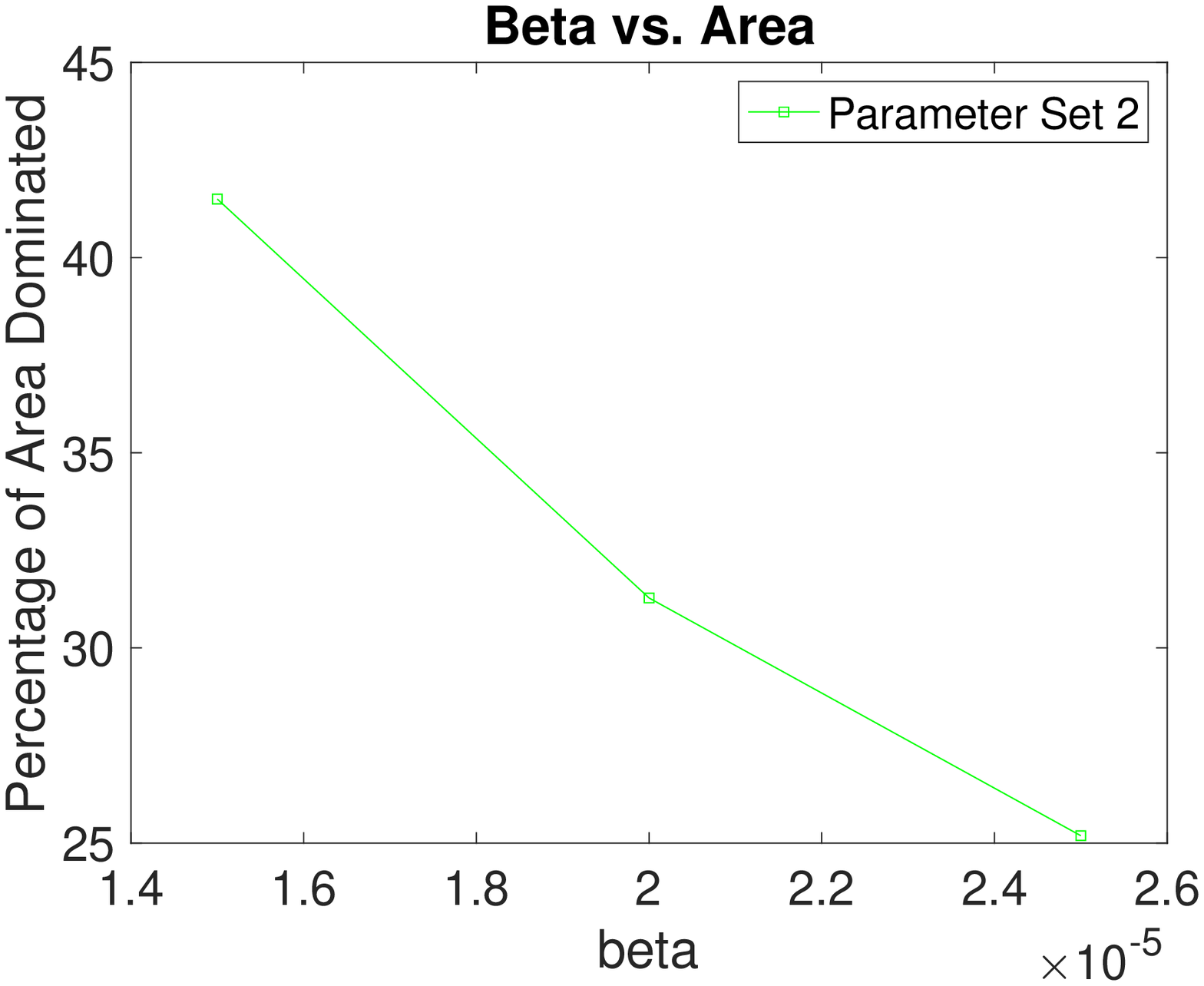}
\includegraphics[width=3.85cm]{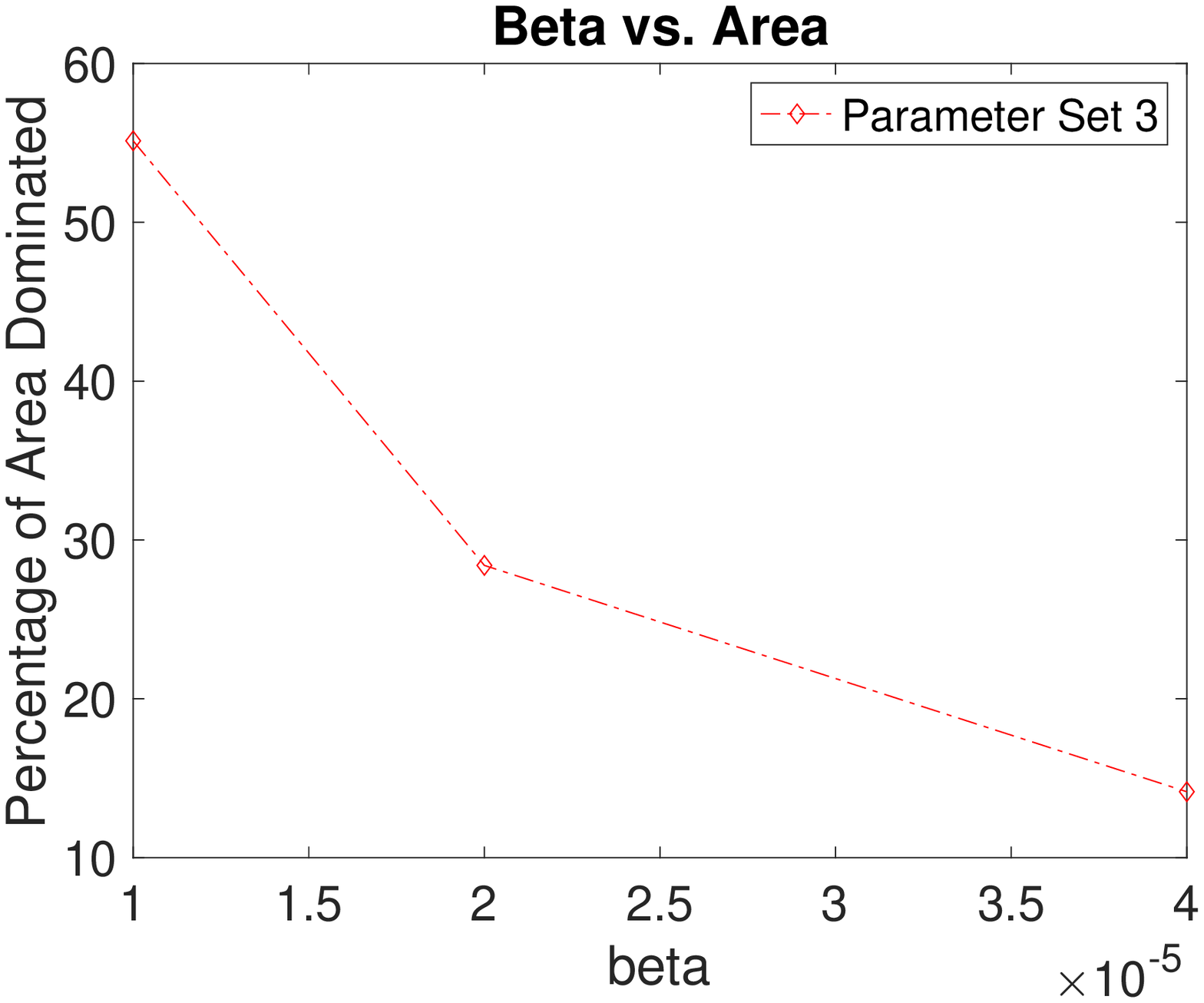}\\
\includegraphics[width=3.85cm]{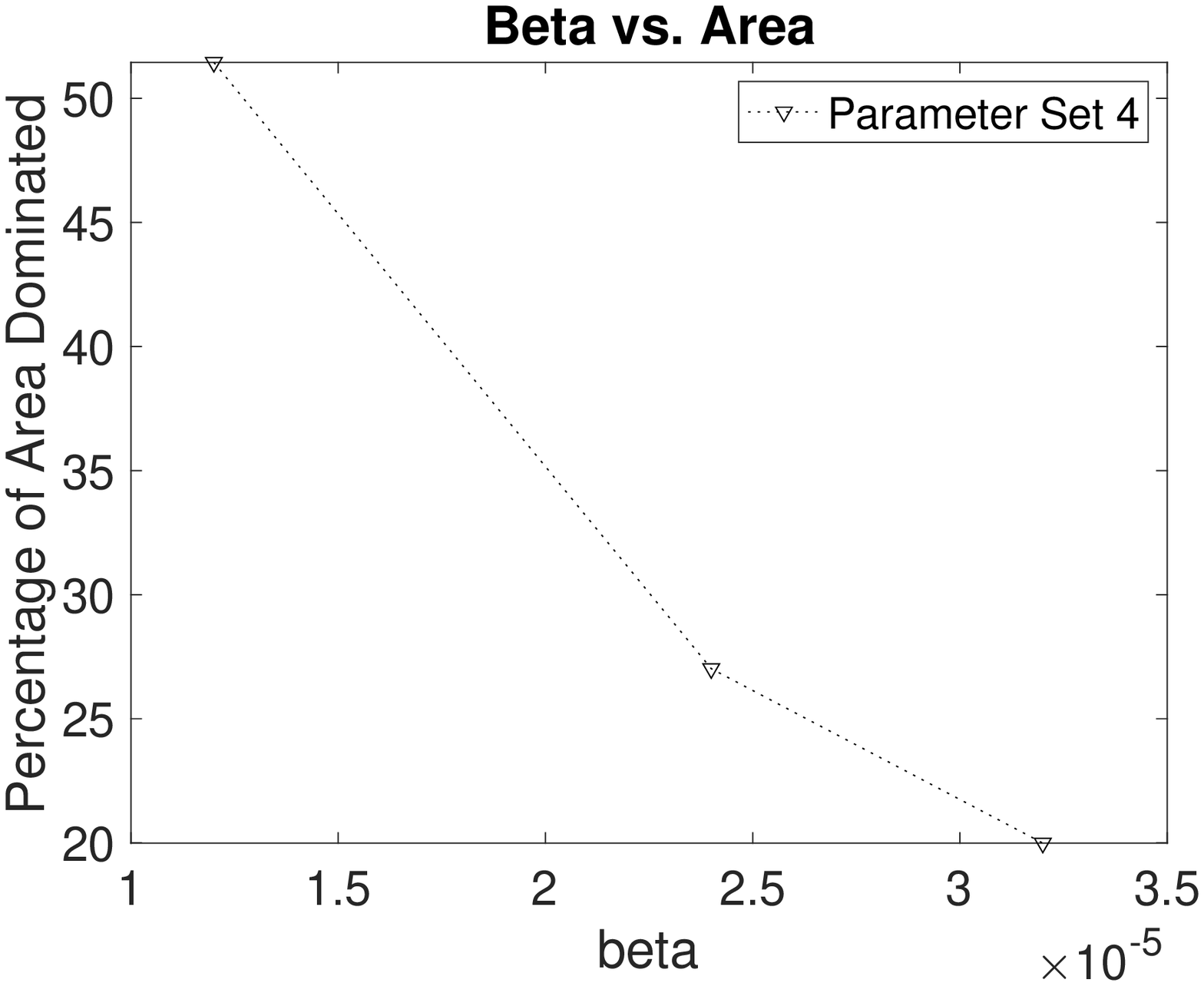}
\includegraphics[width=3.85cm]{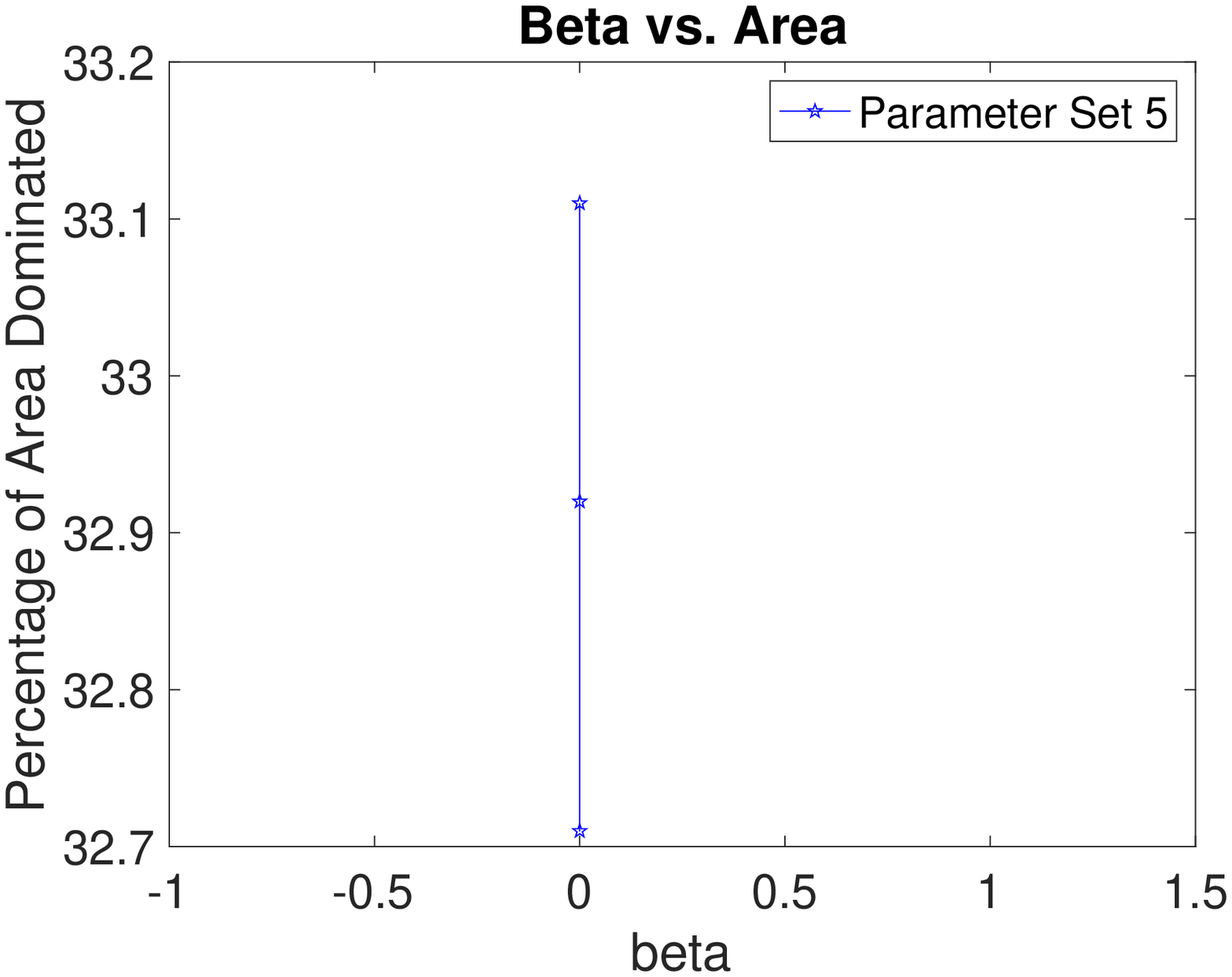}
\includegraphics[width=3.85cm]{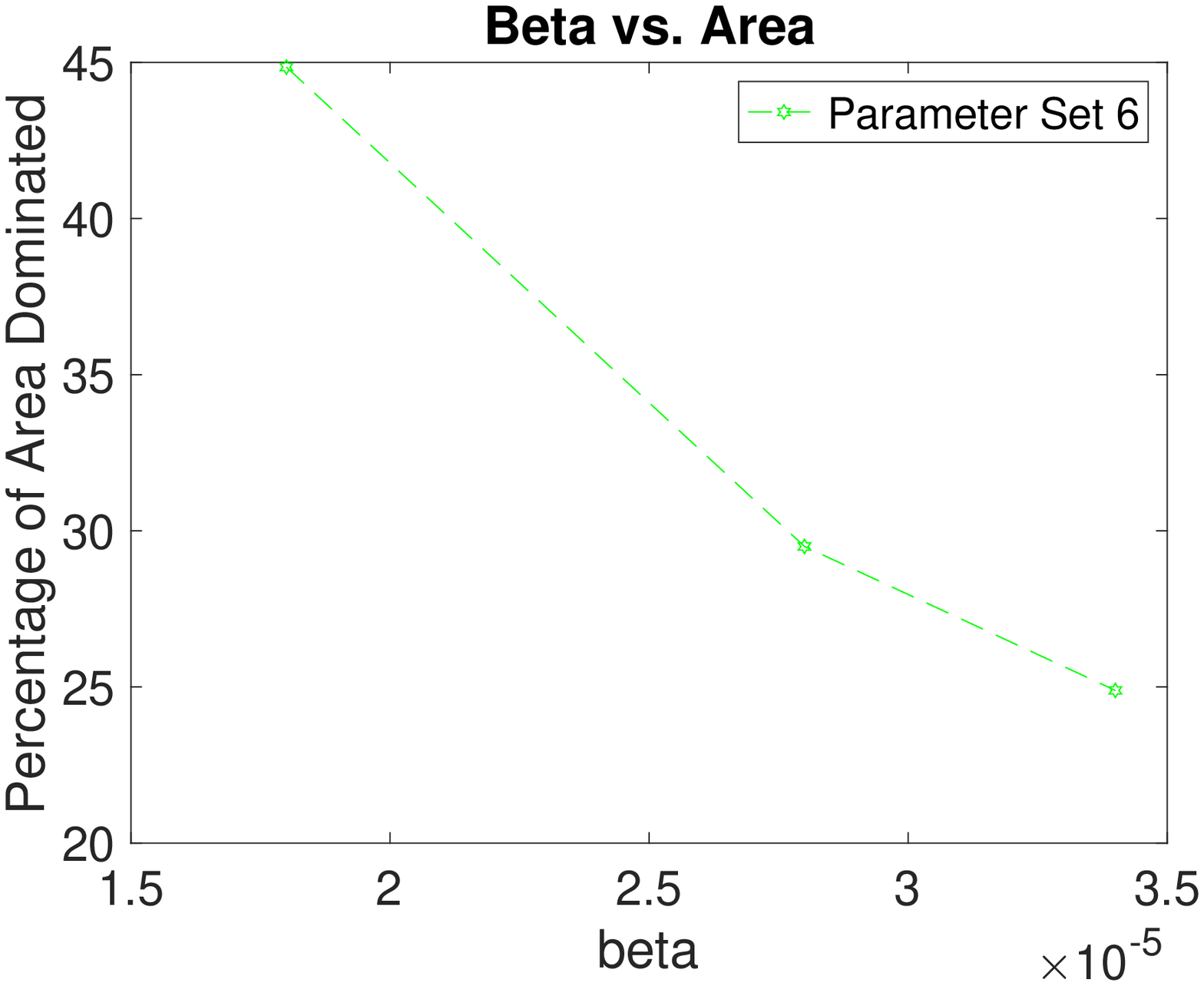}
\caption[Beta_vs_area_model1]{Here, we plot graphs of the beta values $\beta_j$ against the percentage of the area dominated by gang $j$ for the Timidity Model (variation 1). We use six sets of parameters, enumerated in Table \ref{T:paramSets}. }\label{F:betaVsArea_Model1}
\end{figure}

\begin{figure}[!htb]
       \begin{subfigure}[b]{0.495\linewidth}
               \includegraphics[width=5.0cm,,keepaspectratio]{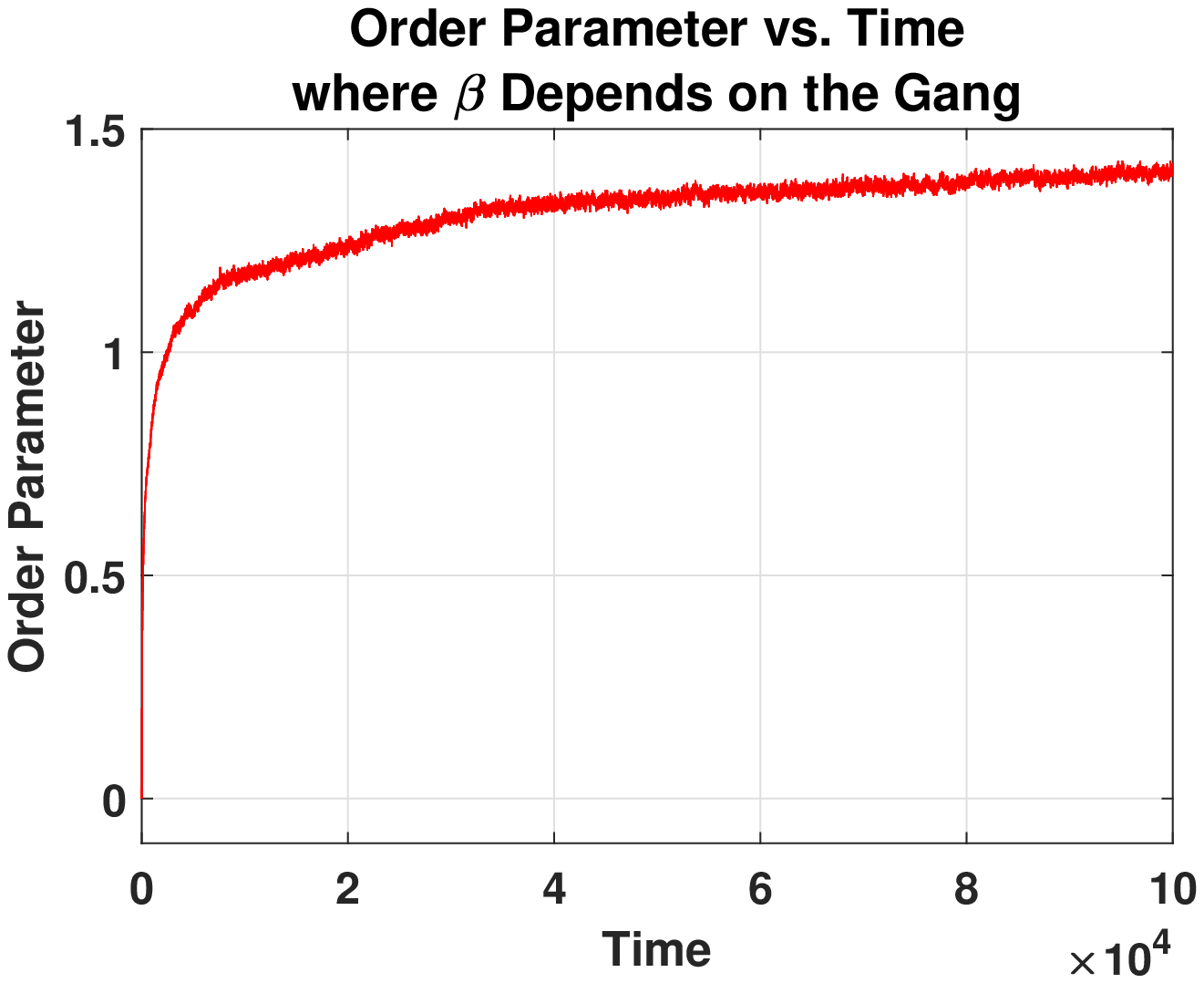}
        \end{subfigure}%
        \begin{subfigure}[b]{0.495\linewidth}
                \includegraphics[width=5.0cm,,keepaspectratio]{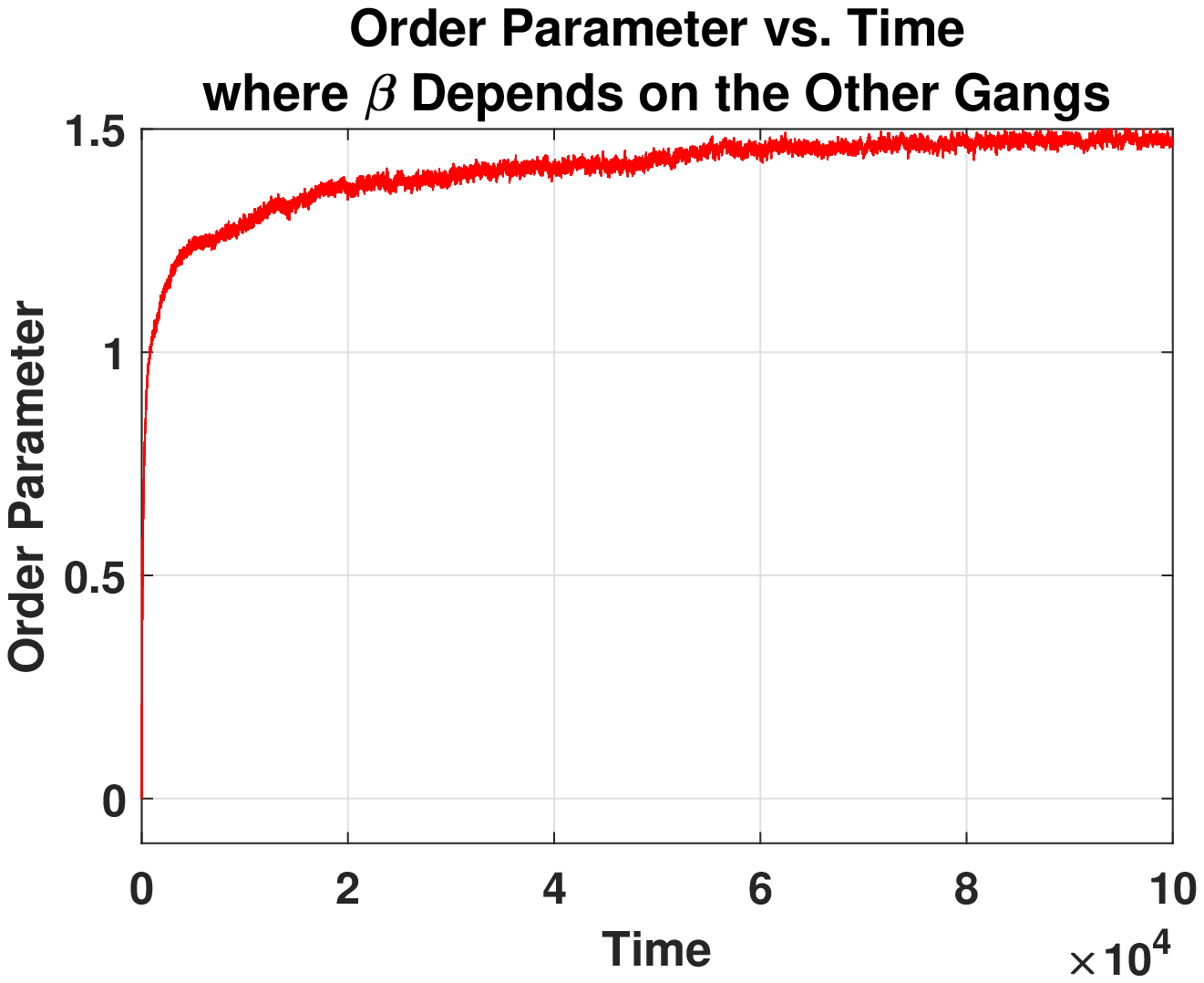}
        \end{subfigure}%
                \caption[Order Parameter for different betas.]{How changing the $\beta$ parameter for different gangs  affects the system. Here we have $N_1 = N_2 = N_3 =50,000$, with $\lambda = \gamma =0.5$ and the lattice size is $100 \times 100$.  We note that we cannot expect the order parameter to approach $1$ even in the case of perfect segregation, since the territories vary in size depending on the values of the $\beta_j$s. \textbf{(Left):} Here, we plot the order parameter for the Timidity Model, using $\beta_1 = 2 \times 10^{-5}, \beta_2 = 3.5 \times 10^{-5}$ and $\beta_3 = 0.5 \times 10^{-5}$. \textbf{(Right):} Here, we plot the order parameter for the Threat Level Model, where the $\beta_i$ parameters represent threat levels; we use $\beta_1 = 2 \times 10^{-5}, \beta_2 = 3.5 \times 10^{-5}$ and $\beta_3 = 0.5 \times 10^{-5}$. As in the original model, we see that our order parameter behaves similarly, increasing quickly before leveling off in both cases as the system segregates.}
        \label{fig:order_parameter_different_beta}
\end{figure}

\begin{figure}[!htb]
        \begin{subfigure}[b]{0.495\linewidth}
               \includegraphics[width=5.0cm,,keepaspectratio]{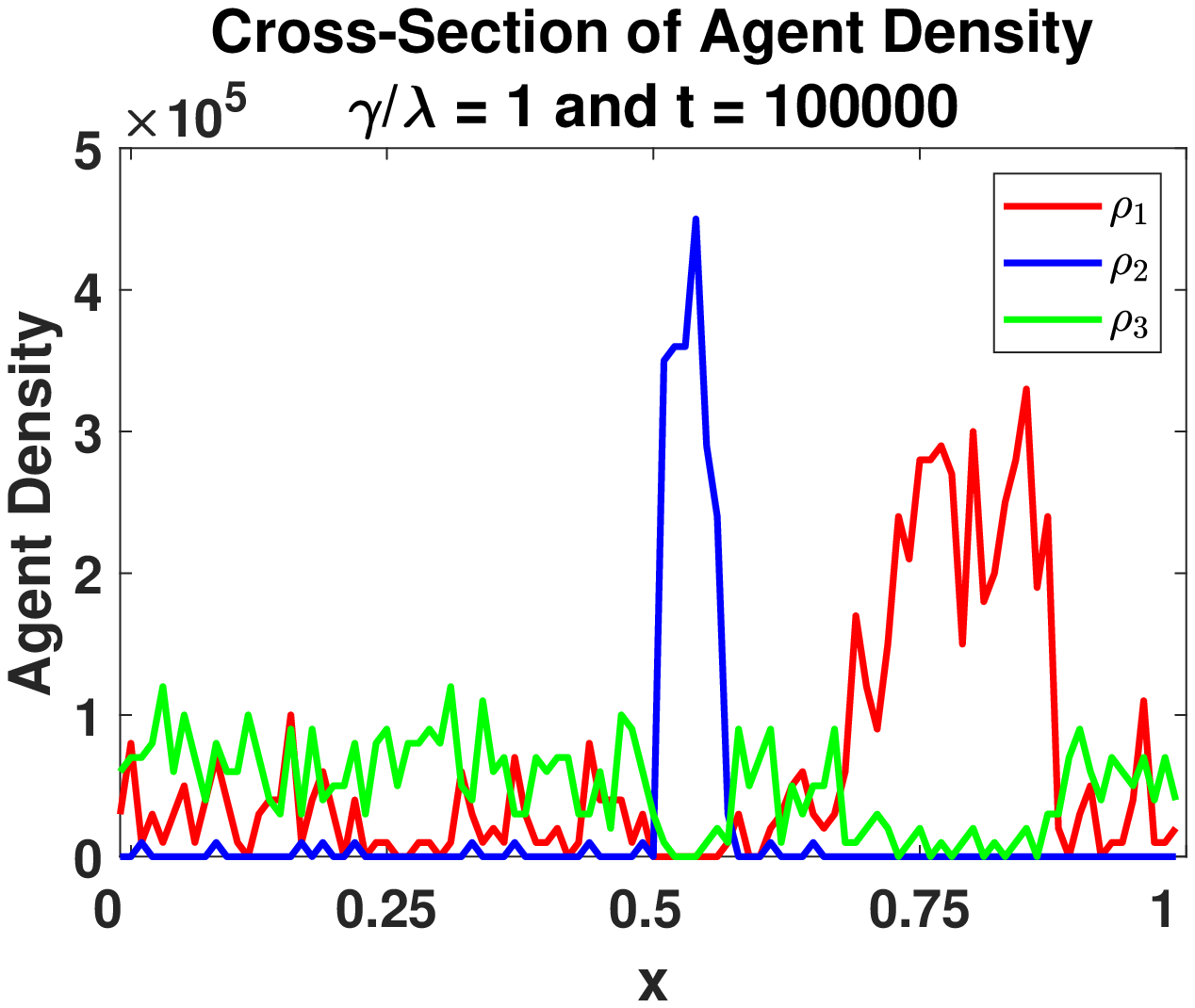}
        \end{subfigure}%
        \begin{subfigure}[b]{0.495\linewidth}
               \includegraphics[width=5.0cm,,keepaspectratio]{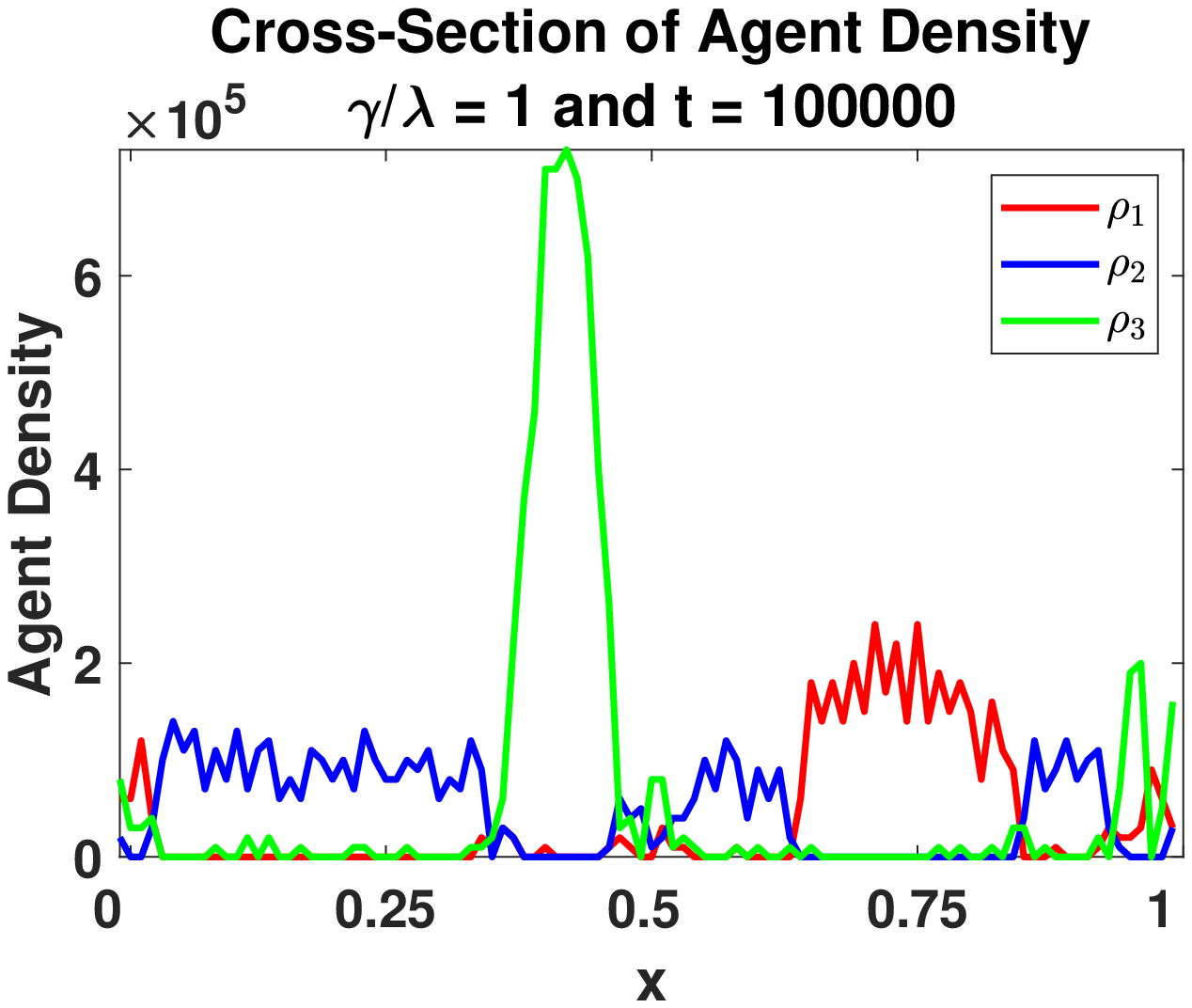}
        \end{subfigure}%
        
        \begin{subfigure}[b]{0.495\linewidth}
               \includegraphics[width=5.0cm,,keepaspectratio]{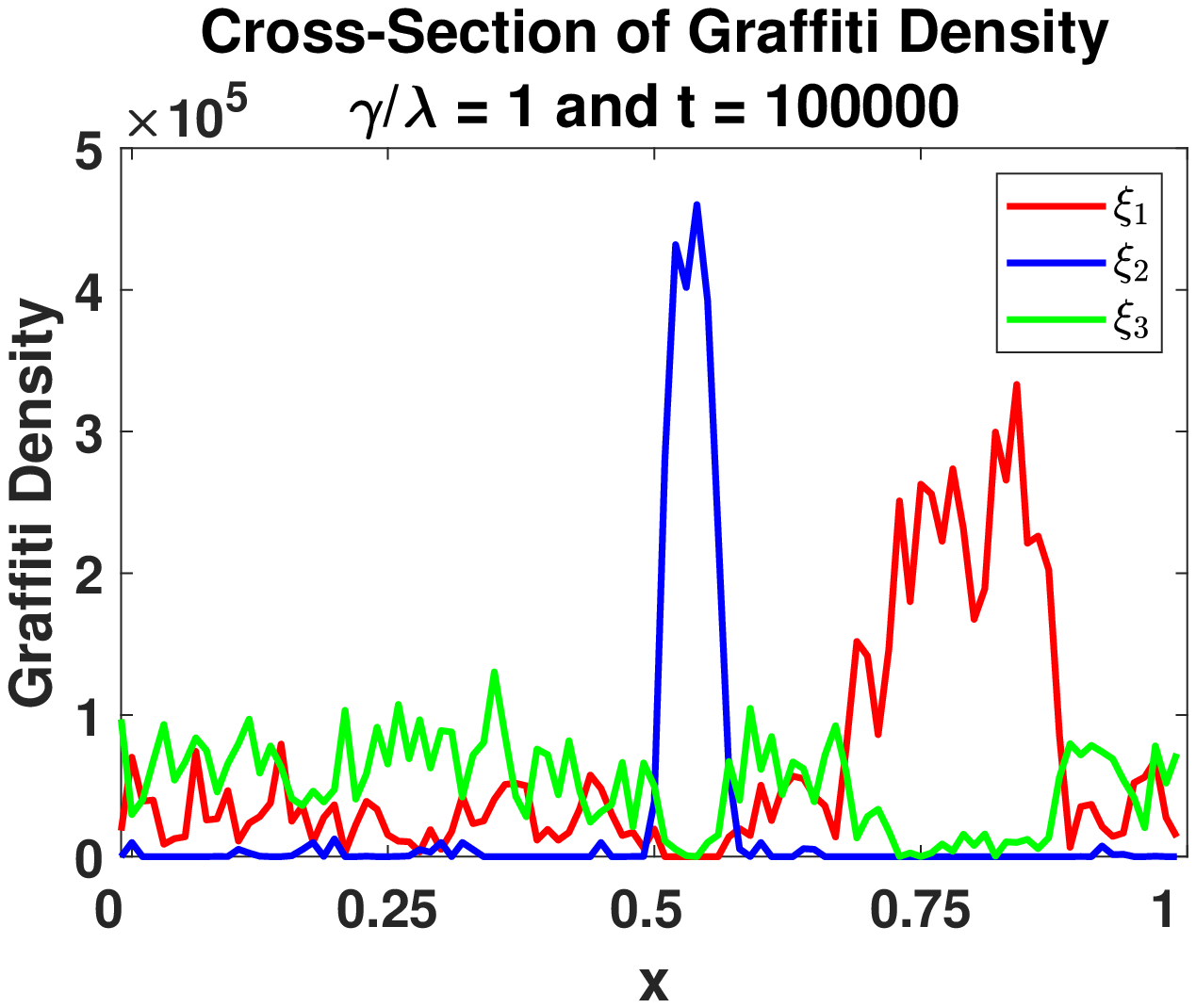}
        \end{subfigure}%
        \begin{subfigure}[b]{0.495\linewidth}
                \includegraphics[width=5.0cm,,keepaspectratio]{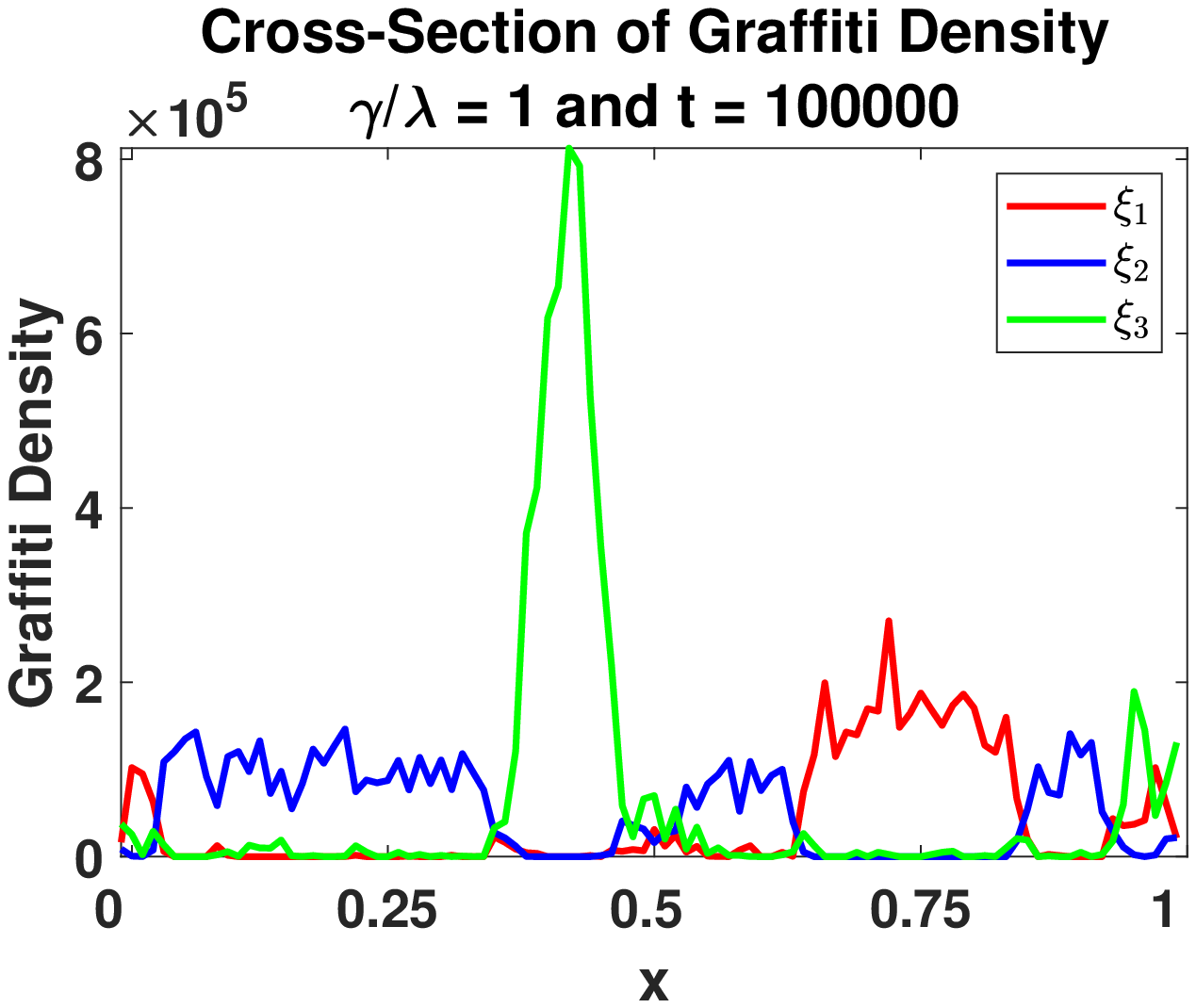}
        \end{subfigure}%
                \caption[beta_extension_slices.]{Cross-sectional slices of the agent and graffiti densities for different $\beta$  extensions at the final time step for a segregated state. Here we have $N_1=N_2=N=3 = 50,000$ with $\delta t = 1$ and the lattice size is $100 \times 100$; in both simulations, $\beta_1 = 2 \times 10^{-5}, \beta_2 = 3.5 \times 10^{-5}$ and $\beta_3 = 0.5 \times 10^{-5}$. \textbf{(Left):} Here, we consider the Timidity variation of the model.  We observe that the territories range from small and very dense, with little incursion from the other gangs, to large and spread out, with other gang members encroaching on the territory, as the gangs' $\beta_j$ value varies from high to low. \textbf{(Right):} Here, we consider the Threat Level variation of the model.  We see that the size of the territory here is correlated with the $\beta_j$ value for the gang, and that all of the territories here seem well-defined, with little of the territorial encroachment seen in the model pictured on the left.}
        \label{fig:beta_extension_slices}
\end{figure}

To test how these changes affect our discrete model, we ran our simulations with three gangs $1, 2$ and $3$; all gangs are assumed to have the identical number of agents $N = 50,000$.  We also assume that the lattice size $L\times L$ is equal to $100 \times 100$, and will use $100,000$ time steps with each step size $\delta t =1$. We assigned $\beta$ values as described above, so that the first gang has $\beta_{1} = 2 \times 10^{-5}$, whereas the second gang $2$ was assigned a larger value of $\beta_{2} = 3.5 \times 10^{-5}$ and the third gang was assigned a low value of $\beta_{3} = 0.5 \times 10^{-5}$. The results of the simulations are presented in Figures \ref{fig:beta_changes_1_Simulations}, \ref{F:betaVsArea_Model1}, \ref{fig:order_parameter_different_beta}, and \ref{fig:beta_extension_slices}, and also in Table \ref{T:paramSets}.

From Figure \ref{fig:beta_changes_1_Simulations}, which shows the temporal evolution of the agent and graffiti densities, we can see that the system does segregate over time, however the segregation differs from the original discrete model simulations in Section \ref{S:Segregated}. We can see that the agents from the gang with the largest $\beta_j$ value, gang $2$, cluster tightly together into small, highly dense spots and do not venture outside these spots.  This is because they are the most strongly avoidant of the other gangs' graffiti, so they are the most timid.  Most of the agents from gang $1$, which has the next highest $\beta_j$ value, also gather into fairly dense groups, motivated by avoiding the graffiti of Gangs $2$ and $3$. However, because $\beta_1$ is less strong than $\beta_2$, a smattering of gang $1$ agents can also be seen spreading roughly evenly over the whole domain aside from the area occupied by gang $2$. The area occupied by gang $2$ is avoided by all other gangs because of the high concentration of graffiti laid down by the strongly localized agents.  Gang $3$'s agents wander more freely but still avoid the areas with denser graffiti,  avoiding gang $2$'s area more strongly than gang $1$'s area due to the higher concentration of graffiti there. But gang $3$'s low $\beta_3$ allows them to spread over much more of the territory, hence dominating more of the lattice than the other two gangs.

Figure \ref{fig:beta_extension_slices} shows cross-sectional slices of the lattice, in order to more clearly show the agent and graffiti density for each gang.  On the left, we see the agent (top) and graffiti (bottom) densities for the Timidity Model. We can again observe that the gang with the highest $\beta_j$ value, gang $2$, has the smallest and densest territory, with a high density of graffiti and little interference from the other gangs inside this territory.  Gang $1$, with the next-largest $\beta_j$ value, has a larger and less distinct territory, with a medium graffiti density, while gang $3$, with the smallest $\beta_j$, is dominating a very large but fairly mixed territory.  We can see agents from all gangs coexisting at different densities in the area dominated by gang $3$ due to the lower graffiti concentration there.

We also use the same order parameter that we employed with the regular discrete model to evaluate this variation on the model, and the results are presented in the plot on the left in Figure \ref{fig:order_parameter_different_beta}. Based on our order parameter, we find that the system does indeed show signs of segregation. We also note that the order parameter does not scale to the value of one; this is because the definition was based on all gangs having approximately the same area in the final segregated state. 

In Table \ref{T:paramSets}, we consider three-gang simulations with six different sets of parameters and tabulate how much of the territory at equilibrium is dominated by each of the gangs.  The $\beta_j$ values are listed in the third column and we focus here on the percentage of the territory is listed in the fourth column (the fifth column contains information on the percentage of territory at equilibrium for the second variation of the model, discussed in the subsequent subsection).  We can see from the table that the percentage of dominated territory has an inverse relationship with the value of $\beta_j$.  

To better examine this relationship, in Figure \ref{F:betaVsArea_Model1}, we plot the $\beta_j$ values against the percentage of territory dominated by the corresponding gang.  We can see that the territory percentage is roughly inversely proportional to the $\beta_j$ value, meaning that, in the parameter regime where territories form, one can expect this model to produce larger territories for those gangs with smaller $\beta_j$.  This is an important feature of this variation at an ecological level.

\begin{table}
\begin{tabular}{ |c|c|c|c|c| }
 \hline
{Parameter set} & Gang & Value of $\beta_j$ & \% Territory, Model 1 & \% Territory, Model 2\\
\hline
\multirow{3}{3em}{Set 1}
& Gang $1$ & $\beta_1 = 0.000005$ & $55.02$\% & $11.27$\% \\ 
& Gang $2$ & $\beta_2 = 0.00002$ & $28.23$\% & $32.48$\% \\ 
& Gang $3$ & $\beta_3 = 0.000035$ & $10.20$\% & $54.10$\% \\ 
\hline
\multirow{3}{3em}{Set 2}
& Gang $1$ & $\beta_1 = 0.000015$ & $41.50$\% & $25.95$\% \\ 
& Gang $2$ & $\beta_2 = 0.00002$ & $31.28$\% & $32.62$\% \\ 
& Gang $3$ & $\beta_3 = 0.000025$ & $25.19$\% & $39.70$\% \\ 
\hline
\multirow{3}{3em}{Set 3}
& Gang $1$ & $\beta_1 = 0.00001$ & $55.13$\% & $16.27$\% \\ 
& Gang $2$ & $\beta_2 = 0.00002$ & $28.40$\% & $28.63$\% \\ 
& Gang $3$ & $\beta_3 = 0.00004$ & $14.15$\% & $53.92$\% \\ 
\hline
\multirow{3}{3em}{Set 4}
& Gang $1$ & $\beta_1 = 0.000012$ & $51.45$\% & $19.39$\% \\ 
& Gang $2$ & $\beta_2 = 0.000024$ & $27.02$\% & $34.85$\% \\ 
& Gang $3$ & $\beta_3 = 0.000032$ & $19.99$\% & $44.58$\% \\ 
\hline
\multirow{3}{3em}{Set 5}
& Gang $1$ & $\beta_1 = 0.000022$ & $33.11$\% & $32.72$\% \\ 
& Gang $2$ & $\beta_2 = 0.000022$ & $32.71$\% & $32.84$\% \\ 
& Gang $3$ & $\beta_3 = 0.000022$ & $32.92$\% & $33.04$\% \\ 
\hline
\multirow{3}{3em}{Set 6}
& Gang $1$ & $\beta_1 = 0.000018$ & $44.85$\% & $23.66$\% \\ 
& Gang $2$ & $\beta_2 = 0.000028$ & $29.50$\% & $34.32$\% \\ 
& Gang $3$ & $\beta_3 = 0.000034$ & $24.89$\% & $41.16$\% \\ 
 \hline
\end{tabular}
\caption{Here, we see the results of both variations of the original model for six different sets of $\beta_j$ in three-gang simulations.  Here, Model 1 refers to the Timidity Model variation, while Model 2 refers to the Threat Level Model variation. The $\beta_j$ values are listed, along with the percentage of the lattice dominated by each gang at equilibrium.  Note that the percentages do not add to $100$\% because in each simulation, a small percentage of the lattice is not clearly dominated by any one of the gangs.}
\label{T:paramSets}
\end{table}

\subsection{Threat Level Model (Variation 2)}

We now consider a different modification of movement dynamics \eqref{E:probability_agent_moves}.  This model is intended to apply in a situation where some gangs are more aggressive or territorial than others. So instead of considering a $\beta$ value which is the same for all gangs, we consider the case where the gangs have varying threat levels. To this end, each gang $i$ has a corresponding threat level encoded by parameter $\beta_i$. This means that gang $j$ will more strongly avoid more threatening gangs, i.e. those gangs with relatively large $\beta$ values.  Based on this, we must modify the opposition sum from equation \eqref{E:graffiti_complement}, so that it becomes

\begin{equation}
\psi_j(x,y,t) := \sum_{\substack{i=1 \\ i\neq j}}^K \beta_i \xi_i(x, y,t).
\label{E:graffiti_complement_beta} 
\end{equation}
Note that the $\beta_i$ parameters can no longer pull out of the sum. The new movement probability then becomes
\begin{equation}
M_{j}(x_1 \rightarrow x_2, y_1 \rightarrow y_2, t) = \frac{e^{- \psi_j(x_2, y_2,t)}}{\sum \limits_{(\tilde x, \tilde y) \sim(x_1,y_1)}e^{-\psi_j(\tilde x, \tilde y, t)}}. \label{E:probability_agent_moves_beta2} 
\end{equation}

Here, every gang then avoids the graffiti of gang $i$ with rate $\beta_i$.  This model applies in the case where the gangs have differing threat levels, so that some gangs are to be avoided more than others. For example, let us suppose that gang $2$ has a large $\beta_2$ value, gang $3$ has a small $\beta_3$ value, and gang $1$ has an intermediate $\beta_1$ value. As $\beta_2$ is large, gang $2$'s territory will be strongly avoided by both gangs $1$ and $3$. Furthermore, since gang $3$ has a small threat level $\beta_3$, its graffiti will not be avoided as much by the other gangs and it will need a higher graffiti density in order to claim territory for itself.

If we follow the same steps used to derive the continuum equations in Section \ref{section:contiuum_background}, now substituting \eqref{E:probability_agent_moves} with \eqref{E:probability_agent_moves_beta2}, it can easily be shown that the resulting system of equations for $j=1,2, \dots, K$ are

\begin{equation}\label{T:continuum_eqns_beta2}
\begin{cases} 
\displaystyle \frac{\partial \xi_j}{\partial t}(x,y,t) = \gamma \rho_j(x,y,t) - \lambda \xi_j(x,y,t) \\
\displaystyle \frac{\partial \rho_j}{\partial t}(x,y,t) =  \frac{D}{4} \nabla \cdot \left[ \nabla \rho_j(x,y,t)  + 2  \left(\rho_j(x,y,t) \nabla \left( \sum_{\substack{i=1 \\ i\neq j}}^K \beta_i \xi_i(x, y,t) \right) \right) \right]
\end{cases}
\end{equation}
\noindent with periodic boundary conditions. Note that the parameters $\beta_i$ now cannot be pulled to the front of the second term of the second equation, and instead must remain inside the sum.

\begin{figure}[!htb]
        \begin{subfigure}[b]{0.249\linewidth}
               \includegraphics[width=1.75cm,,keepaspectratio]{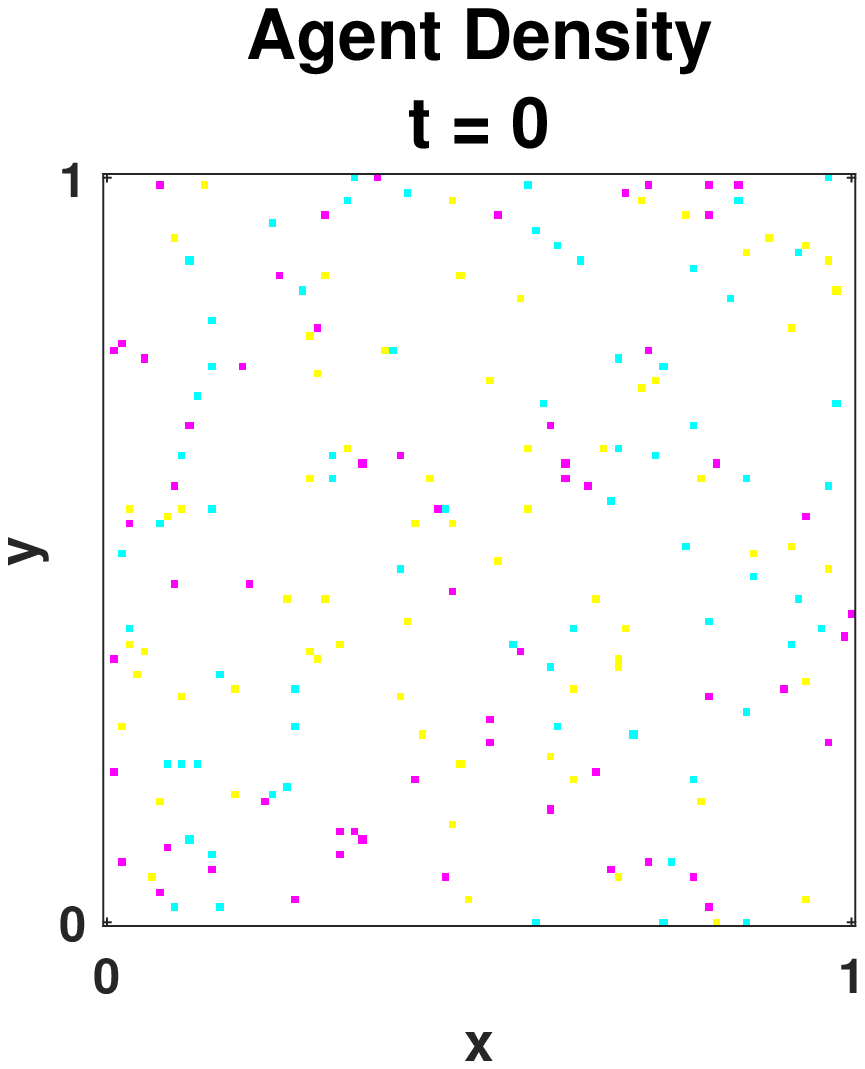}
        \end{subfigure}%
        \begin{subfigure}[b]{0.249\linewidth}
                \includegraphics[width=1.75cm,,keepaspectratio]{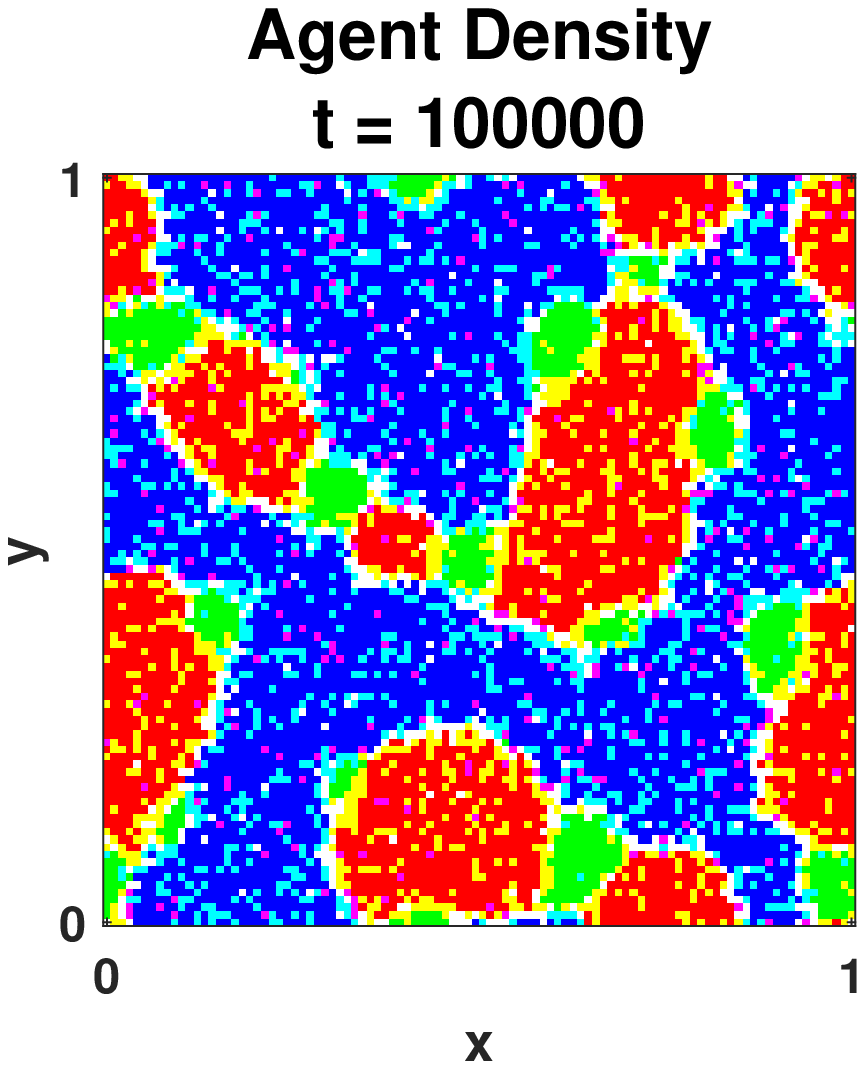}
        \end{subfigure}%
        \begin{subfigure}[b]{0.249\linewidth}
              \includegraphics[width=1.75cm,,keepaspectratio]{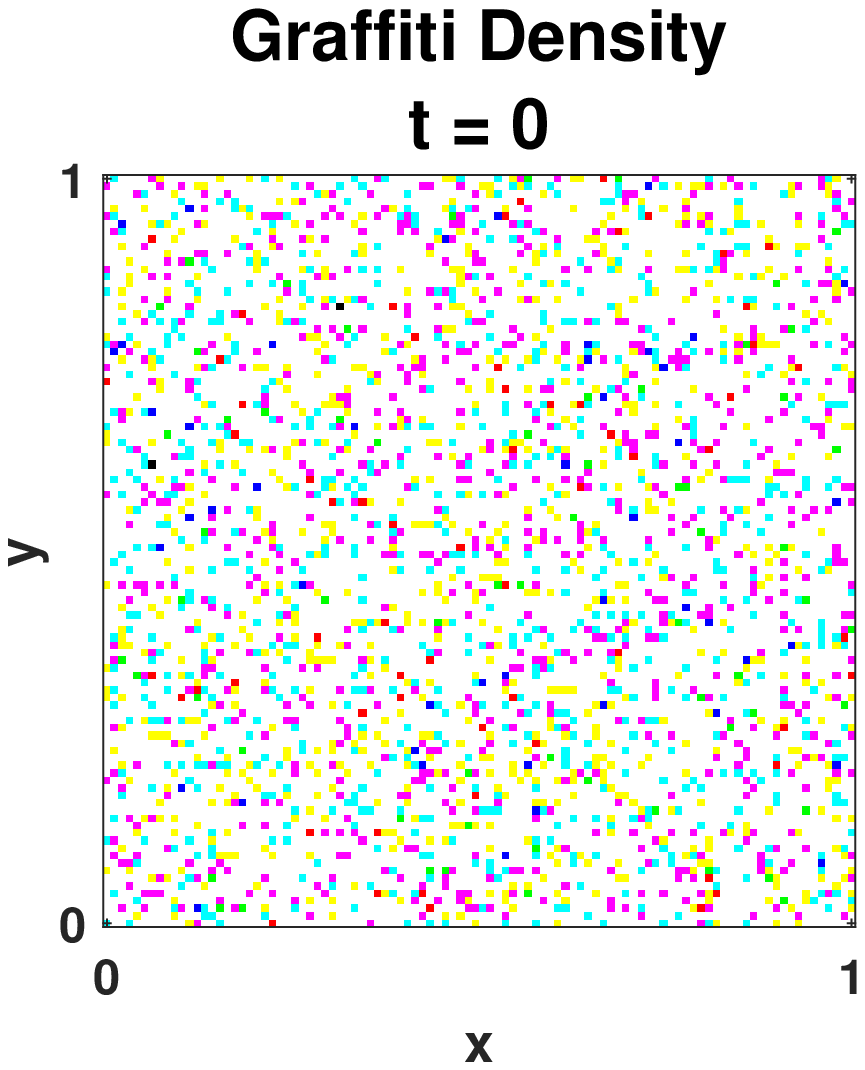}
        \end{subfigure}%
        \begin{subfigure}[b]{0.249\linewidth}
               \includegraphics[ width=1.75cm,keepaspectratio]{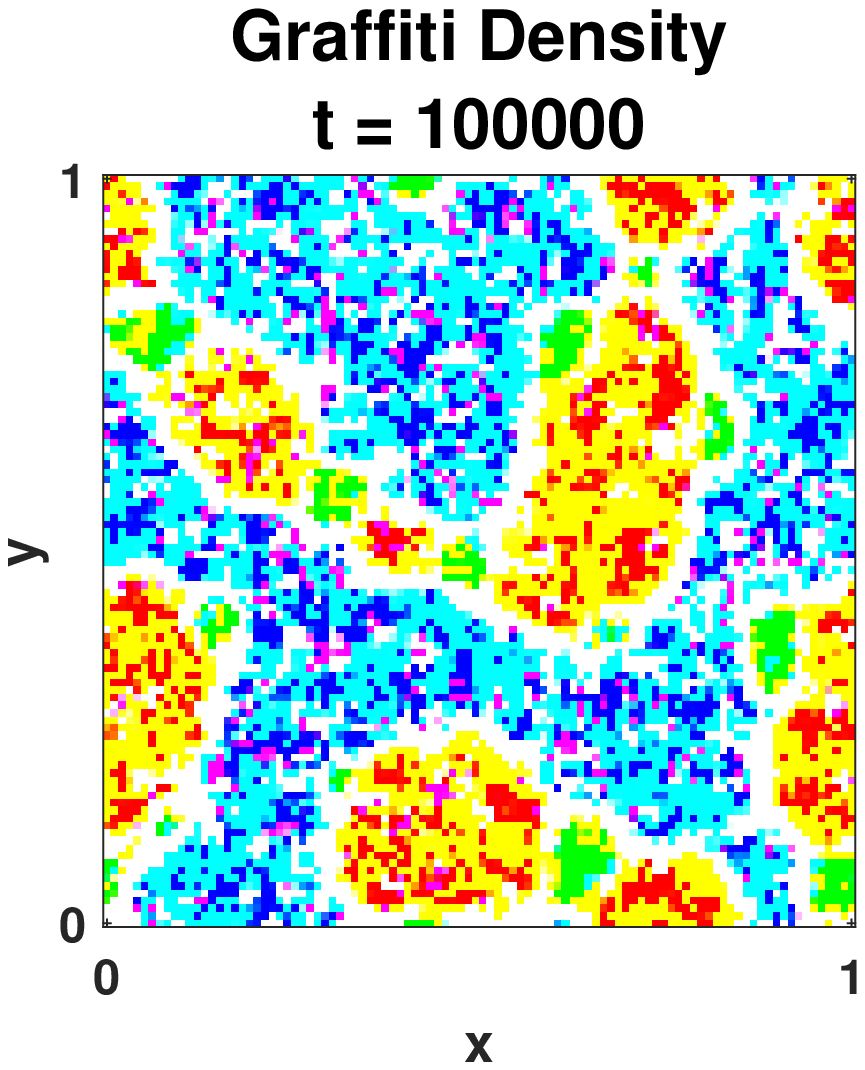}
        \end{subfigure}
        \begin{subfigure}[b]{0.33\linewidth}
               \includegraphics[width=3.05cm,,keepaspectratio]{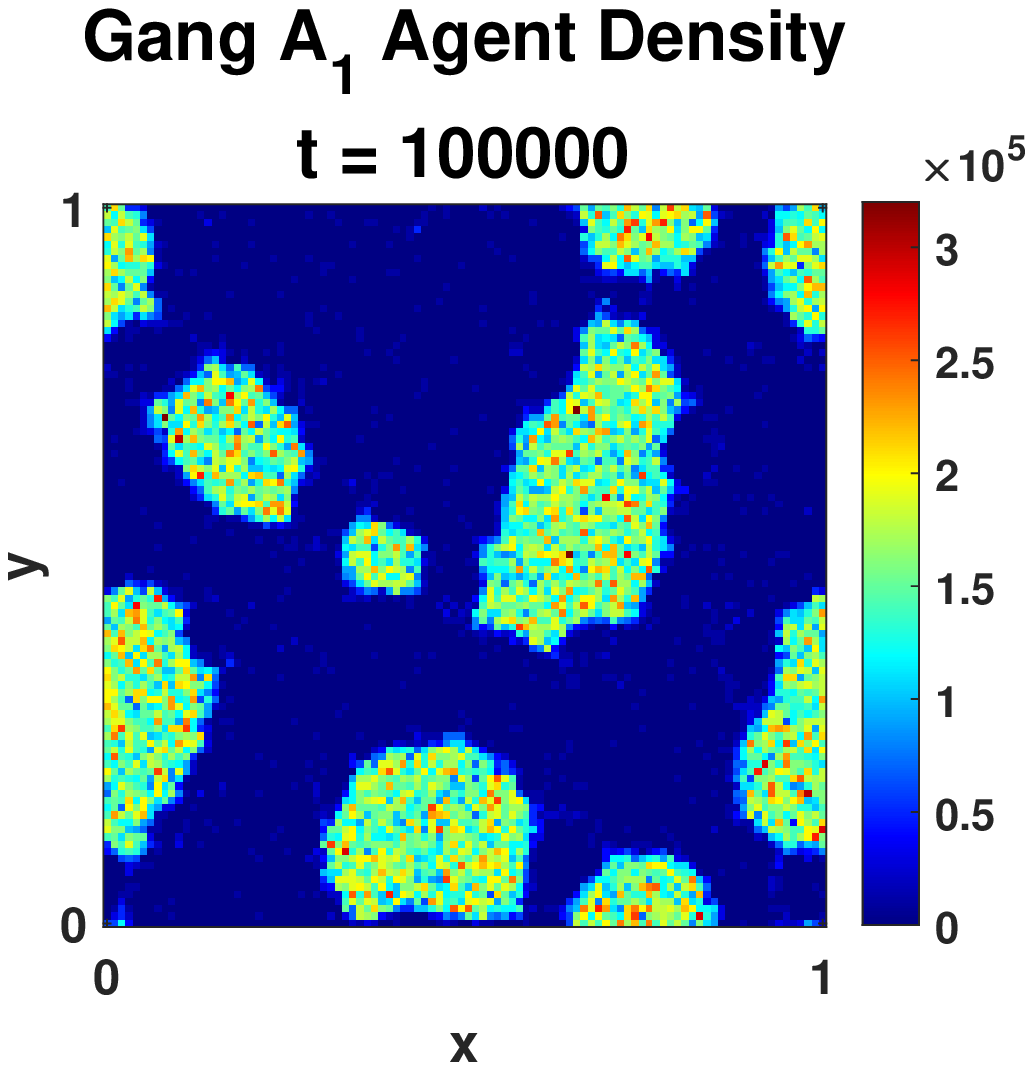}
        \end{subfigure}%
           \begin{subfigure}[b]{0.33\linewidth}
               \includegraphics[width=3.05cm,,keepaspectratio]{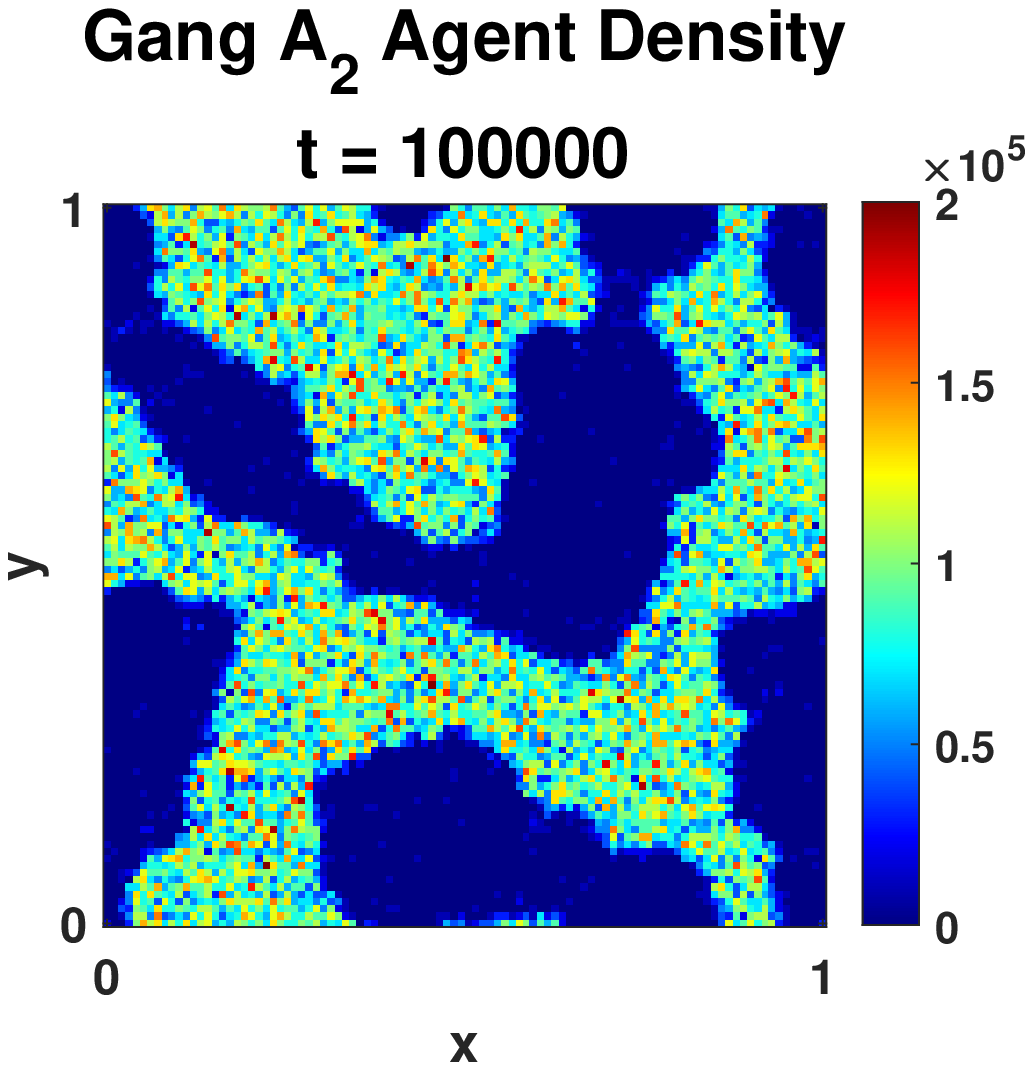}
        \end{subfigure}%
         \begin{subfigure}[b]{0.33\linewidth}
               \includegraphics[width=3.05cm,,keepaspectratio]{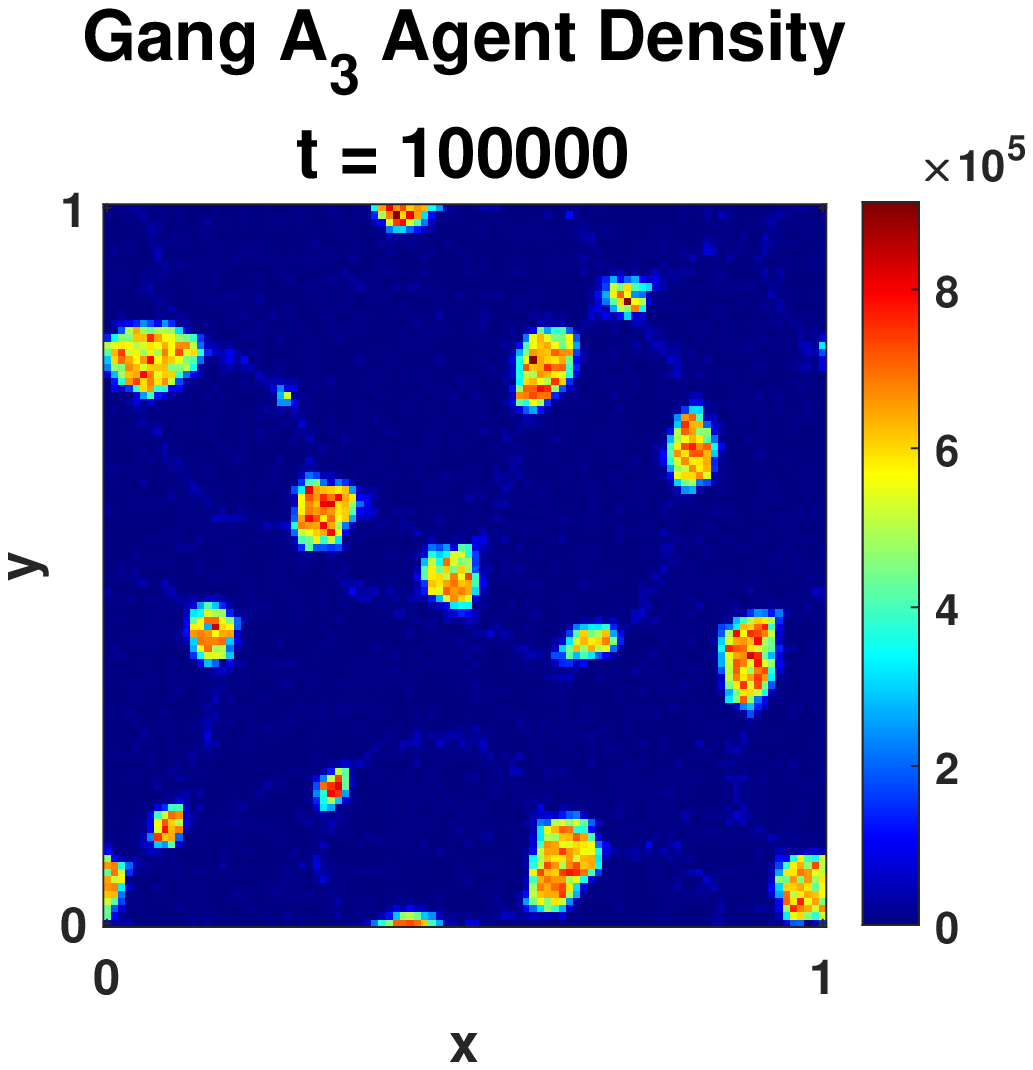}
        \end{subfigure}%
                \caption[Temporal evolution of the agent and graffiti densities lattice for a well-mixed state when $\beta$ depends on the other gangs.]{\textbf{Top row:} Temporal evolution of the agent and graffiti densities for the Threat Level Model. Here $\beta_1 = 2 \times 10^{-5}, \beta_2 = 3.5 \times 10^{-5}$ and $\beta_3 = 0.5 \times 10^{-5}$. We also have $N_1 = N_2 = N_3 =50,000$, with $\lambda = \gamma =0.5$, $\delta t = 1$ and the lattice size is $100 \times 100$. It is clearly seen that the agents segregate over time.  \textbf{Bottom row:} The agent and graffiti densities for gangs 1 (left), 2 (middle), and 3 (right) can be seen after $100,000$ time steps.}           
        \label{fig:beta_changes_2_Simulations}
\end{figure}

\begin{figure}
\begin{center}
\includegraphics[width=7cm]{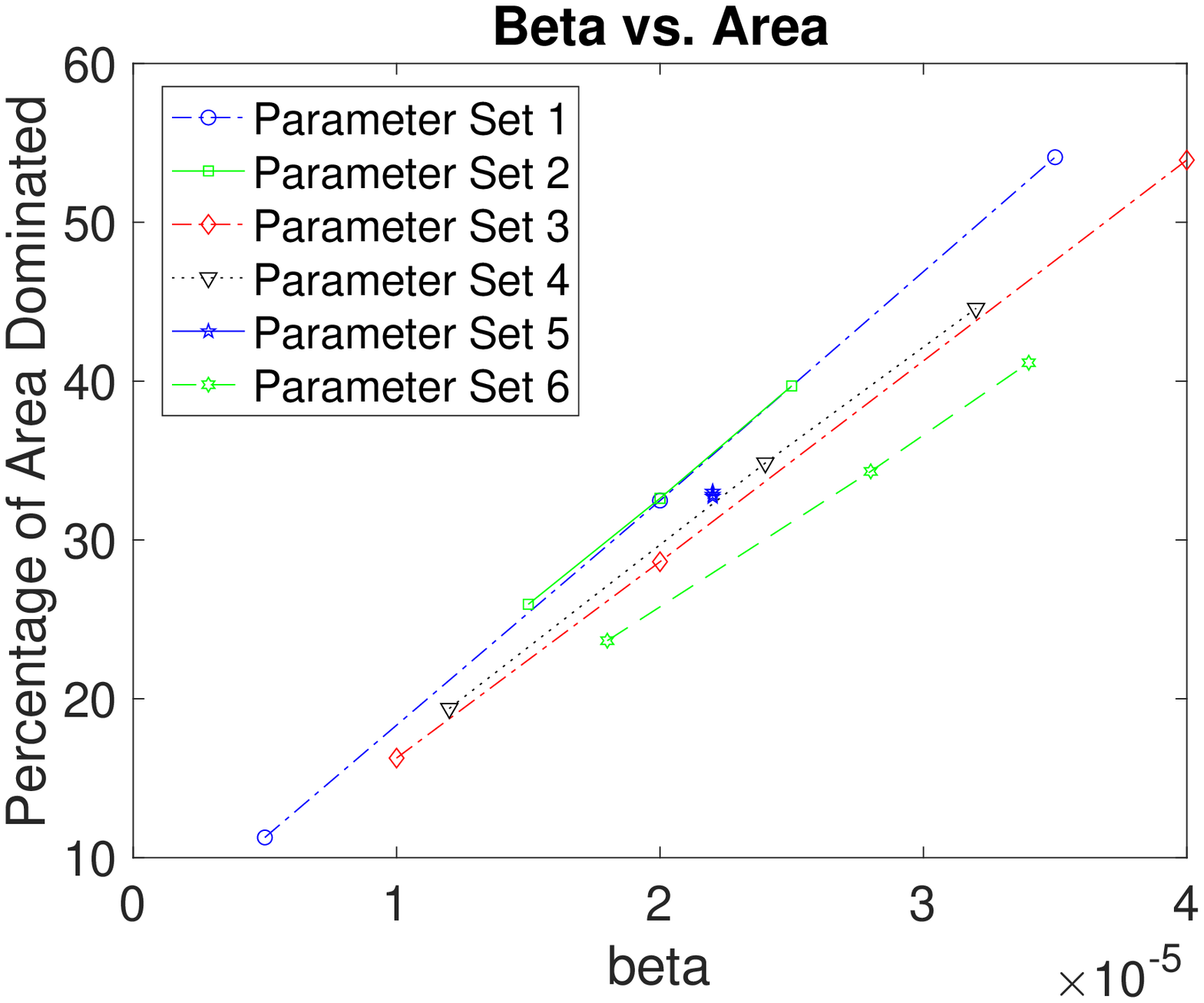}
\end{center}
\includegraphics[width=3.5cm]{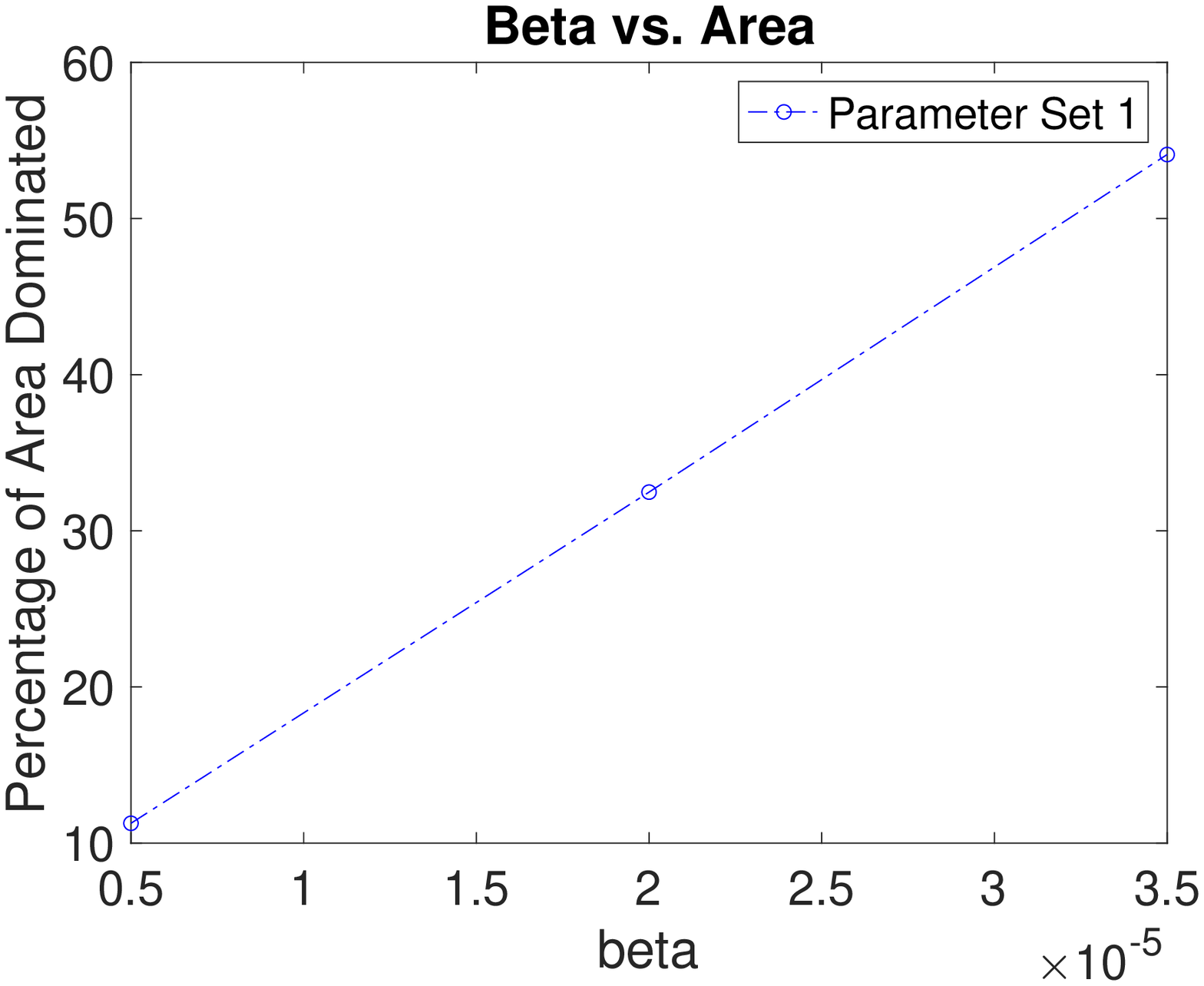}
\includegraphics[width=3.5cm]{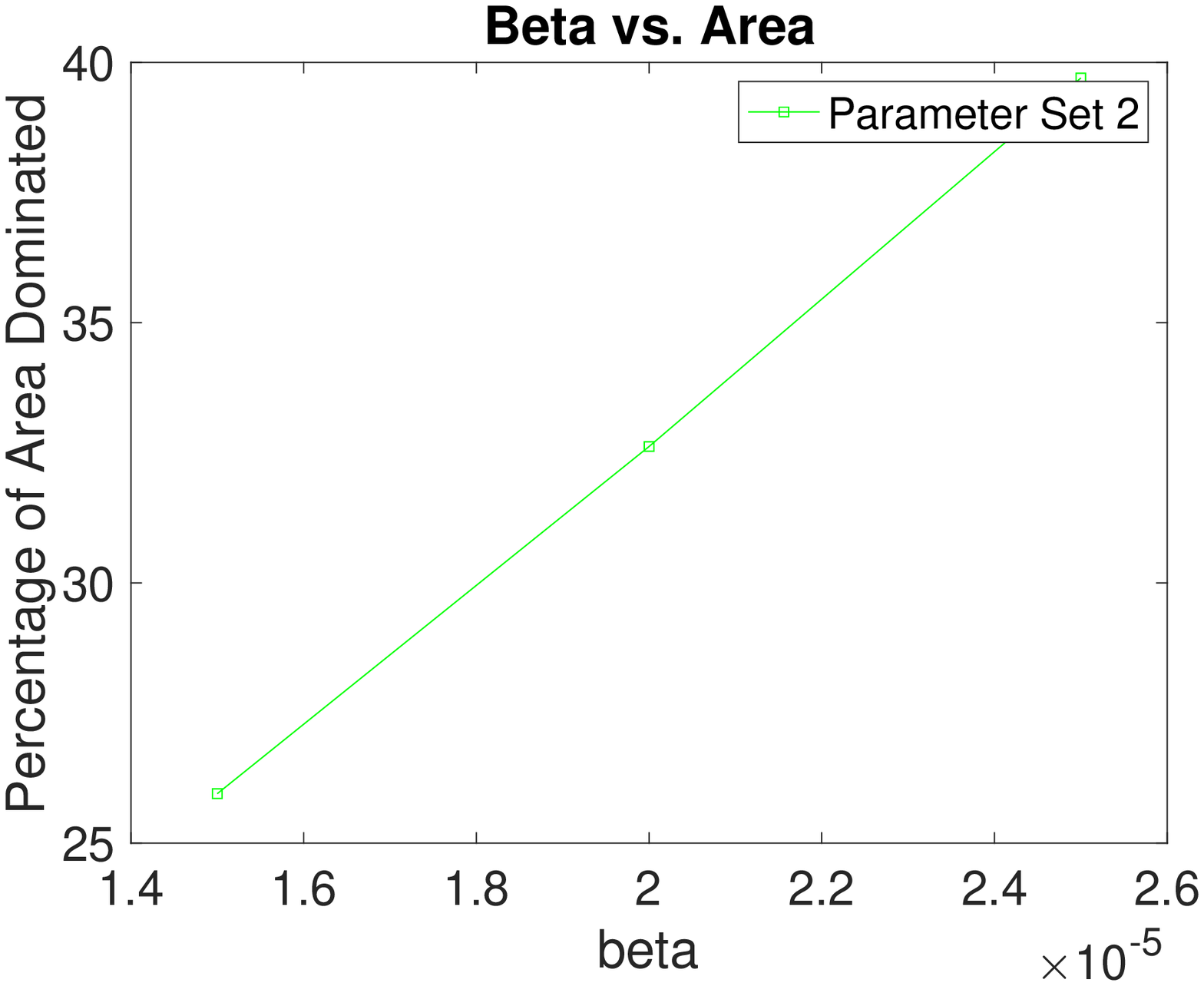}
\includegraphics[width=3.5cm]{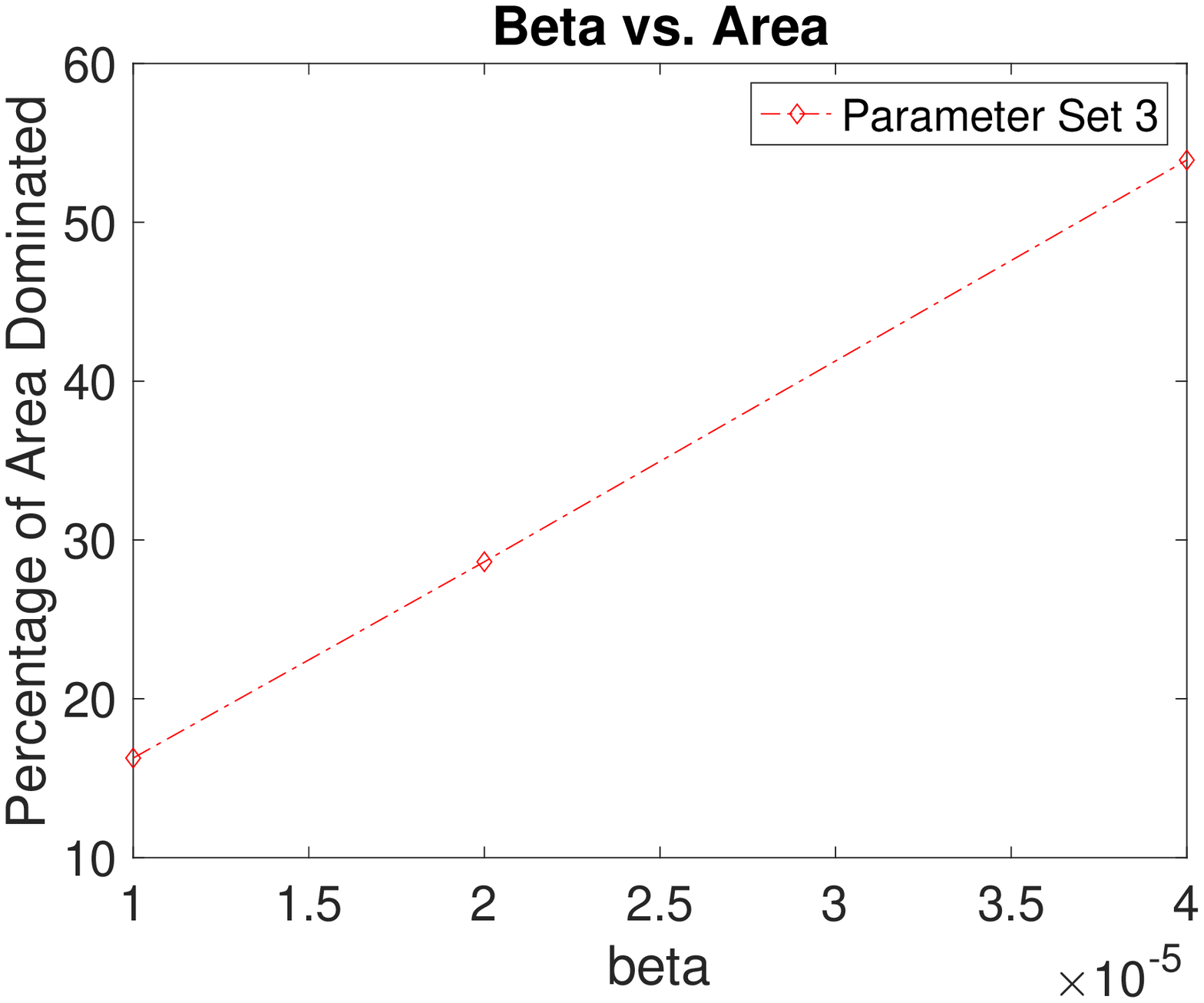}\\
\includegraphics[width=3.5cm]{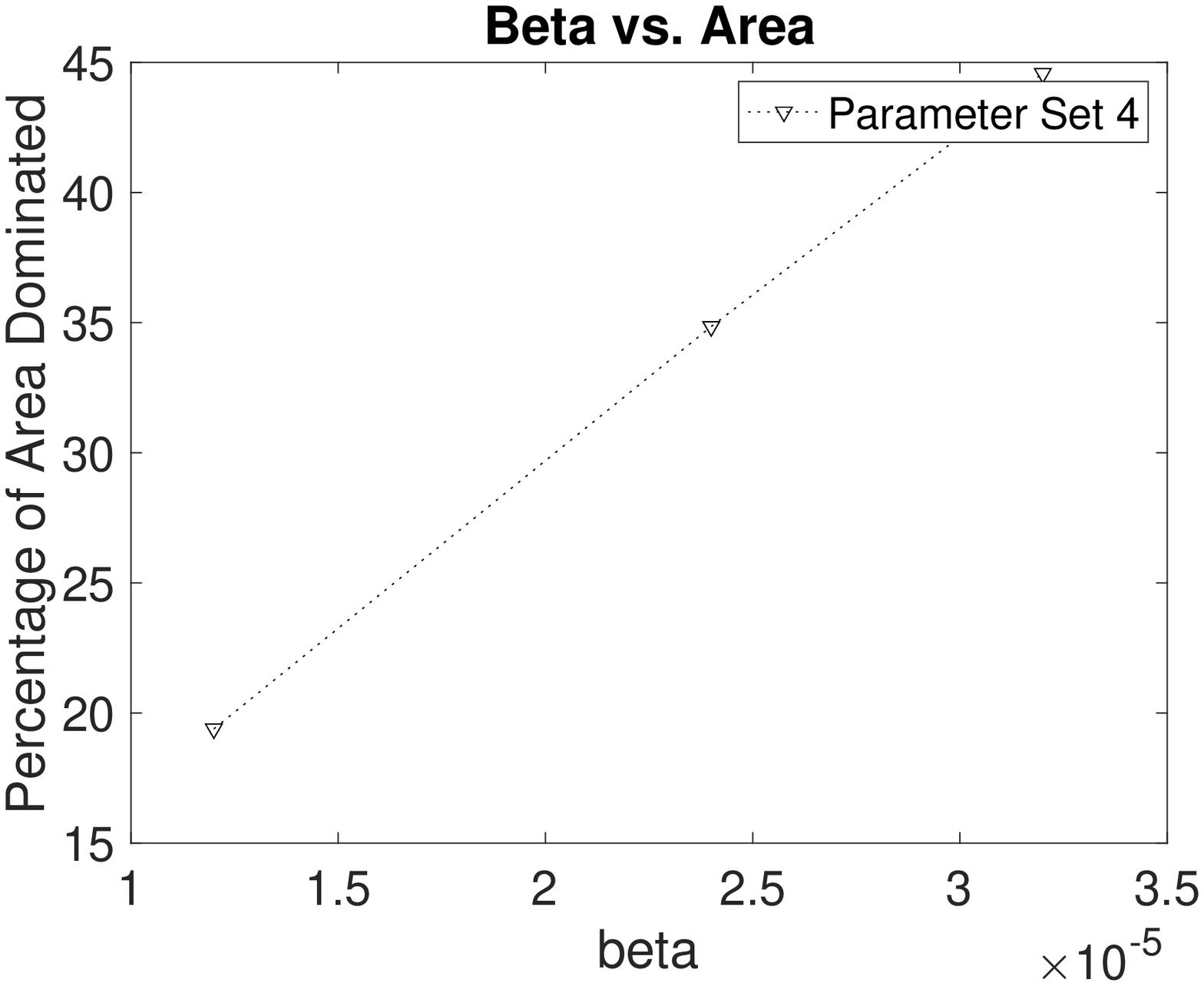}
\includegraphics[width=3.5cm]{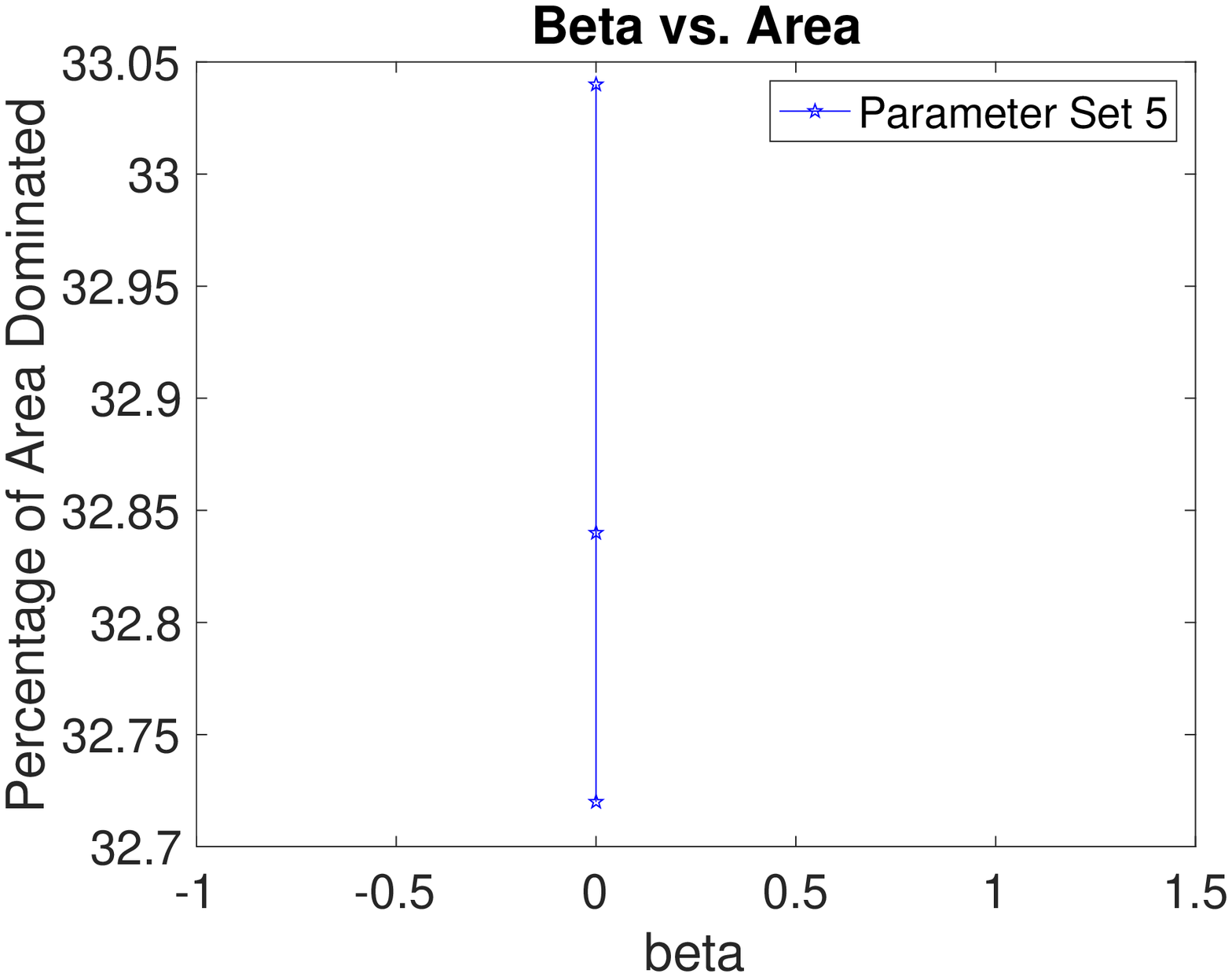}
\includegraphics[width=3.5cm]{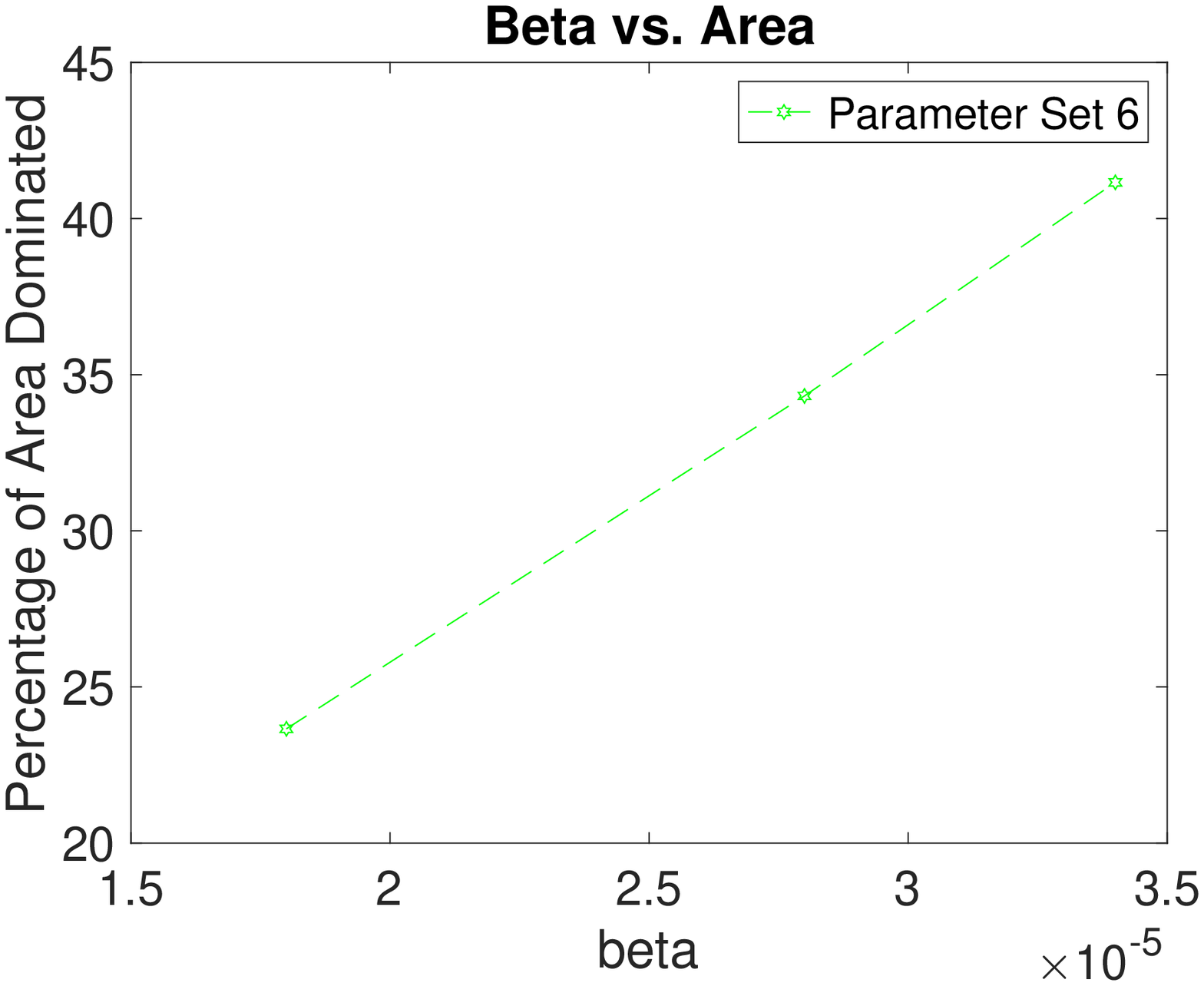}
\caption[Beta_vs_area_model2]{Here, we plot the $beta_j$ values against the percentage of the area dominated by gang $j$ for the Threat Level variation of the model. We use the six sets of parameters enumerated in Table \ref{T:paramSets}. }\label{F:betaVsArea_Model2}
\end{figure}

To test these changes with our discrete model, we ran our simulations with three gangs $1, 2$ and $3$,  where all gangs are assumed to have $50,000$ agents.  We assume that the lattice size $L\times L$ is equal to $100 \times 100$, and use $100,000$ time steps with each step size $\delta t =1$. We assigned the first gang to have $\beta_{1} = 2 \times 10^{-5}$, the second gang $2$ to have a larger value of $\beta_{2} = 3.5 \times 10^{-5}$, while the third gang is assigned a low value of $\beta_{3} = 0.5 \times 10^{-5}$. The results of these simulations are presented in Figures \ref{fig:order_parameter_different_beta}, \ref{fig:beta_extension_slices}, \ref{fig:beta_changes_2_Simulations}, \ref{F:betaVsArea_Model2}, as well as Table \ref{T:paramSets}.

From Figure \ref{fig:beta_changes_2_Simulations}, we can see that the system can segregate over time, in the right parameter regime.  This segregation, however, differs both from that of the discrete model in Section \ref{S:Segregated} and from that of the previous subsection. Here, we see that the gang with the largest $\beta$ value, whose territory appears in blue in the top row of Figure \ref{fig:beta_changes_2_Simulations}, has the largest and least dense territory. This is reasonable since the other gangs avoid the graffiti of gang 1 quite strongly; therefore, the gang does not need to put down as much graffiti to maintain a territory.  They can then spread over more space and still maintain their territory. The gang with the smallest $\beta$ value, on the other hand, whose color is green in the top row of Figure \ref{fig:beta_changes_2_Simulations}, clearly has the smallest and most dense territory. This makes sense, since the other gangs are not avoiding the territory of gang $3$ very strongly; gang $3$ then has to put down a much higher density of graffiti to force the other gangs to avoid it, and it can only do this by limiting its gang members to a smaller area.

We also tested segregation using the same order parameter to that we used previously; the evolution of the order parameter for this model is presented on the right in Figure \ref{fig:order_parameter_different_beta}. We note that, as in the previous variation of the model, we can no longer expect the order parameter to tend to $1$ in the fully segregated case.  However, we do still see segregation over time.

Figure \ref{fig:beta_extension_slices} shows cross-sectional slices of the lattice, to show the agent and graffiti density for each gang.  On the right, we see the agent (top) and graffiti (bottom) densities.  From this figure, we can see that the territories formed in this variation are much more distinct than in the last variation; there is very little overlap inside the territories.  This is in contrast to the first variation on the model.  We can also observe that the $\beta_j$ value seems to be proportional to the territory size. Traveling outside an agent's own territory seemingly happens only along the boundaries of other gangs' territories.

In Table \ref{T:paramSets}, as described in the previous subsection, we see the results of this model run with three gangs. We ran the simulation with six different sets of $\beta_1, \beta_2,$ and $\beta_3$ and, in the right-hand column of the table, we see the percentage of the lattice occupied at steady-state by each of the three gangs.  We can see that in this variation of the model, in contrast to the last variation, the size of the territory in each simulation seems to be directly proportional to the values of $\beta_i$.  

We further examine this result in Figure \ref{F:betaVsArea_Model2}, where we plot the values of $\beta_i$ for each simulation against the percentage of the lattice occupied by each of the gangs.  We see in this figure that the $\beta_i$ and the percentage of occupied areas are indeed very nearly directly proportional. It is an interesting open question why this is the case.

\section{Discussion} \label{section:conclusion}

In this work, we have presented an extension of a previous agent-based system that models gang territorial development ``motivated'' by graffiti tagging \cite{alsenafi2018convection} to now include a finite number $K$ of gangs as opposed to only two. In the special case of three gangs, we have shown by using numerical simulation that our model also undergoes a phase transition as we change the value of different parameters. We formally derived the continuum limit for our model, giving us a set of $2\times K$ convection-diffusion equations with cross-diffusion. By using linear stability analysis on the continuum equations,  we showed that there is a bifurcation point in which the well-mixed state becomes linearly unstable.  Furthermore, we have numerically shown that the bifurcation point matches the critical parameter found in the numerical simulations for the case of $K=3$ for the discrete model.  This generalization from two to $K$ gangs makes the model much more flexible.  In the form presented in this paper, the model can be applied to many coexisting gangs or many packs of animals, and this is important in practice, since it can rarely be assured that there are only two. 

We have also presented two novel variations of the model, each of which exhibits different segregation dynamics from the original model and from the other variation. These variations allow for further flexibility.  For the Timidity model (variation 1), each gang is allowed a different value of the $\beta$ parameter, allowing some more timid gangs (with large $\beta$) to be more sensitive to the existence of graffiti and some (with small $\beta$) to be less sensitive.  Assuming the gangs have identical membership, this resulted in the more timid gangs having smaller and more distinct territories, while the less timid gangs had larger and less distinct territories where members of other gangs were also occasionally present.  For the Threat Level model (variation 2), each gang $i$ has a threat level $\beta_i$ associated to their graffiti, so that other gangs react more strongly to the graffiti of gangs with a large $\beta_i$ and less strongly to those with a small $\beta_i$.  When gangs have identical membership, this variation results in larger territories for gangs with higher threat level $\beta_i$ and smaller territories for gangs with lower threat levels.  In contrast to the Timidity model, all of the territories are distinct, with very little overlap from other gangs' agents.  These two variations could prove useful in ecological applications where more is known about the traits of the groups.

The model is also intriguing from the perspective of pattern formation.  The segregation dynamics for the system with constant $\beta$ and the two variations give three different dynamics for the territory formation.  These new models open the possibility of further studies, such as comparing pattern formation with similarly segregating systems such as Cahn-Hilliard \cite{cahn1958free}. Additionally, this model exhibits a phase transition from non-segregating populations to segregating populations as $\beta$ changes; it is highly likely that a phase transition would also occur as $\lambda$ increases.  An open problem with significant ecological consequences would be to look for this phase transition, since it would provide an indication that climate change, in particular increased precipitation, could have an effect on the territorial dynamics for animals such as wolves and coyotes.  

The system of PDEs derived in this paper also are interesting in their own right.  The form is reminiscent of Patlak-Keller-Segel model \cite{patlak1953random, keller1970initiation}, with chemo-repellent rather than chemo-attractant and no diffusion of the chemical. The graffiti densities evolve in response only to the agent and graffiti densities of the corresponding gang, while the agent densities evolve only in response to the corresponding gang's agent density and the graffiti densities of all the other gangs.  This leads to a system's cross-diffusion form. Originating in spatial ecology \cite{morisita1950population, morisita1952habitat, gurtin1984note}, cross-diffusion is widely recognized as a mechanism for pattern formation \cite{vanag2009cross}.  Recent interest in cross-diffusion has led to advances in analytical understanding of these systems \cite{burger2020segregation, di2018nonlinear, carrillo2018zoology, bruna2017cross}.  Since this paper offers three variations on a novel cross-diffusion system, new avenues are opened for further numerical and analytical study to better understand the properties and behavior of these systems, such as the analytical work done on the two-gang system \cite{BRYZ2020}.

\section{Acknowledgements}
The authors would like to thank Nancy Rodriguez, Havva Yoldas, and Nicola Zamponi for helpful discussions of the original model upon which this paper is based.

\clearpage

\bibliography{references}
\bibliographystyle{unsrt}

\clearpage

\end{document}